# A megathrust earthquake genesis process observed by a Global Positioning System

Fumihide Takeda[1, 2,*]

[1] Takeda Engineering Consultant Inc., Hiroshima, Japan

[2] Earthquake Prediction Institute, Imabari, Japan

[*] Email: f_takeda@tec21.jp

Japan's GPS observed the fifteen-month megathrust earthquake genesis process of the 2011 Tohoku M9, suggesting the real-time predictability of such events.

## Content

# 1 Abstract

This article presents updated Global Positioning System (GPS) observations on the 2011 Tohoku M9 earthquake and tsunami genesis processes, building upon a megathrust earthquake prediction patent. Tohoku is in the subduction zone at the eastern edge of an overriding continental plate. The daily displacement observed at every GPS station in Tohoku and the Northwest Pacific Ocean is inherently noisy and non-differentiable in time due to various environmental factors, specifically, the daily vertical displacement noise level of ± 20 mm. By using mathematical operators called Physical Wavelets, we can analyze the GPS displacements and observe the genesis process of the 2011 Tohoku megathrust earthquake. The observations reveal that the earthquake resulted from complex interactions between the subducting and overriding plates, causing crustal bulge-bending deformation of the Tohoku crust. The bulge-bending deformation refers to the gradual bending of the Tohoku crust by the continental plate-driving eastward force over fifteen months, causing the crust to bulge downwards and upwards by a few millimeters. This process reveals three distinctive phases that evolved from the regular deformation expected for the past three hundred years: the initial, transitional, and final.

1. The initial phase of bulge-bending deformation began in January 2010 and lasted for six months, during which the bulge over Tohoku gradually subsided by 1 to 2.8 mm.

2. In June 2010, the east coast entered the transition phase with a further gradual subsidence of 3.3 mm. After a month, the east coast began to pull the subducting Pacific Plate westward, nine months before the M9 earthquake on March 11, 2011. The entire Tohoku in this phase experienced a subsidence of 0.6 to 3.3 mm over six months, with the motion gradually gaining westward speed. The speed reached an abnormal 0.69 mm/day on December 22, 2010, three times higher than usual. Approximately one month before reaching the highest speed observed, the east coast began the final deformation phase, with an upheaval growth.

3. During the final phase, the Tohoku region experienced an upheaval growth of 1.2 to 2.5 mm over three months, generating the lifting force along the Tohoku subduction zone. The east coast underwent an upheaval growth of 1.2 mm over 115 days while decelerating, stopping, and two weeks before the event reversing the pulling action of the subducting plate's westward motion. The buildup of lifting force on the east coast gradually reduced the static frictional strength of the subduction interface, eventually causing the shear stress to exceed the frictional strength and leading to the decoupling of the overriding and subducting plates.

The lifting force on the entire Tohoku region contributed to the decoupling process, releasing a massive recoil force of the compressed west coast against the eastward-plate-driving force. This recoil rapidly restored the bulge-bent deformation of the east and west coasts, which had been elastically compressed by the overriding plate-driving eastward force. This restoration process led to the Tohoku M9 earthquake and tsunami.

It is important to note that the continental plate-driving eastward force bulge-bent the east coast westward across Tohoku, not the subducting plate-driving westward force. Thus, the Tohoku M9 earthquake and tsunami were not caused by the commonly suggested elastic rebound of the east coast compressed by the subducting plate-driving force coupled with the over-riding plate through the fault.

The fifteen-month genesis process of the megathrust earthquake and the last three-month tsunami genesis process suggest that such events can be predicted in real-time.

As of May 22, 2021, Tohoku is still experiencing the effects of the 2011 M9 event. The dynamic state of the crust indicates that the slow restoring process to the expected crust state, free of deformation, is ductile. By monitoring the dynamic states using Physical Wavelets, we can predict when the ductile restoring process will end and when the subsequent regular deformation will begin.

As of March 11, 2023, there are currently no signs of an imminent megathrust event in Japan's subduction zones based on the Pacific Plate motions observed at the Chichijima-A and Hahajima GPS stations, as well as the Philippine Sea Plate observed at the Minamidaitojima GPS station. However, effectively communicating the risks and potential impacts of future megathrust earthquakes and tsunamis to local communities and decision-makers remains a significant challenge for disaster mitigation efforts.

## 2 Summary

This article presents updated Global Positioning System (GPS) observations on the 2011 Tohoku M9 earthquake and tsunami genesis processes, building upon the megathrust earthquake (EQ) prediction patent [1]. The update suggests that the fifteen-month EQ genesis process can predict an anticipated megathrust EQ and tsunami [2] in real-time, mitigating the impact of disasters.

Japan has 1,300 GPS stations [3] and a 20 km-mesh Seismograph Network [4], which covers the entire Tohoku at the eastern edge of a continental plate, overriding the subducting northwestern Pacific Plate. The networks recorded the Tohoku M9 EQ on March 11, 2011, showing the geographical distribution of the foreshocks, aftershocks, and vertical co-seismic displacements in Fig. 3. The distribution estimates the Tohoku EQ's fault size as 500 km in length and 200 km in width.

The GPS observations on the 2011 Tohoku M9 EQ are the daily displacements at the GPS stations in Tohoku and the Northwest Pacific Ocean. Each displacement time series is inherently noisy due to various environmental causes [3]; specifically, the daily vertical displacement noise level is ± 20 mm. Thus, the daily displacement series is non-differentiable in time. Physical Wavelets (Appendix A) are mathematical operators used to define the equations of motions for such non-differentiable time series data to analyze the paths in the displacement and velocity phase plane, having a displacement resolution of 0.1 mm, four orders of magnitude greater than the daily vertical noise level of ± 20 mm, and the rate change resolution of 0.01 to 0.0001 mm per day. The paths in Fig. 6-1 to Fig. 7-3 suggest that a cross-sectional deformation process at the northern latitude of about 38 degrees in Fig. 3 is the simplified schematic of Fig. 4, illustrating the genesis process of the 2011 Tohoku EQ.

Figure 4a illustrates the regular deformation caused by tectonic plate-driving forces, resulting in Tohoku's west and east coasts moving upward and downward, respectively. This deformation has been expected for the past three hundred years, as discussed in section 7.4. The west coast has been experiencing an upward deformation rate of 1.5 mm per year, while the east coast has been undergoing a subsidence rate of 6 mm per year. These rates were expected over an approximate distance of 500 km along the shores of Tohoku. Over ten years, the subducting northwestern Pacific Plate moved westward at an average rate of 0.1 mm per day and an average northward movement of 0.03 mm per day. The rate of westward movement is approximately equivalent to the growth rate of a fingernail.

Bulge-bending deformation refers to the gradual bending of the overriding plate of Tohoku by the eastward continental plate-driving force over fifteen months, causing the over-riding crust of Tohoku to bulge upwards and downwards by a few millimeters. This bulge process is a transition from the expected regular deformation, as shown in Fig. 4a. The GPS observation detected the onset of the bulge-bending deformation, which grew in three distinct phases: an initial phase of 6 months with gradual subsidence of 1 to 2.8 mm across the Tohoku region, followed by a transition phase of 6 months with a further gradual subsidence of 3.3 mm on the east coast, and a final phase of 3 months with an upheaval growth of 1 to 3 mm across Tohoku until the March 2011 M9 EQ event, as summarized in section 7.4.

In January 2010, the initial phase of bulge-bending deformation began, creating a significant westward restoring force on the west coast that had been compressed eastwardly by the overriding plate-driving force.

During the transition phase, the east coast began to pull the subducting plate by fault coupling on July 11, 2010, nine months before the Tohoku M9 EQ. The dotted arrow over the fault in Fig. 4b indicates the pulling action by the overriding eastern edge. This pulling accelerated the westward motion of the plate, as seen in Figs. 6-1e, 6-1f, 6-1h, and 13-2-3f - 13-2-3i (Appendix B). The motion at an island's GPS station on the plate gained an abnormally increased speed of 0.69 mm/day by December 22, 2010, approximately three times higher than that on July 11, 2010, as shown in Fig 6-1.

With an upheaval growing along the east coast (section 7.1), the final phase, began one month before the highest speed observation on December 22, 2010. As the upheaval grew, the oceanic plate's westward motion decelerated and stopped by February 21, 2011. After four days, the motion reversed direction, and by March 8, 2011, the eastward speed had reached 0.06 mm/day. On March 10, 2011, one day before the Tohoku M9 EQ, the upheaval had grown to 1.2 mm over 115 days from November 14, 2010.

During the final phase, the entire Tohoku experienced an upheaval growth of 1.2 to 2.5 mm, accumulating lifting force (section 7.1) along the Tohoku subduction zone. The GPS observations indicate that the lifting force buildup eventually caused the shear stress to exceed the static frictional strength of the subduction interface weakened by the lifting force, decoupling the overriding and subducting plates. The decoupling process released a massive recoil force of the compressed west coast against the continental plate-driving eastward force. The recoil rapidly restored the bulge-bent deformation of the east and west coasts that had been elastically compressed by the overriding plate-driving eastward force. The restoration process led to the megathrust earthquake and tsunami generations, as illustrated in Fig. 4c.

It is crucial to note that the continental plate-driving eastward force bulge-bent (compressed) the east coast westward across Tohoku, not the subducting plate-driving westward force. Thus, the Tohoku M9 earthquake and tsunami-generating processes did not result from the commonly suggested elastic rebound of the east coast compressed by the subducting plate-driving force coupled with the fault. Instead, the generating processes resulted from a massive recoil of the compressed west coast while rapidly restoring the bulge-bent deformation of the east and west coasts, which were elastically compressed by the overriding plate-driving eastward force.

The restored geographical amounts are the co-seismic shifts. The geophysical processes are detailed in sections 7.1 through 7.4. Table 1 summarizes the abnormal oceanic plate motion, and Table 2 outlines the bulge-bending deformation. Table 3 summarizes the co-seismic shifts and current dynamic states of the Tohoku crust.

The GPS observed that the fifteen-month Tohoku M9 EQ genesis process is the bulge-bending deformation in three phases, coupling with the subducting oceanic plate motion over the last nine months. The final three-month process is a tsunami genesis process. Based on the observations, the megathrust EQ events can be predicted in real-time. The tsunami genesis process can offer real-time disaster prevention warnings and hazard mitigation measures leading up to the event. However, effectively informing societal decisions to mitigate the impact of disasters remains a significant challenge (Appendix C).

As of March 11, 2021, the Tohoku crust motion is still experiencing the influence of the M9 EQ, as summarized in Table 3. The rapid restoring processes of the bulge-bent crust triggered the co-seismic shifts of the crust across Tohoku. The current dynamic state of the crust indicates that the slow restoring process to the expected crust state, free of deformation, is ductile. Monitoring the dynamic states using Physical Wavelets, as detailed in section 7, can determine when the ductile restoring process will end and when the subsequent regular deformation generated by the continental and oceanic plate-driving forces coupled with the unbroken barriers (faults) will begin in the Tohoku subduction zone.

Based on the motions of the Pacific Plate observed at the Chichijima-A and Hahajima GPS stations and the Philippine Sea Plate observed at the Minamidaitojima GPS station as of March 11, 2023, there are currently no signs of an imminent megathrust event in Japan's subduction zones (Appendix B). However, there are some issues with applying the prediction method based on the Tohoku megathrust EQ genesis process to the imminent Nankai-trough megathrust EQs (Appendix B). It should be noted that slow-slip events, commonly suggested as precursory phenomena to megathrust events, were not observed in the Tohoku M9 EQ genesis process (Appendix B). This observation suggests that slow-slip events may be independent of megathrust EQs in subduction zones.

## 3 The foreshocks, aftershocks, and co-seismic shifts of the Tohoku M9

The foreshocks, aftershocks, and vertical co-seismic shifts of the Tohoku M9 EQ are shown in Fig. 3a [1]. The GPS stations depicted in Fig. 3a captured the co-seismic downward and upward displacement along the east and west coasts, as well as no displacement along a ridge on Tohoku. The distribution of these observations suggests that the rectangular fault surface of the earthquake was about 500 km long and 200 km wide. The earthquake's occurrence decoupled the overriding Tohoku crust (the eastern edge of the overriding continental plate) from the subducting western edge of the northwestern Pacific Plate.

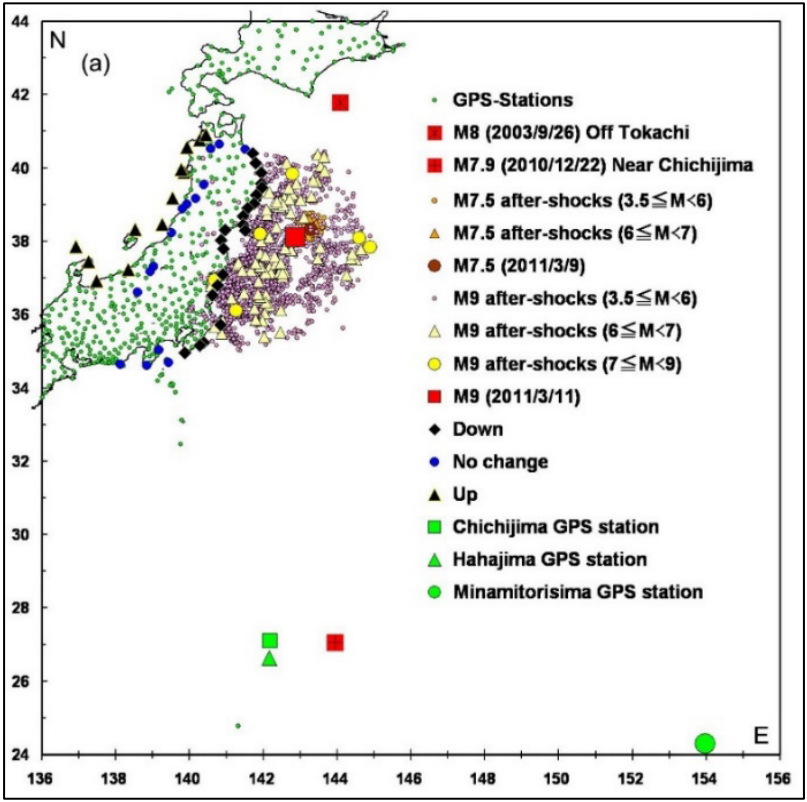

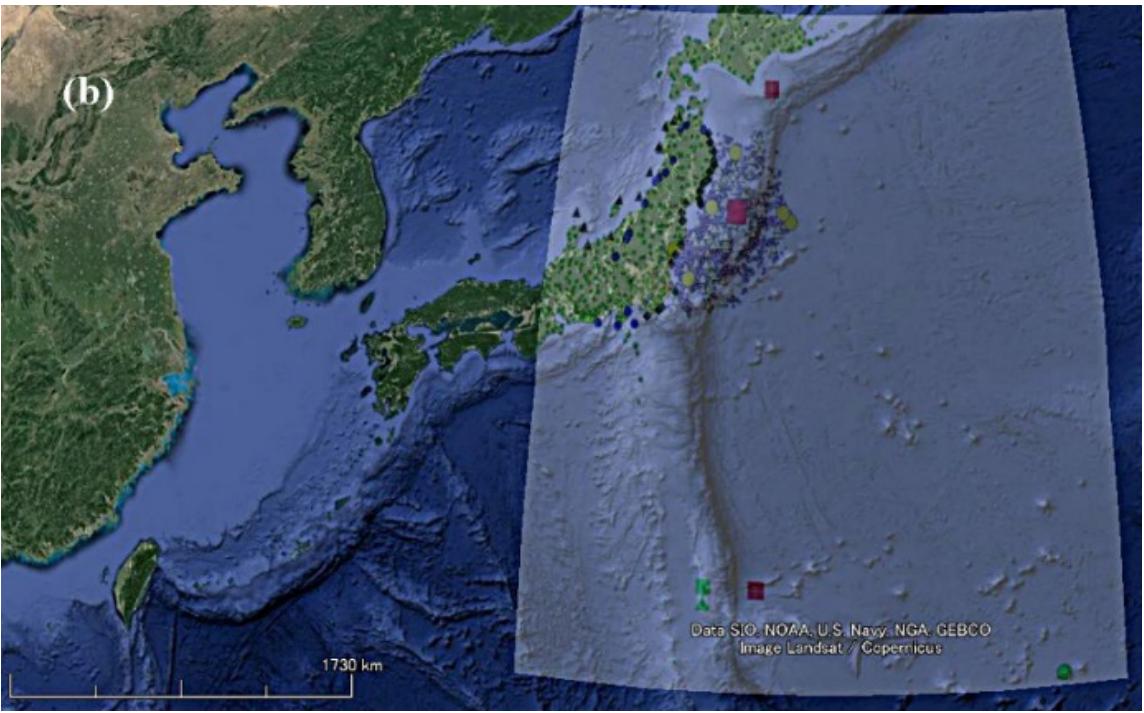

**Fig.3. GPS stations, vertical co-seismic displacement, foreshocks, and aftershocks of the M9 EQ.**

(a) The EQ source parameters are from the JMA's unified hypocenter catalogs [4]. The EQ's magnitude M is JMA's magnitude [5]. The M7.5 (2011/3/9) EQ is a foreshock (Appendix C) of the Tohoku M9 EQ (2011/3/11). The M9's hypocenter and CMT solution were (38.1006°N, 142.8517°E, 24 km) and the reverse faulting of (STR = 193°, DIP = 10°, SLIP = 79°) [5]. The near Chichijima M7.9 (2010/12/22) EQ [1] triggered by the Pacific Plate's abnormal westward motion is in section 6. The off Tokachi M8 (2003/9/23)

EQ [7] is in Fig. 9. The vertical co-seismic displacements over the 500 km distance are the downward (Down), upward (Up), and no change (No change) displacement at each GPS station. Three GPS stations above 38 degrees north latitude-line, used in sections 4 and 7, are Ryoutsu2 (Up) at (38.0633° N, 138.4717° E, west coast), Murakami (No change) at (38.2307° N, 139.5069° E), and Onagawa (Down) at (38.4492° N, 141.4412° E, east coast). The M9 EQ (2011/3/11) is on the 38-degree N line. The GPS stations in the Northwest Pacific Ocean are Chichijima, Hahajima, and Minamitorishima. Chichijima and Hahajima stations are on the subducting western edge of the northwestern Pacific Plate, whereas Minamitorishima station is on the northwestern Pacific Plate (far right below). (b) A google earth map with Fig.3a overlaid.

The normal and abnormal oceanic plate motions observed at the Chichijima, Hahajima, and Minamitorishima GPS stations in the Northwest Pacific Ocean were identical (section 6) and coupled with the bulge-bending deformation along the Tohoku east coast (section 7.1). In Fig. 4, we illustrate the coupling at the Tohoku crust cross-section along approximately 38 degrees north latitude and the subducting oceanic plate. Note that the hypocenter of the Tohoku M9 EQ is located on the 38-degree N line.

## 4 A megathrust EQ and Tsunami genesis processes

The Tohoku M9 EQ and the resulting tsunami were not caused by the elastic rebound of the east coast subsiding with the subducting oceanic plate coupled with the fault. Instead, they resulted from complex interactions between the subducting and overriding plates, which caused bulge-bending deformation over the Tohoku subduction zone. The bulge-bending deformation refers to the gradual bending of the overriding plate of Tohoku by the eastward continental plate-driving force, causing the over-riding crust of Tohoku to bulge upwards and downwards by a few millimeters. This deformation evolved from the expected regular deformation that had been occurring for the past three hundred years (section 7.4), as shown in Fig. 4a. The complex interactions are summarized in section 7.4.

### 4.1 Distinctive phases of the Tohoku crustal deformation

The GPS observation detected the onset of the bulge-bending deformation, which grew in three distinct phases: an initial phase of 6 months with gradual subsidence of 1 to 2.8 mm across the Tohoku region, followed by a transitional phase of 6 months with further gradual subsidence of 3.3 mm on the east coast, and a final phase of 3 months with an upheaval growth of 1 to 3 mm across Tohoku until the 2011 EQ event.

### 4.1.1 Regular deformation (Before January 2010)

Figure 4a illustrates the regular deformation of the dotted line over Tohoku with the subducting oceanic plate moving westward, causing the east coast to subside (represented by a westward arrow) and the overriding continental plate moving eastward, causing the west coast to move upward (represented by an eastward arrow). For instance, the Onagawa station (in Fig. 3) had a subsidence rate of 6 mm per year, while the Ryoutsu2 station (in Fig. 3) had an upward displacement rate of 1.5 mm per year. These expected movements have been occurring for the past three hundred years, as discussed in section 7.4.

### 4.1.2 Initial phase (January 2010 to June 2010)

One month prior to January 2010, the deformation property of the west coast changed from non-elastic to elastic, leading to the generation of a restoring force on the compressed west coast. This property change of the west coast marks the onset of an underlying bulge-bending deformation that persisted for 15 months, resulting in gradual subsidence and upheaval growth of a few millimeters over a 500 km stretch along the east and west coasts.

In January 2010, the westward-restoring force of the west coast, compressed by the overriding continental plate-driving force, caused a gradual subsidence of 1 to 2.8 mm across the Tohoku region. By June 2010, the subsidence had reached 1 mm on the west coast, 1.5 mm on the top, and 2.8 mm on the east coast. The east coast subsidence was a pre-transition process from regular to bulge-bending deformation.

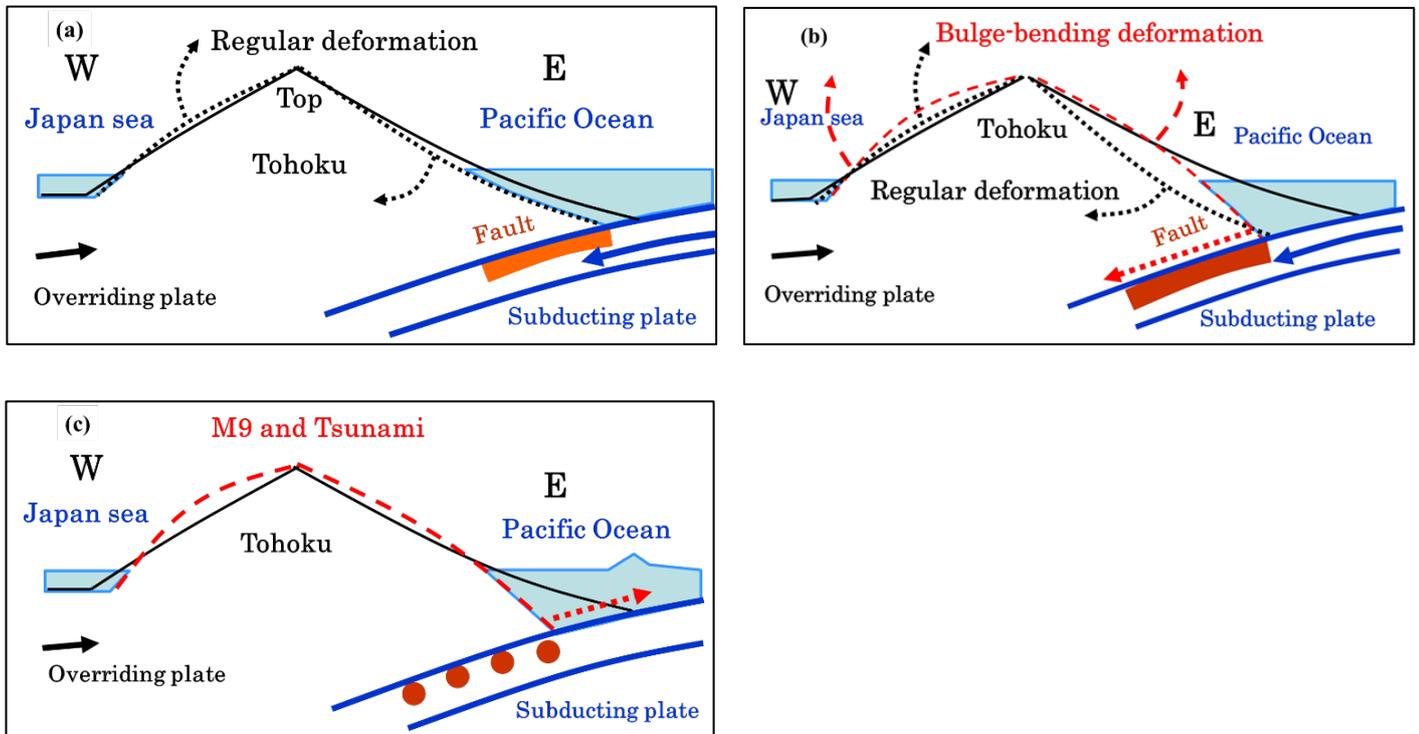

**Fig. 4. Three phases of crustal deformation (a megathrust EQ genesis process).**
The labels E, W, and Top correspond to the East coast (Onagawa station), West coast (Ryoutsu2 station), and Top (Murakami station). The east and west coasts of the Tohoku region face the Pacific Ocean and the Sea of Japan, respectively. The Top is the location of the no-displacement-ridge line in Fig. 3. (a) A regular slow deformation began changing the west coast deformation from non-elastic to elastic, generating a restoring force of the compressed west coast with gradual subsidence of 1 to 2.8 mm across Tohoku starting in January 2010. (b) The transition was from the regular to a bulge-bending deformation, which pulled the subducting Pacific Plate by coupling with the fault. As indicated by the eastward arrow, the overriding plate-driving force compressed the elastic west coast, causing the east coast to pull down the subducting plate, as indicated by a dotted arrow on the fault. The east coast's pulling action with the westward displacement began in July 2010. (c) The final phase began in November 2010 with the gradual upheaval growth of 1.2 mm on the east coast, generating the lifting force along the Tohoku subduction zone. The buildup of lifting force on the east coast eventually caused the shear stress to exceed the static frictional strength of the subduction interface weakened by the lifting force, causing the overriding and subducting plates to decouple. The lifting force on the entire Tohoku region helped the decoupling process, releasing a massive recoil force of the compressed west coast against the eastward-plate-driving force. The recoil rapidly restored the bulge-bent deformation of the east and west coasts elastically compressed by the overriding plate-driving eastward force. This rapid restoration led to the Tohoku M9 earthquake and tsunami on March 11, 2011.

### 4.1.3 Transitional phase (June 2010 to November 2010)

In June 2010, the transitional phase began with a further gradual subsidence of 3.3 mm on the east coast until the final phase, as illustrated in Fig. 4b. The subsidence firmly grasped the fault and started to pull down the subducting Pacific Plate on July 11, 2010, represented by the dotted arrow over the fault. The pulling action, caused by compressing the west coast, accelerated the westward movement, reaching the highest speed of 0.69 mm/day on December 22, 2010.

About a month before reaching the highest speed, the east coast bulge shifted to the final phase of upheaval growth. The linear upheaval growth of 1.2 mm over 115 days began on November 14, 2010.

The Top began the upheaval growth on August 26, 2010, reaching 2.5 mm before the 2011 M9 EQ event. The gradual upheaval growth of 2.3 mm on the west coast began on October 29, 2010.

### 4.1.4 Final phase (November 2010 to March 2011)

In November 2010, the final phase began with the gradual upheaval growth of 1.2 mm on the east coast. The bulge-bending force with this growth decelerated the subducting Pacific Plate's westward movement, which halted by February 21, 2011, and remained motionless for another four days. On February 25, the bulge force reversed the westward plate motion, with the eastward speed reaching 0.06 mm/day just three days before the M9 EQ on March 11, 2011.

The final upheaval growth of 1.2 mm rapidly released a massive restoring force in the west coast, compressed elastically by the overriding plate-driving eastward force. This recoil restored the bulge-bent west and east coasts, ultimately decoupling the overriding and subducting plates and leading to the megathrust EQ and tsunami on March 11, 2011, as depicted by the dotted arrow in Fig. 4c. Throughout the transitional and final phases, Tohoku experienced an elastic compression of 10.0 mm on the west coast and a pulling movement of 13.2 mm on the east coast in the same westward direction as the subducting oceanic plate-driving westward force, persisting until the 2011 Tohoku EQ event on March 11, 2011.

## 4.2 Precursory microgravity anomaly observations

The continental plate-driving eastward force compressed the west coast eastward and simultaneously caused the east coast to bulge-bend, initiating the process of pulling down the subducting oceanic plate in July 2010. The plate-driving force continued to bulge-bend the east coast until the halt of pulling action in February 2011. The timeline of this pulling action by the enormous continental plate-driving force across Tohoku is consistent with a pre-seismic microgravity anomaly observed by the GRACE satellite data from July 2010 to February 2011 [8]. The upheaval growth that began in November 2010 near the Tohoku M9 epicenter area has a timeline that agrees with the uplift growth suggested by the sea-surface gravity change observation from November 2010 to February 2011 [9].

## 4.3 Tsunami genesis process

As discussed in section 7.4, the relationship between averaged accelerations (forces) and displacements on the Tohoku subduction zone indicates that the compression of the west coast by the overriding plate-driving eastward force is elastic. The continental plate-driving eastward force simultaneously bulge-bent the east coast westward across Tohoku, not the subducting plate-driving westward force. Thus, the Tohoku M9 earthquake and tsunami were not caused by the commonly suggested elastic rebound of the east coast compressed by the subducting plate-driving force coupled with the over-riding plate through the fault.

The overriding plate-driving eastward force compressed the west coast by +18.4 mm (eastward until the Tohoku M9 EQ on March 11, 2011), initiating the underlying bulge-bending deformation during the initial phase and generating the east coast's pulling action during the transitional and final phases with -13.2 mm (westward) movement. During the final phase, the upheaval growth of 1.2 mm resulted in the accumulation of lifting force along the east coast, which weakened the static frictional strength of the subduction interface. This weakening eventually caused the shear stress to exceed the frictional strength, decoupling the overriding and subducting plates. The decoupling triggered the sudden release of a massive elastic potential energy stored in the compressed west coast, resulting in the rapid restoration of the bulge-bent west and east coast deformation to the regular as an elastic rebound of the entire Tohoku. This entire elastic rebound generated the megathrust EQ and tsunami, as illustrated in <u>Fig. 4c</u>. Thus, the Tohoku M9 earthquake and tsunami-generating processes were not a result of the commonly suggested elastic rebound of the east coast compressed by the subducting plate-driving force coupled with the fault.

## 5 The crustal displacement and the equations of motion

### 5.1 Displacement time series

We utilized the Geospatial Information Authority of Japan (GSI)'s F3 solutions [3] to obtain the daily positions of GPS stations. GSI updates these positions weekly with a two-week latency and makes them publicly available [3]. The displacement vector represents the change in position of a GPS station relative to a reference point, with its components along the geographic axis $c$ denoted as $E$ (west to east), $N$ (south to north), and $h$ (down to up) in right-handed coordinates ($E$, $N$, $h$). The temporal variation of each component is represented by a time series, $\{c\} = \{d(c, 0), d(c, 1), d(c, 2), \ldots, d(c, j), \ldots\}$. The index $j$ represents the chronological event of time in days. If the first day's position is used as the reference, the first component is zero, i.e., $d(c, 1) = 0$, where the first day is represented by $j = 0$. However, the reference position can be arbitrary, in which case the first day is represented by $j = 1$, and $\{c\}$ starts with $d(c, 1)$. The daily GPS displacement, $d(c, j)$, fluctuates due to environmental causes [3], as shown in Fig. 5-1. These fluctuations make it difficult to define the time rate of change of $d(c, j)$ over time. However, mathematical operators named Physical Wavelets (see <u>Appendix A</u>) can provide the time rate of change of $d(c, j)$ over time, which defines the equations of motions to describe the underlying dynamics of the time series data $\{c\}$ as discussed in the next section.

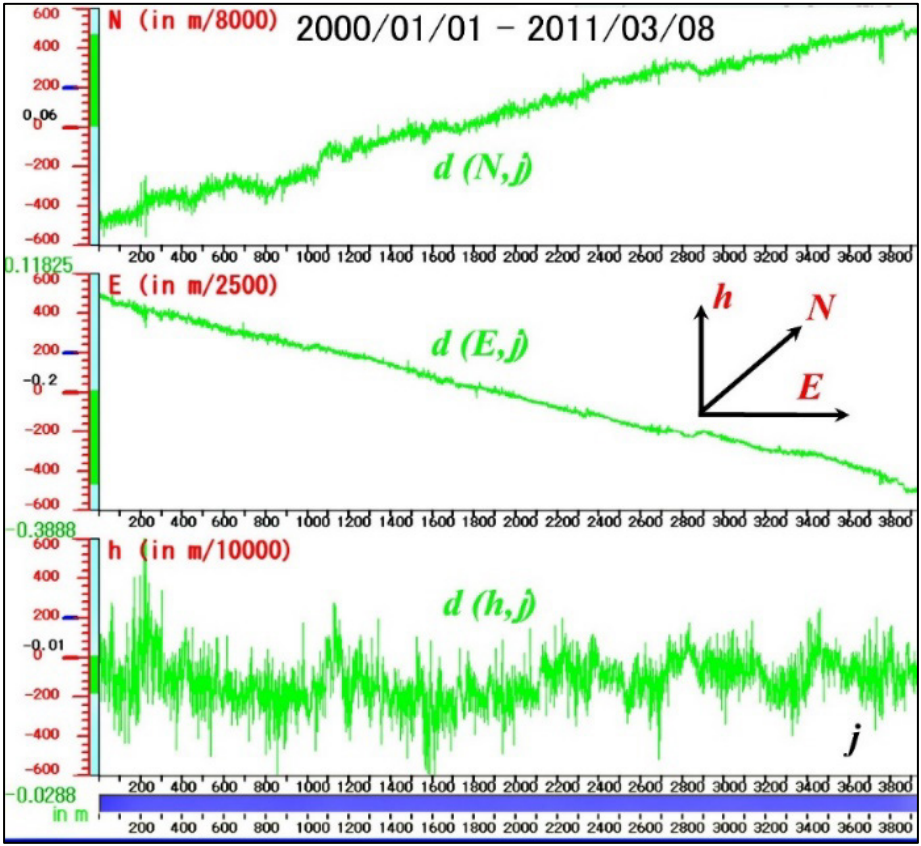

**Fig. 5-1. Displacement $d(c, j)$ at Chichijima station.**

Chichijima station started its operation on March 21, 1996, and ended on March 8, 2011, three days before the M9 EQ. The GSI replaced Chichijima with Chichijima-A [3]. The displacement series $\{c\}$ at the Chichijima station was highly noisy for the first four years, with a four-month data deficiency in 1999. To obtain a more reliable data set, we considered January 1, 2000, the first day for the Chichijima $\{c\}$. We removed several scattered and two-week-long data deficiency days and chronologically re-indexed time $j$ in $\{c\}$. The two-week deficiency occurred at $j = 3078$ in October 2008. The figure shows the northward, eastward, and upward displacements $d(N, j)$, $d(E, j)$, and $d(h, j)$ in the first, second, and third windows from the top, respectively. The abscissa represents time $j$, ranging from 0 (January 1, 2000) to 3640 (March 8, 2011). The ordinate represents each displacement $d(c, j)$ in meters. Since each displacement is zero at $j = 0$, each graphical origin has the offset value from zero for the graphics of $\{c\}$. The offset amounts are 0.06 m, $-0.2$ m, and $-0.01$ m from the top window. The northward, eastward, and upward displacements from the origin are positive. The magnifications on each offset scale are 8000, 2500, and 10000 times, respectively. Each column height times the magnification factor represents its last $d(c, j)$ at $j = 3940$ from the reference with its value. Thus, the net displacement is the $d(N, j)$ at $j = 3940$, 0.11825 m. The averaged northward-moving rate of the subducting Pacific Plate over 3940 days is 0.03 mm per day (1.1 cm per year). The $d(E, j)$ has a net downward (westward) amount of $-0.3888$ m. The average westward-moving rate over 3940 days is 0.1 mm per day (3.60 cm per year). Small variations in $d(N, j)$ and $d(E, j)$ are indistinguishable from various GPS environmental noises whose possible sources are listed [3]. The noise level for $d(h, j)$ is approximately between $\pm 200$ ($\pm 20$ mm) by its magnification of m/10000 (0.1 mm).

## 5.2 The equations of motion

As described in [Appendix A](#), Physical Wavelets (P-Ws) are powerful operators for defining position (displacement), velocity, and acceleration on time-varying stochastic time series data $\{c\}$. The operators are the displacement-defining operator $DDW(t - \tau)$, velocity-defining operator $VDW(t - \tau)$, and acceleration-defining operator $ADW(t - \tau)$ at time $\tau$. They satisfy the time reversal property for displacement and their derivatives, enabling the definition of respective fundamental physical quantities. By taking the cross-correlation (or inner product in the case of vector representations) of each operator with the time series data $\{c\}$ that is non-differentiable in time, we can define displacement $D(c, \tau)$, velocity $V(c, \tau)$, and acceleration $A(c, \tau)$ at time $\tau$ as follows.

$$D(c,\tau) = \int_{-\infty}^{+\infty} \{c\} \, DDW(t-\tau)\mathrm{d}t = [1/(2w+1)] \sum_{j=-w}^{w} d(c, \tau + j), \qquad (1)$$

$$V(c,\tau) = \int_{-\infty}^{+\infty} \{c\} \, VDW(t-\tau)\mathrm{d}t = [D(c,\tau + s/2) - D(c,\tau - s/2)]/s \qquad (2)$$

and

$$\begin{aligned} A(c,\tau) &= \int_{-\infty}^{+\infty} \{c\} \, ADW(t-\tau)\mathrm{d}t = [V(c,\tau + s/2) - V(c,\tau - s/2)]/s \\ &= [D(c,\tau + s) - 2D(c,\tau) + D(c,\tau - s)]/s^2. \end{aligned} \qquad (3)$$

In Eq. (1), $D(c, \tau)$ has $d(c, j)$ averaged over a time interval of $j = \tau \pm w$. The time-reversal operation changes $\tau$ to $-\tau$ and confirms that $D(c, -\tau) = D(c, \tau)$, $V(c, -\tau) = -V(c, \tau)$, and $A(c, -\tau) = A(c, \tau)$. Thus, they exhibit the time differential properties for the differentiable $D(c, \tau)$ and $V(c, \tau)$ obtained from the non-differentiable $d(c, j)$ of $\{c\}$, which has periodically fluctuating components and trends.

The correlation between P-Ws and $\{c\}$ should be strong to accurately extract specific periodicities and trends. A strong correlation ensures that the P-Ws effectively capture the most fundamental physical quantities to describe the underlying dynamics of the time series data. Equations (1), (2), and (3) represent low pass filtered, bandpass filtered, and another bandpass filtered physical quantity, respectively. The parameters $w$ and $s$ can be any integer with which to filter out the selected frequency components and trends in $\{c\}$ to define $D(c, \tau)$, $V(c, \tau)$, and $A(c, \tau)$. The relations between $D(c, \tau)$, $V(c, \tau)$, and $A(c, \tau)$ represent the equations of motion for the observed displacement series $\{c\}$.

We define a time-rate change of the kinetic energy (velocity squared) as the product of $V(c, \tau)$ and $A(c, \tau)$, which is the power, $PW(c, \tau)$. We reassign them as $V(c, j)$ and $A(c, j)$ at time $j (= \tau + w + s)$, defined for $A(c, \tau)$. The power is then calculated as $PW(c, j) = V(c, j) \times A(c, j)$. Monitoring the oceanic plate and the Tohoku crust's motions with $PW(c, j) \geq$ a predetermined threshold can detect any unexpected motion and its onset. The threshold level is set based on the observed $PW(c, j)$'s maximum amplitude during periods of expected standard motions or the statistical values obtained from the observation during periods of expected standard motions.

## 6 The abnormal motion of the subducting northwestern Pacific Plate

As shown in Figs. [3](#) and [9](#), four GPS stations are located near the Japan Trench in the Northwest Pacific Ocean: Minamitorishima, Hahajima, Chichijima, and Chichijima-A. The Chichijima and Chichijima-A stations are on the same island, while the Minamitorishima station is on the northwestern Pacific Plate. In contrast, the Chichijima, Chichijima-A, and Hahajima stations are located on the western edge of the Ogasawara Plateau, as shown in Fig. 13-1-1 (Appendix B) [3]. The qualitatively identical GPS displacement time series $\{c\}$ from these stations indicates that the Chichijima, Chichijima-A, and Hahajima stations are under the western edge motion of the subducting northwestern Pacific Plate. In contrast, the Minamitorishima station is under the motion of the northwestern Pacific Plate.

The daily displacement $\{c\}$ of the subducting oceanic plate motion, where the geographic axis $c$ is denoted as $E$ (west to east), $N$ (south to north), and $h$ (down to up) in right-handed coordinates $(E, N, h)$, is constantly subjected to the lunar tidal force loading due to the gravitational attraction between the Moon and Earth. To observe the responses of the oceanic plate motion to the lunar synodic loading of period 30 (29.5) days, we define displacement $D\,(c, \tau)$, velocity $V\,(c, \tau)$, and acceleration $A\,(c, \tau)$ with $w = 7$ and $s = 20$. Any external force coupling with the overriding eastern edge (the east coast) may alter the periodic lunar responses in amplitudes and phases on the $D\,(c, \tau) - V\,(c, \tau)$ and the $D\,(c, \tau) - A\,(c, \tau)$ paths.

As shown in Figs. 6-1 to 6-4, the $D\,(E, \tau) - V\,(E, \tau)$ path at every GPS station draws the plate motion exhibiting an unexpected trend change followed by an abnormally increasing westward speed caused by the bulge-bending pulling of the east coast (section 7). The east coast pulling action was the external force coupled with the eastward continental plate-driving force pushing the west coast facing the Sea of Japan. The external force coupling first appeared as the trend change on $D\,(E, \tau)$ in July 2010, preceding the abnormal motion. The abnormal westward speed $V\,(E, \tau)$ at every station reached its highest value on December 22, 2010, 76 days before the Tohoku M9 EQ with values of $-0.69$ mm/day for Chichijima, $-0.78$ mm/day for Chichijima-A, $-0.84$ mm/day for Hahajima, and $-1.15$ mm/day for Minamitorishima. These values were approximately three times higher than each westward speed at the time of the trend change, as shown in Table 1. A rapid deceleration followed, stopping the westward motion by the bulge-bending external force on the east coast around February 21, 2011. In about four days, the moving direction reversed, and the subducting plate moved eastward at $+0.08$ mm/day (Chichijima-A) until the Tohoku megathrust M9 EQ on March 11, 2011.

The trend change and abnormal motion were initially detected using a window size of $w = 7$ and a separation size of $s = 20$ for monitoring the power $PW\,(c, j) = V(c, j) \times A(c, j)$ with $c = E$ (the eastward axis) at time $j$, as depicted in Figs. 6-1d and 6-1e. To quantify the observed motion under the external force $F\,(E, \tau)$, the $D\,(E, \tau) - A\,(E, \tau)$ paths were employed. Specifically, a segment equation, $F\,(E, \tau) \approx A\,(E, \tau) \approx K \times D\,(E, \tau)$, with a time and synodic-cycle dependent constant $K$ (positive or negative), was used, as illustrated in Figs. 13-2-3f to 13-2-3i in Appendix B. This segmented equation provides a detailed understanding of the dynamic state of the plate motion.

Detecting long-term trends and anomalies in plate motion necessitates a low-frequency selection using a window size of $w = 15$ and a separation size of $s = 40$ for $PW\,(E, j)$ monitoring, as demonstrated in Fig. 6-1f. Although the low-frequency selection with $w = 15$ and $s = 40$ introduces a delay in detecting the onset of anomalies in plate motion, the $PW\,(E, j)$ monitoring can still detect the same anomaly onsets in plate motion, as shown in Figs. 6-1g and 6-1h. The delay time increases with a larger separation size.

Equations with a window size of $w \approx 200$ and separation size $s \approx 300$ are crucial to quantify the Tohoku's bulge-bending deformation in a three-phase process by minimizing yearly and seasonal variations and environmental noises in $\{c\}$ $(c = h)$, as explained in section 7. The first-phase bulge-bending pulling action by the east coast generated every abnormal westward movement (Table 1) with fault coupling, as illustrated schematically in Fig. 4.

The variations of daily vertical displacement $d(h, j)$ in $\{h\}$ measures the up and down movement of the Earth's surface caused by lunar fortnightly tidal force loading. However, due to environmental noises, $d(h, j)$ fluctuates about ± 20 mm, making it challenging to observe the dynamic crustal stress state appearing on the Earth's surface under the lunar fortnightly tidal force loading while in a state of frictional failure. Therefore, a smaller window size ($w \approx 2$) and separation size ($s \approx 7$) are necessary for $PW(h, j)$ monitoring to detect unusual subtle loading patterns, which may be associated with the precursors to significant EQs in the region. Based on previous studies [6, 7], the $PW(h, j)$ monitoring detected unusual loading two weeks before EQs of magnitude larger than about five in the region. The GPS observations suggest that monitoring the crustal vertical-displacement responses to lunar tidal force loadings can determine if the regional crust is at risk of a significant EQ. The $PW(c, j)$ monitoring of crustal responses to periodic lunar tidal force loadings is a powerful observational and analytical tool for detecting subtle anomalies in crustal motion buried in noisy displacement time series $\{c\}$.

## 6.1 Chichijima station (on the western edge of the subducting northwestern Pacific Plate)

Chichijima station (27.0956° N, 142.1846° E) made its last observation on March 8, 2011, just three days before the Tohoku M9 EQ on March 11, 2011. Figure 6-1 illustrates the observation at the Chichijima station on the unexpected trend change followed by the abnormally increasing westward speed coupled with the bulge-bending pulling of the Tohoku east coast.

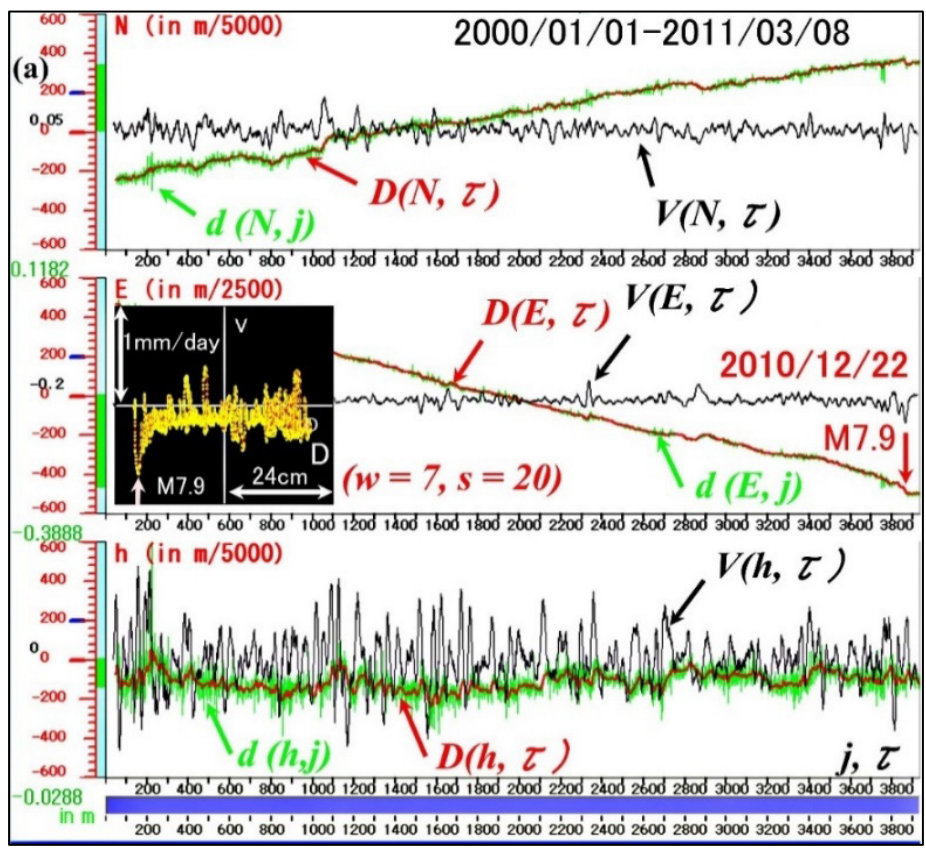

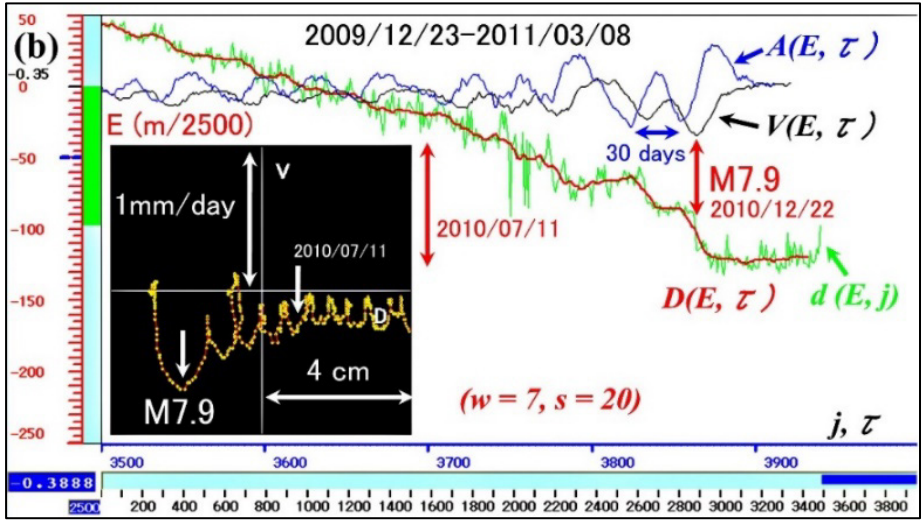

(b) 2009/12/23−2011/03/08

(c) 2009/12/23 − 2011/03/08

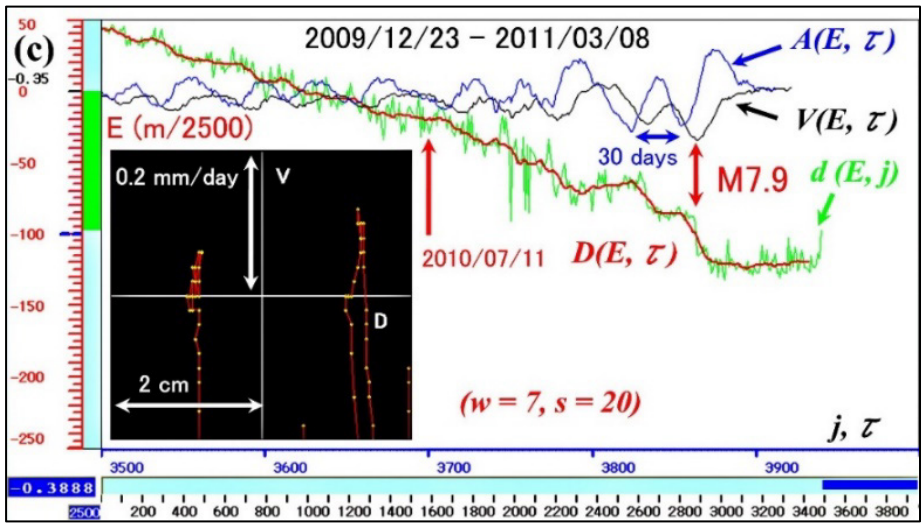

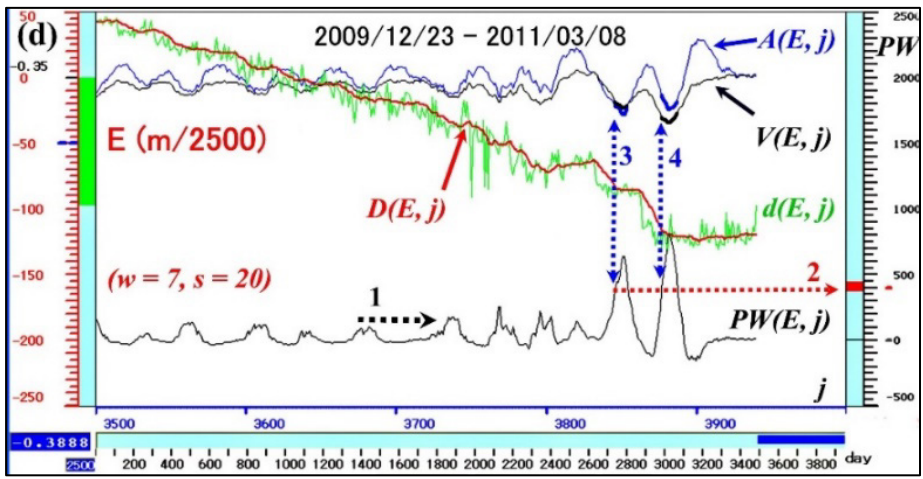

(d) 2009/12/23 − 2011/03/08

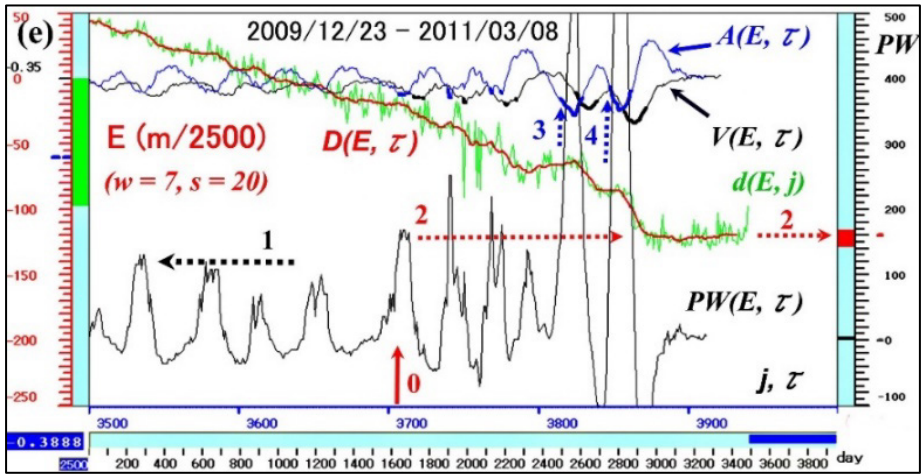

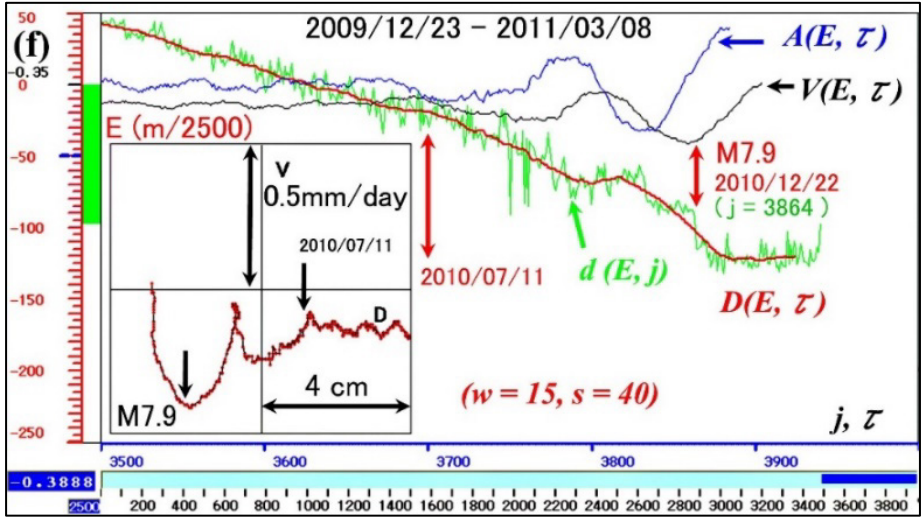

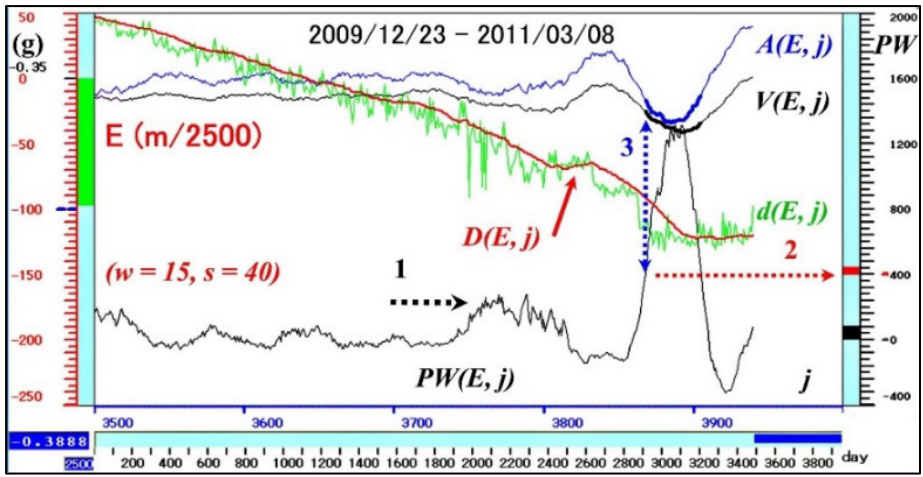

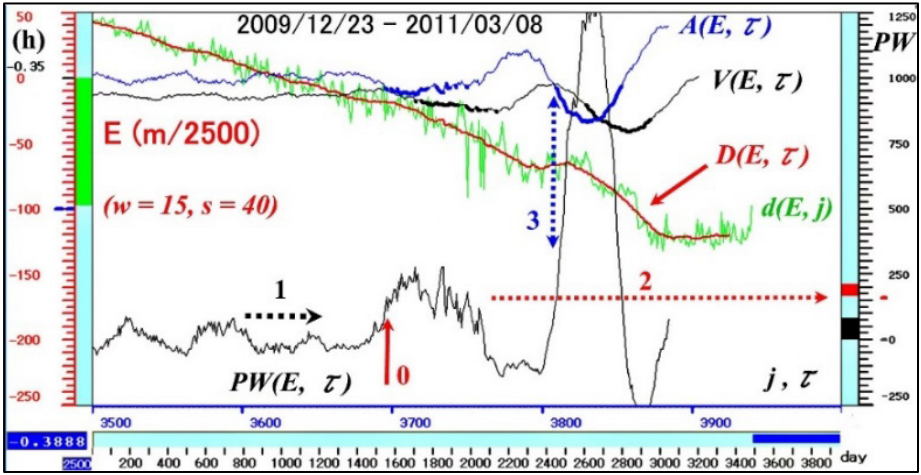

**Fig. 6-1. The abnormal westward motion at Chichijima station.**

(a) The series $\{c\}$ covers the period from $j = 0$ (January 1, 2000) to $j = 3940$ (March 8, 2011), with parameters $w = 7$ and $s = 20$ used for $D(c, \tau)$, $V(c, \tau)$, and $A(c, \tau)$. The $D(E, \tau) - V(E, \tau)$ plane has a scale of (24 cm, 1 mm/day), with an origin at $D(E, \tau) = -0.2$ m, which is the offset reference ($-0.2$ m) at scale 0 from the first-day position ($j = 0$). The $V(E, \tau)$ origin is 0 mm/day. The right half of the $D(E, \tau) - V(E, \tau)$ plane is east of the offset origin. The upper half is eastward and positive $V(E, \tau)$, while the lower half is westward and negative. The M7.9 EQ (2010/12/22) is located on $D(E, \tau)$ at $\tau = 3864$, and the (24 cm, 1 mm/day) path. The $d(c, j)$ is in green, $D(c, \tau)$ in red, and $V(c, \tau)$ in black. The $V(c, \tau)$ is in a relative scale from the graphical origin 0. (b) The expanded time window is from $j = 3500$ (December 23, 2009) to $j = 3940$ (March 8, 2011). The (4 cm, 1 mm/day) phase plane has an offset origin at the blue-line-scale $-50$, $D(E, \tau) = -0.22$ m ($-0.2 - 50/2500$). The $V(E, \tau)$ and $A(E, \tau)$ are in relative scales from the same graphical origin 0. A blue line representing the lunar synodic tidal force loading (29.5-day period) is shown as acceleration $A(E, \tau)$ with a period of 30 days. The date label 2010/07/11 corresponds to the $D(E, \tau)$ trend change at $\tau = 3700$ (July 11, 2010). (c) The window is the same as Fig. part b, with a magnified (2 cm, 0.2 mm/day) plane having the offset origin at the blue-line-scale $-100$, $D(E, \tau) = -0.39$ m ($-0.35$ m $- 100 \times$ m/2500). The path ends at $V(E, \tau) = +0.06$ mm/day. (d) The window is the same as Fig. part b with a $PW(E, j)$ monitoring. The relative power scales are on the right column. The monitoring with $PW(E, j) \geq 400$ detected an anomalous lunar synodic loading on $\{E\}$. The predetermined 400 was about twice the expected standard power level at arrow 1. Level 400 is at the red scale at arrow 2. Arrow 3 was the first anomaly detection at $j = 3847$ (December 5, 2010). At the detection, $PW(E, j)$ rose from 378 (at $j = 3846$) to 440 (at $j = 3847$), showing the red column height. Arrow 4 was the second detection at $j = 3877$ (January 4, 2011). The anomalous $V(E, j)$ and $A(E, j)$ were highlighted in bold under $PW(E, j) \geq 400$. (e) The window is the same as Fig. part b with $PW(E, j) \geq 160$, finding the westward trend change at time $j = 3735$ on August 15, 2010. In time $\tau$, the change was at $\tau = 3708$ (at arrow 0) on July 19, 2010. The detecting level 160 at arrow 2 was adopted from the standard power level at arrow 1. By shifting time $j$ back to time $\tau$, the displays are $D(E, \tau)$, $V(E, \tau)$, $A(E, \tau)$, and $PW(E, \tau)$. (f) The time window is the same as Fig. part b, with parameters $w = 15$ and $s = 40$, removing the 30-day-period oscillation. The (4 cm, 0.5 mm/day) plane has the offset origin, $D(E, \tau) = -0.37$ m ($-0.35 - 0.02$) at the blue-line scale $-50$. The $D(E, \tau)$ trend change has the date label 2010/7/11. (g) The window is the same as Fig. part b with $w = 15$ and $s = 40$. Arrow 3 was the anomaly detection at $j = 3869$ (December 27, 2010) by $PW(E, j) \geq 400$. The threshold 400 is at the red scale pointed by arrow 2. The threshold of 400 was chosen based on the standard power level indicated by arrow 1. The power column height change in red is from 400 (at $j = 3868$) to 442 (at $j = 3869$) at the anomaly detection. The black column height shows the last $PW(E, j)$ (just above level 0) on March 8, 2011 ($j = 3940$). (h) The window is the same as Fig. part b for the $D(E, \tau)$ trend change detection. Up-arrow 0 indicates the trend change detection at time $j = 3751$ (August 31, 2010) by $PW(E, j) \geq 160$ with $w = 15$ and $s = 40$. In time $\tau$, the change was at up-arrow 0, $\tau = 3696$ on July 7, 2010. Arrow 1 is the standard power level, for which the unexpected level 160 was about twice the standard level.

In Fig. 6-1a, the $D(E, \tau)$ and $V(E, \tau)$ relation of the time series $\{E\}$ is plotted as a path on a phase plane with a scale of (24 cm, 1 mm/day) in the second window, overlaying the analyzed data from Fig. 5-1. At the highest westward (downward) speed, the M7.9 EQ in <u>Fig. 3</u> occurred in the Pacific about 187 km away from the station. The event had a normal faulting of (STR = 340°, DIP = 57°, SLIP = − 56 °) <u>[5]</u>, suggesting that the abnormal westward motion triggered the event. This relationship is further illustrated in <u>Fig. 13-2-3i</u> (Appendix B) <u>[1]</u>.

The plots in Fig. 6-1b illustrate the impact of lunar synodic tidal force loading on the subducting Pacific Plate at Chichijima station. The $D(E, \tau) - V(E, \tau)$ path moves eastward while protruding by 0.2 mm/day from the westward speed at − 0.23 mm/day, which suggests that a periodic oscillation may be due to synodic tidal loading. The westward speed can be calculated as − 0.2 mm/day by dividing the 6 mm separation between the two protruding peaks by the 30 days, which qualitatively confirms that the periodic oscillation is due to the synodic tidal loading. The westward trend of $D(E, \tau)$ changed at around $\tau = 3700$ (July 11, 2010), indicated by the labeled arrow, 2010/07/11, on the $D(E, \tau)$. The trend change onset occurred because of insufficient synodic tidal loading on the roughly linear segments of $D(E, \tau)$, $V(E, \tau)$, and $A(E, \tau)$. The onset divided the $D(E, \tau) - V(E, \tau)$ path into two small linear segments. The first linear segment before the trend change has an explicit segment equation $A(E, \tau) \approx K \times D(E, \tau)$ with a positive constant $K$ on the $D(E, \tau) - A(E, \tau)$ path-segment, as shown in <u>Figs. 13-2-3f - 13-2-3i</u> (Appendix B). The constant $K$ is positive. However, $K$ is negative under effective lunar synodic tidal loading, obeying the oscillatory motion. After the trend change, the motion became anomalous and reached the highest westward speed of $V(E, \tau) = − 0.69$ mm/day at $\tau = 3864$ on December 22, 2010, approximately three times faster than $V(E, \tau) = − 0.25$ mm/day at $\tau = 3700$ on July 11, 2010. The westward motion rapidly decelerated until it stopped at $\tau = 3908$ on February 4, 2011, after which it began moving eastward.

In Fig. 6-1c, a magnified path is displayed, which shows a reversal in motion towards the east, with $V(E, \tau)$ changing positively (i.e., becoming eastward) on February 8. The eastward motion reached a speed of $V(E, \tau) = + 0.06$ mm/day and a displacement of 1.6 mm at time $\tau = 3918$ on February 14, 2011. In terms of time $j$, this was March 8, 2011, which was the last operational day at the GPS station, three days before the March 11 M9 EQ. The time $j$ is ahead of $\tau$ by 17 days ($\tau = j − s/2 − w$).

In Fig. 6-1d, the $PW(E, j) \geq 400$ monitoring detected the anomaly with a predetermined threshold level of 400, which is twice the standard $PW(E, j)$ amplitudes at dot-arrow 1. Level 400 is an unexpected power level, and the threshold adoption can be automatic or manual during power monitoring. Arrow 3 and Arrow 4 represent the first and second anomaly detections at $j = 3847$ (December 5, 2010) and $j = 3877$ (January 4, 2011), respectively. At each detection, $V(E, j)$ and $A(E, j)$ become bold and remained bold as long as $PW(E, j) \geq 400$. The negative values indicate that the anomalous acceleration and movement were towards the west.

Figure 6-1e illustrates that the unexpected westward motion began at the upward arrow 0. The $PW(E, j) \geq 160$ monitoring found the westward trend change at time $j = 3735$ on August 15, 2010. In time $\tau$ ($\tau = j − s − w$),

it was 3708 on July 19, 2010. By sifting time $j$ back to time $\tau$, we can draw $D(E, \tau)$, $V(E, \tau)$, $A(E, \tau)$, and $PW(E, \tau)$, satisfying each time reversal property. For anomalies 3 and 4, the negative $A(E, \tau)$ precedes negative $V(E, \tau)$, indicating the anomalous acceleration (oceanic plate-driving force) and the motion that follows are westward.

In Fig. 6-1f, $D(E, \tau)$, $V(E, \tau)$, and $A(E, \tau)$ are defined with $w = 15$ and $s = 40$, which masks the lunar synodic loading. The $D(E, \tau) - V(E, \tau)$ path shows the abnormal $V(E, \tau)$ more clearly than in Fig. 6-1d.

Figure 6-1g displays the result of the $PW(E, j) \geq 400$ monitoring with $w = 15$ and $s = 40$, which detected the abnormal motion at arrow 3 ($j = 3869$ on December 27, 2010). The red column represents the power change. The $V(E, j)$ and $A(E, j)$ in bold are negative, indicating that the abnormal motion was westward.

In Figure 6-1h, the $PW(E, \tau) \geq 160$ monitoring with $w = 15$ and $s = 40$ detects the westward trend change independent of the effective synodic tidal force loading, supporting the conclusion that the geophysical origin of the trend change was a transition from regular subsidence deformation on the east coast of Tohoku to bulge-bending deformation (pulling), as explained in [section 7.1]. The abnormal westward motion of the subducting northwestern Pacific Plate followed the trend change, as the pulling action progressed.

The $PW(E, j)$ monitoring for the low-frequency $D(E, \tau)$, $V(E, \tau)$, and $A(E, \tau)$ introduces an anomaly-onset-detection delay in time $j$, which depends on the detecting objectives. For instance, the unexpected $V(E, j)$ and $A(E, j)$ were detected on December 5, 2010, at $j = 3847$, which could have been useful to predict the imminent M7.9 event on December 22. However, in Fig. 6-1g, with $w = 15$ and $s = 40$, the detection was on December 27, 2010, at $j = 3869$, which followed the trend change in $D(E, \tau)$ and was useful for predicting the Tohoku M9 EQ on March 11, 2011. An automated power monitoring that includes multiple frequencies and thresholds for abnormal event detections is always available.

## 6.2 Chichijima-A station (on the western edge of the subducting northwestern Pacific Plate)

Chichijima-A station (27.0675° N, 142.1950° E), located on the western edge of the subducting northwestern Pacific Plate, began observations on December 4, 2007, and shows similar abnormal motion to the Chichijima station, as illustrated in Fig. 6-2.

Figure 6-2a displays observations from December 4, 2007, to June 6, 2020, with the M9 EQ occurring on March 11, 2011. The M9 EQ appears as an environmental spike noise on the $d(E, j)$ at $j = 1175$. The path indicates that the abnormal westward motion occurred only before the M9 EQ.

Figure 6-2b shows the M9 EQ on the expanded time window with the path, and the highest westward speed was recorded as $V(E, \tau) = -0.78$ mm/day at $\tau = 1096$ on December 22, 2010.

Figure 6-2c displays the same time window with the M9 spike removed as a data deficiency. Events 1 and 2 on the magnified path are before and after the M9 EQ, and $V(E, \tau)$ is $+0.12$ mm/day at $\tau = 1174$ (March 10, 2011, label 1), and $+0.10$ mm/day at $\tau = 1175$ (March 12, 2011, label 2).

In Fig. 6-2d, the observed $\{c\}$ is only from December 4, 2007 ($j = 1$) to March 10, 2011 ($j = 1174$). The expanded time window shows the lunar synodic tidal loading on the $V(E, \tau)$ and $A(E, \tau)$, labeled as 30 days. The westward trend change label (2010/07/11) notes the similar phase relationship among $D(E, \tau)$, $V(E, \tau)$, and $A(E, \tau)$ at $\tau = 932$, as in Figs. [6-1b - 6-1d] at Chichijima station. The magnified path shows $V(E, \tau) = +0.12$ mm/day at $\tau = 1157$ (on February 21, 2011) with the last available data of $d(E, j)$ at $j = 1174$ on March 10, 2011.

On March 8, 2011, the Chichijima-A station observed $V(E, \tau) = +0.08$ mm/day, while Chichijima station recorded the last available $V(E, \tau)$ as $+0.06$ mm/day.

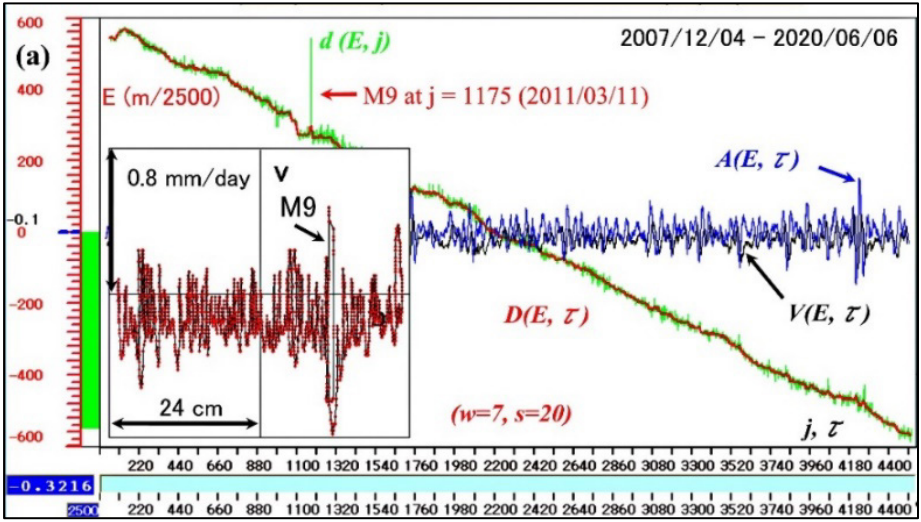

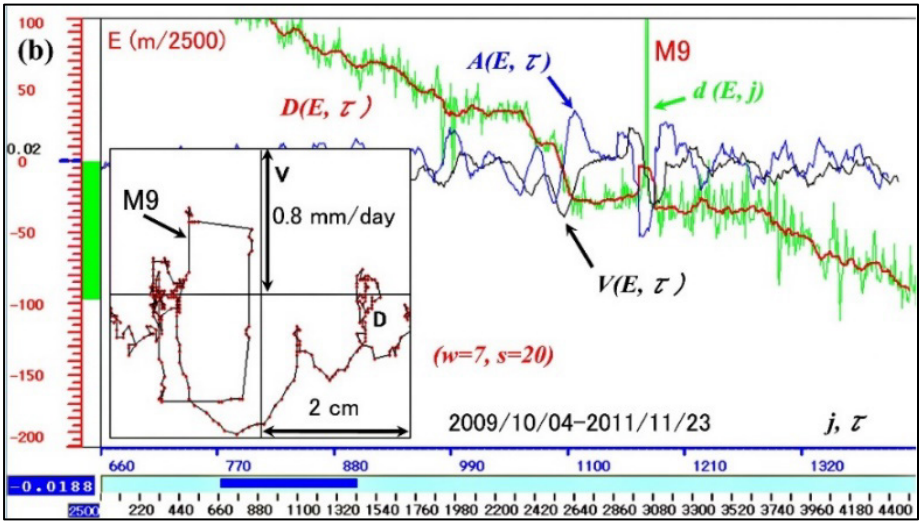

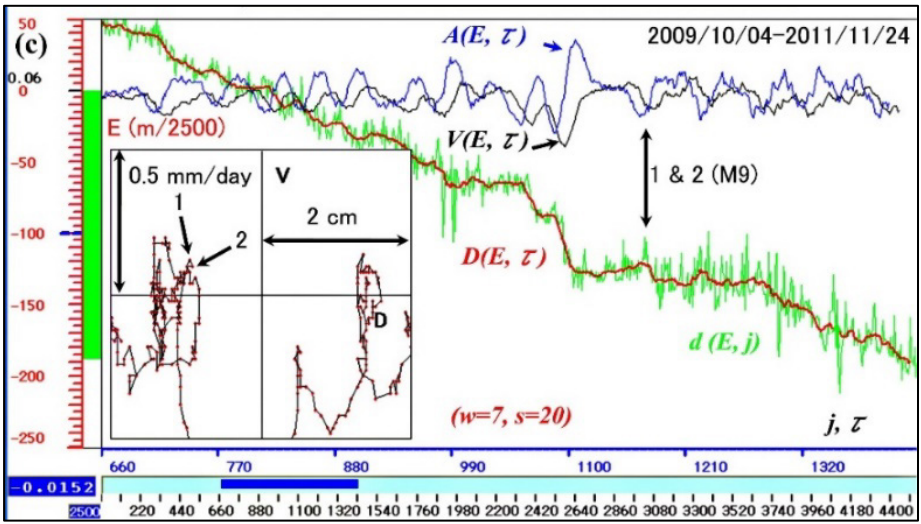

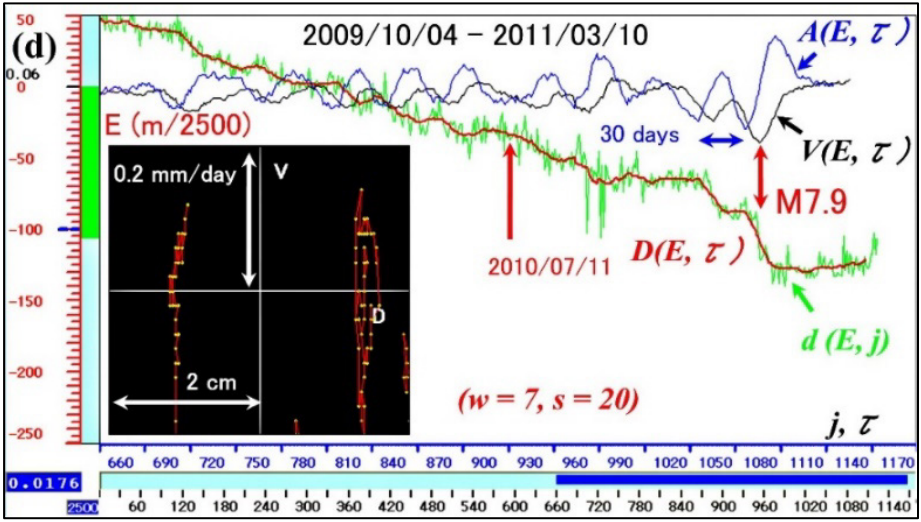

**Fig. 6-2. The abnormal westward motion at Chichijima-A station.**
(a) The series {$c$} covers the period from $j = 1$ (December 4, 2007) to $j = 4533$ (June 6, 2020, with an environmental spike noise caused by the M9 EQ on the $d(E, j)$ at $j = 1175$ (March 11, 2011). Parameters $w = 7$ and $s = 20$ are used for $D(E, \tau)$, $V(E, \tau)$, and $A(E, \tau)$. The $D(E, \tau) - V(E, \tau)$ plane has a scale of (24 cm, 0.8 mm/day). The path is labeled with the M9 EQ. The $V(E, \tau)$ and $A(E, \tau)$ are shown in relative scales from the graphical offset origin. (b) The expanded time window is from $j = 660$ (October 4, 2009) to $j = 1430$ (November 23, 2011), with the $D(E, \tau) - V(E, \tau)$ plane of a magnified scale of (2 cm, 0.8 mm/day). The path shows the highest westward speed was $V(E, \tau) = -0.78$ mm/day at $\tau = 1096$ on December 22, 2010. (c) The window is from $j = 660$ (October 4, 2009) to $j = 1430$ (November 24, 2011), with the M9 spike removed as a data deficiency at 1 & 2 (M9) on $d(E, j)$. Two events labeled 1 and 2 on the (2 cm, 0.5 mm/day) path are the events before and after the M9. Label 1 has $V(E, \tau) = +0.12$ mm/day at $\tau = 1174$ (March 10, 2011), and label 2 has $V(E, \tau) = +0.10$ mm/day $\tau = 1175$ (March 12, 2011). (d) The series {$c$} covers the period from $j = 1$ (December 4, 2007) to $j = 1174$ (March 10, 2011), and the expanded window is from $j = 660$ (October 4, 2009) to $j = 1174$ (March 10, 2011). Label 2010/07/11 is on $d(E, j)$, $D(E, \tau)$, $V(E, \tau)$, and $A(E, \tau)$ at $\tau = 932$ on July 11, 2010. The path shows $V(E, \tau) = +0.12$ mm/day at $\tau = 1157$ (February 21, 2011), which corresponds to $j = 1174$ on March 10, 2011 (due to $j = \tau + 17$). It also shows $V(E, \tau) = +0.08$ mm/day on March 8, 2011.

## 6.3 Hahajima station (on the western edge of the subducting northwestern Pacific Plate)

Hahajima station (26.6352° N, 142.1628° E) experienced data deficiencies from June 16, 2004, to September 23, 2005. Therefore, the time series {$c$} only covers data from September 24, 2005 ($j = 1$).

In Fig. 6-3a, a spike is observed on $d(E, j)$ at $j = 1987$ due to the M9 EQ on March 11, 2011. The path shows abnormal motion only before the M9 EQ.

Figure 6-3b displays the path of the M9 spike event. Prior to the M9 EQ, the path exhibited the highest westward speed, $V(E, \tau) = -0.84$ mm/day at $\tau = 1908$ on December 22, 2010, similar to Chichijima and Chichijima-A stations.

Figure 6-3c shows the same expanded time window from $j = 1040$ to $j = 2340$, with the M9 spike removed as deficit data. Two events 1 and 2 on the path are the events before and after the M9 EQ. The $V(E, \tau)$ is $+0.18$ mm/day at $\tau = 1986$ (March 10, 2011, label 1), and $+0.20$ mm/day at $\tau = 1987$ (March 12, 2011, label 2).

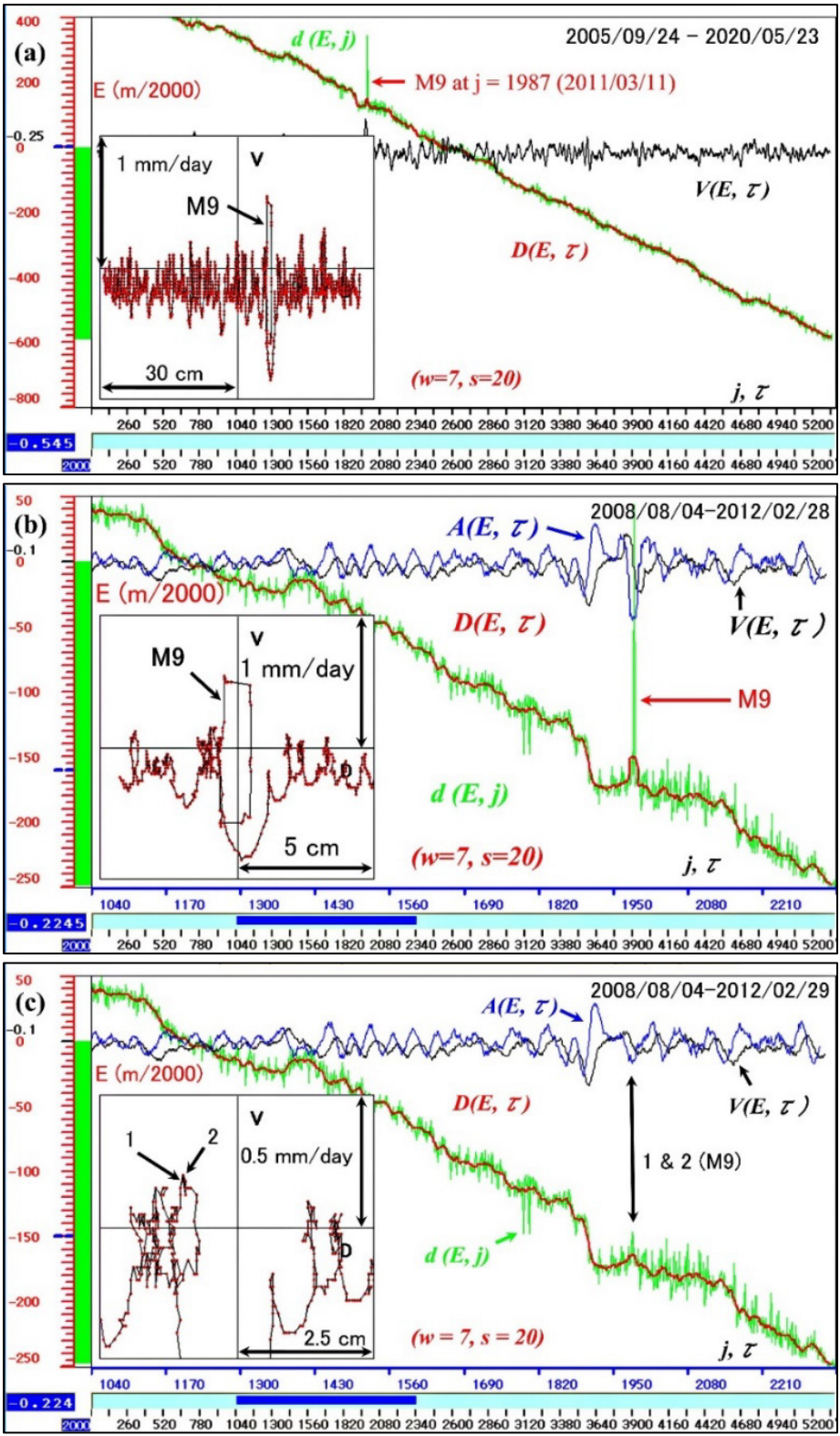

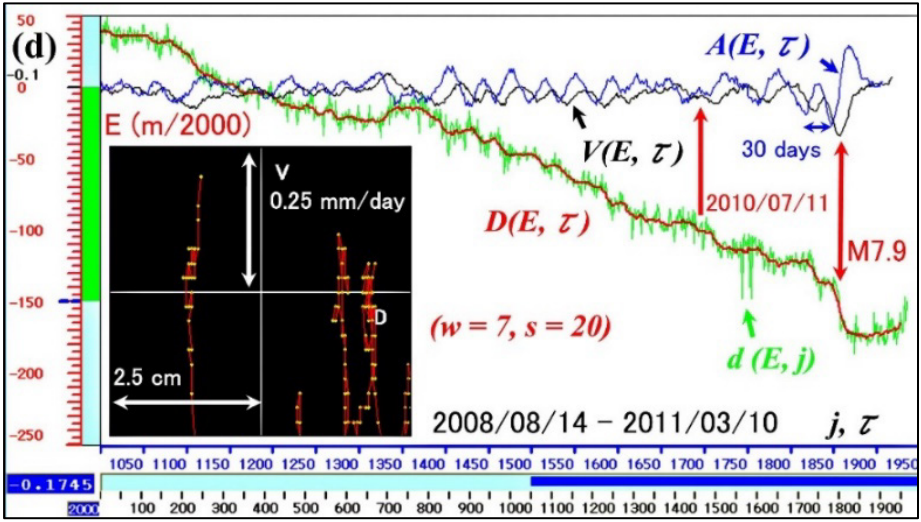

**Fig. 6-3. The abnormal westward motion at Hahajima station.**

(a) The series {$c$} covers the period from September 24, 2005 ($j = 1$) to May 23, 2020 ($j = 5345$), with parameters $w = 7$ and $s = 20$ used for $D(E, \tau)$, $V(E, \tau)$, and $A(E, \tau)$. The $D(E, \tau) - V(E, \tau)$ plane has a scale of (30 cm, 1 mm/day), with a spike noise caused by the M9 EQ on March 11, 2011, at $j = 1950$. (b) The time window is from $j = 1040$ (August 4, 2008) to $j = 2340$ (February 28, 2012), and the magnified path with a scale of (5 cm, 1 mm/day) shows a sudden change due to the M9 spike. (c) The window is the same as Fig. part b with the M9 spike noise removed as a data deficiency at 1 & 2 (M9) on the $d(E, j)$. Two events 1 and 2 on the (2.5 cm, 0.5 mm/day) path are the events before and after the M9 EQ. (d) The series {$c$} covers the period from September 24, 2005 ($j = 1$) to March 10, 2011($j = 1986$), one day before the Tohoku M9 EQ. The expanded time window is from $j = 1050$ (August 14, 2008) to $j = 1986$ (March 10, 2011), with label 2010/7/11 indicating the trend change observed at Chichijima stations is on $D(E, \tau)$ at $\tau = 3744$. The (2.5 cm, 0.25 mm/day) path ends at $V(E, \tau) = 0.20$ mm/day, which has the last $d(E, j)$ on March 10, 2011. It also shows $V(E, \tau) = 0.13$ mm/day on March 8, 2011.

In Fig. 6-3d, the time series {$c$} covers data from September 24, 2005 ($j = 1$) to March 10, 2011 ($j = 1986$), which is one day before the Tohoku M9 EQ. The highest speed observed was $V(E, \tau) = -0.84$ mm/day at $\tau = 1908$ on December 22, 2010. The lunar synodic period of 30 days is displayed on $A(E, \tau)$ in the expanded time window from $j = 1050$ to $j = 1986$ (March 10, 2011). The path exhibited an eastward speed of $V(E, \tau) = 0.12$ mm/day on March 8, 2011, and 0.20 mm/day on March 10, 2011. The eastward speed of $V(E, \tau) = 0.12$ mm/day on March 8, 2011, was twice that of Chichijima (Table 1).

The label 2010/07/11 indicates the trend change observed at Chichijima station. The onset displayed a similar phase relationship among $D(E, \tau)$, $V(E, \tau)$, and $A(E, \tau)$ at $\tau = 1744$ on July 11, 2010, similar to those in Figs. 6-1b - 6-1d, and 6-2d.

The highest speed was $V(E, \tau) = -0.84$ mm/day at $\tau = 1908$ on December 22, 2010, and then the westward motion nearly stopped by February 21, 2011 (in day $\tau$). The subducting plate reversed the westward motion.

### 6.4 Minamitorishima station (on the northwestern Pacific Plate)

Minamitorishima station (N 24.2901°, E 153.9787°) in the Northwest Pacific Ocean (Figs. 3 and 9) began its operation on July 1, 2004, but had many observational deficits in {$c$}. Specifically, there were data gaps during the following periods: 1) May 10, 2005, to May 29, 2005 (at $j = 314$). 2) August 3, 2005, to September

14, 2005 (at $j = 379$). 3) October 5, 2006, to November 23, 2006 (at $j = 765$). 4) July 16, 2008, to August 20, 2008 (at $j = 1362$). 5) September 30, 2008, to November 26, 2008 (at $j = 1401$). 6) June 9, 2009, to November 17, 2009 (at $j = 1595$). 7) August 17, 2010, to September 25, 2010 (at $j = 1865$). 8) October 13, 2010, to January 1, 2011 (at $j = 1882$). 9) October 22, 2016, to March 11, 2017 (at $j = 3996$). As of July 25, 2020, the station has stopped observing since March 18, 2020.

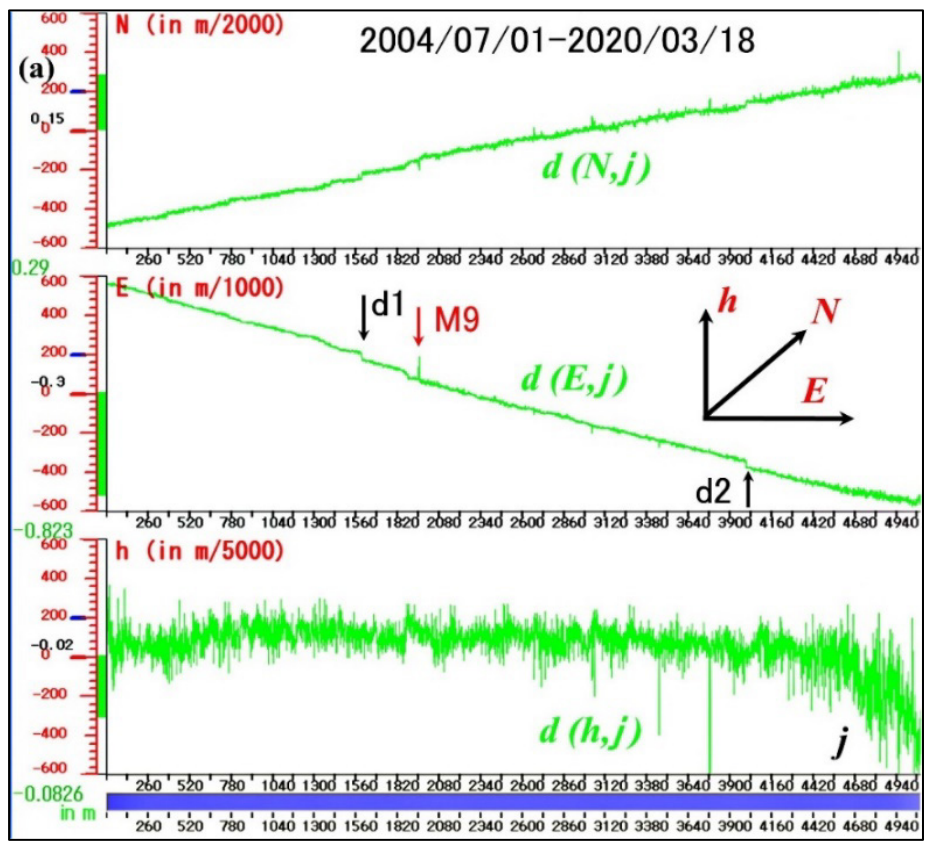

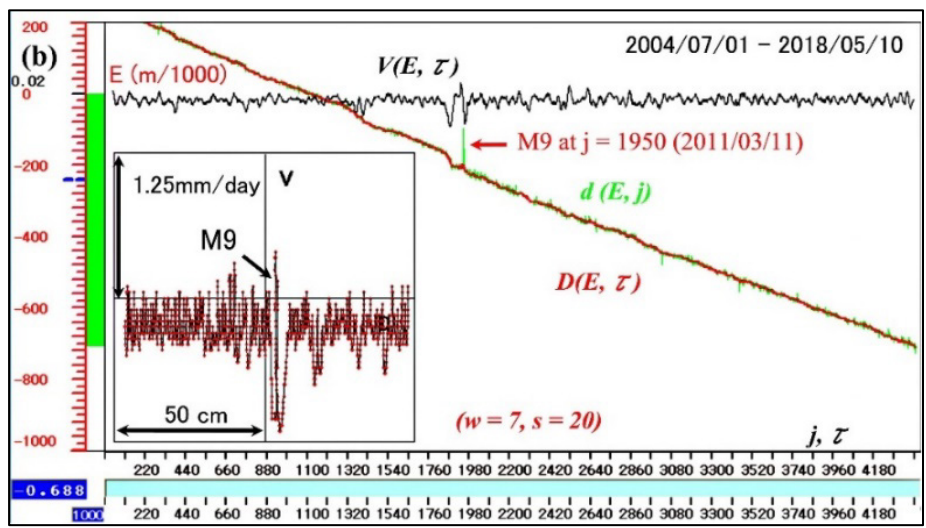

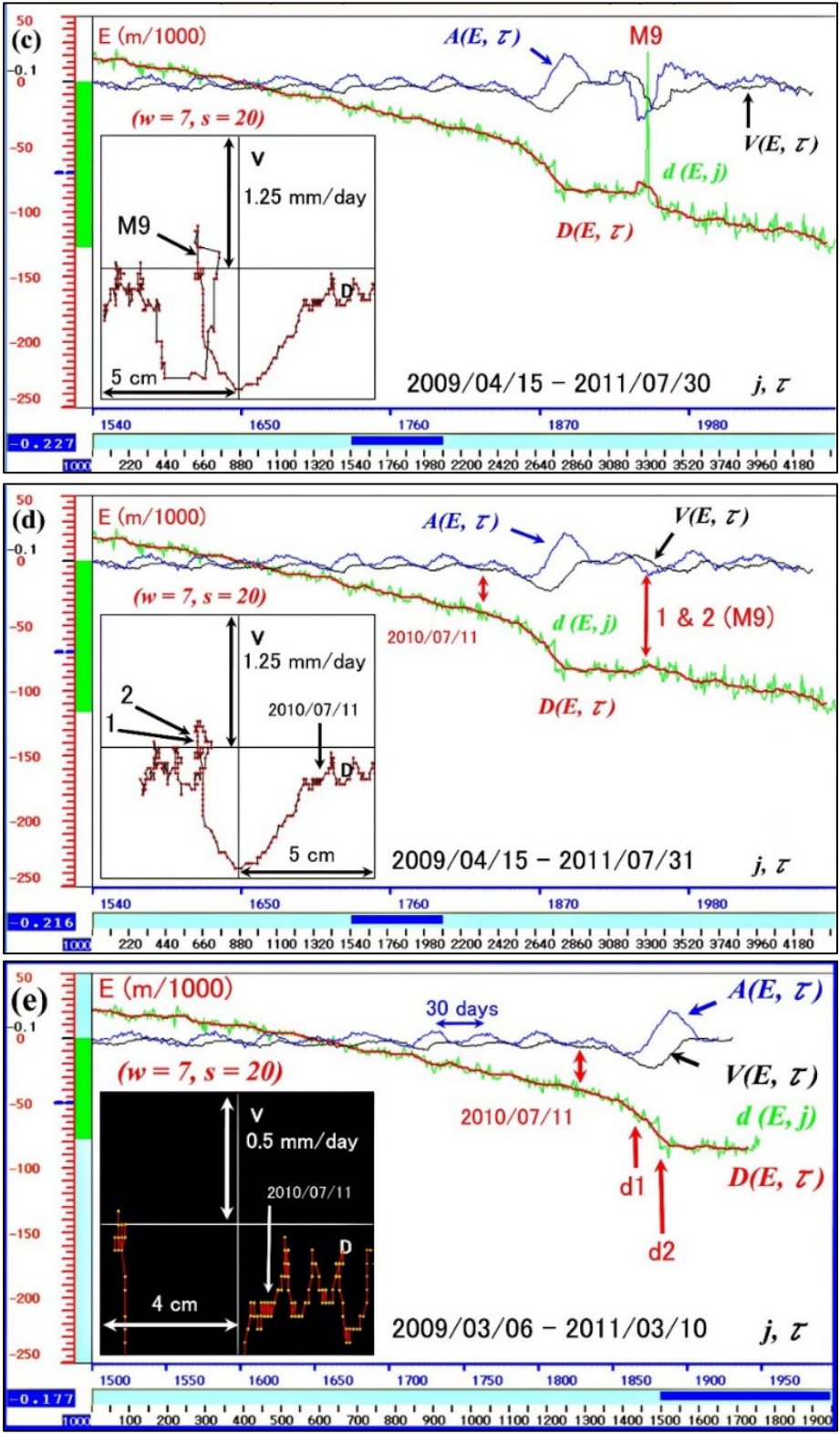

**Fig. 6-4. The abnormal westward motion at Minamitorishima station.**
(a) The series {$c$} covers the period from $j = 1$ (July 1, 2004) to $j = 5096$ (March 18, 2020), with time gaps at d1 and d2. The d1 gap is from June 9, 2009, to November 17, 2009 ($j = 1595$), and the d2 gap is from October 22, 2016, to March 11, 2017 ($j = 3996$). The M9 Tohoku earthquake on March 11, 2011, caused a spike at $j = 1950$. The {$h$} component becomes very noisy after $j = 4420$ (May 10, 2018) so that the analyzing period is from $j = 1$ to $j = 4420$. (b) The period is from $j = 1$ to $j = 4420$, with smoothed offsets caused by d1 and d2.

Parameters $w = 7$ and $s = 20$ are used for $D(E, \tau)$, $V(E, \tau)$, and $A(E, \tau)$. The $D(E, \tau) - V(E, \tau)$ plane has a scale of (50 cm, 1.25 mm/day) with the spike noise caused by the M9 EQ at $j = 1950$. (c) The expanded time window is from $j = 1540$ (April 15, 2009) to $j = 2090$ (July 30, 2011), showing an abrupt change due to the M9 EQ on the (5 cm, 1.25 mm/day) path. (d) The time window is from $j = 1540$ (April 15, 2009) to $j = 2090$ (July 31, 2011), with the M9 spike event removed as deficient data at 1 & 2 (M9) on $d(E, j)$. Events 1 and 2 on the (5 cm, 1.25 mm/day) path are the events before and after the M9 EQ. (e) The $\{c\}$ covers the period from July 1, 2004, to March 10, 2011 ($j = 1949$), one day before the Tohoku M9 EQ. The expanded time window is from j = 1500 (March 6, 2009) to j = 1949 (March 10, 2011). The lunar synodic loading oscillation of 30 days is present on $A(E, \tau)$, $V(E, \tau)$, and the (4 cm, 0.5 mm/day) path. The time gaps from August 17, 2010, to September 25, 2010 (at j = 1865), and from October 13, 2010, to January 1, 2011 (at j = 1882) are respectively indicated by d1 and d2 on $d(E, j)$ and $D(E, \tau)$. The trend change observed at Chichijima stations on July 11, 2010, is labeled as 2010/7/11 on $D(E, \tau)$ at $\tau = 1830$, showing qualitative agreement.

In Fig. 6-4a, all observed time series data$\{c\}$ are shown, with an average moving rate of 7.1 cm per year to the west and 2.3 cm/year to the north in $\{E\}$ and $\{N\}$, respectively.

Figure 6-4b displays the single M9-related-abnormal westward motion on the phase-plane path.

Figure 6-4c shows the M9 spike noise of $d(E, j)$, causing a stepwise motion on the magnified path.

In Fig. 6-4d, the M9 spike event is removed as a data deficiency in $\{c\}$, resulting a smooth path connecting events before (event 1) and after (event 2) the M9, with $V(E, \tau) = + 0.05$ mm/day at event 1, and $V(E, \tau) = + 0.10$ mm/day at event 2. Label 2010/07/11 denotes the westward trend change observed at Chichijima station, with a similar phase relationship among $D(E, \tau)$, $V(E, \tau)$, and $A(E, \tau)$ at $\tau = 1830$ on July 11, 2010, similar to those in Figs. 6-1b - 6-1d, 6-2d, and 6-3d. The trend change triggered the anomalous $D(E, \tau) - V(E, \tau)$ path, reaching the maximum speed of $- 1.15$ mm/day at $\tau = 1877$ (October 8, 2010). It was not December 22, as in Chichijima and Hahajima, because there was a data deficiency period from October 13, 2010, to January 1, 2011 ($j = 1882$). The highest eastward speed was $V(E, \tau) = + 0.25$ mm/day five days after event 2 (March 12, 2011). The plate moved eastward by 4 mm, returning to the standard lunar synodic loading.

In Fig. 6-4e, the path and $V(E, \tau)$ show that the anomalous motion started at $V(E, \tau) = - 0.34$ mm/day on July 11, 2010. The eastward speed on March 10, 2011, was $V(E, \tau) = + 0.05$mm/day. Overall, the Pacific Plate's $\{E\}$ path on the M9 EQ is in good harmony with the $\{E\}$ path of Chichijima and Hahajima on the Pacific Plate's western edge (Fig. 3). However, it is worth noting that the speed $V(E, \tau) = + 0.05$mm/day on March 10, 2011, increased by five times to $V(E, \tau) = + 0.25$mm/day on March 17. The increase after the megathrust EQ suggests some property difference between the Pacific Plate (Minamitorishima with increase) and its subducting western edge (Chichijima and Hahajima without increase).

## 6.5 Summary of the abnormal motion

Table 1 summarizes the GPS observations with $w = 7$ and $s = 20$ in sections 6.1- 6.4. The observed data used for the P-Ws analyses in Table 1 are until March 10, 2011, one day before the Tohoku M9 EQ event. Chichijima station terminated the operation after March 8. Each movement $D(E, \tau)$ on March 8 and 10 is the corresponding net displacement from $D(E, \tau)$ at $V(E, \tau) = 0.00$ mm/day to $D(E, \tau)$ on each date.

**Table 1 (Abnormal motion of the subducting northwestern Pacific Plate)**

| GPS station | Westward(−) and Eastward(+) speed $V(E,\tau)$ mm/day | | | | | | |
|---|---|---|---|---|---|---|---|
| | $D(E,\tau)$ trend change | | Max $V(E,\tau)$ | | $V(E,\tau)$ and $D(E,\tau)$ mm on March 8 and 10 | | |
| | yyyy/mm/dd | $V(E,\tau)$ | yyyy/mm/dd | $V(E,\tau)$ | March 8 | | March 10 |
| Chichijima | 2010/07/11 | −0.25 | 2010/12/22 | −0.69 | +0.06 | +1.6 mm | |
| | No observation after 2011/03/08 | | | | | | |
| Chichijima-A | 2010/07/11 | −0.26 | 2010/12/22 | −0.78 | +0.08 | +2.0 mm | +0.12 +2.4 mm |
| Hahajima | 2010/07/11 | −0.28 | 2010/12/22 | −0.84 | +0.12 | +2.0 mm | +0.20 +2.5 mm |
| Minamitorishima | 2010/07/11 | −0.34 | 2010/10/08 | −1.15 | 0.00 | 0.0 mm | +0.05 0.0 mm |
| | No data available; 2010/08/07～2010/09/25, 2010/10/13～2011/01/01 | | | | | | |

## 7 A bulge-bending deformation observation

Bulge-bending deformation refers to the gradual bending of the overriding plate of Tohoku by the eastward continental plate-driving force over fifteen months, causing the over-riding crust of Tohoku to bulge upwards and downwards by a few millimeters. This deformation evolved from the expected regular deformation that had been occurring for the past three hundred years (section 7.4), as shown in Fig. 4a. The GPS observation detected the onset of the bulge-bending deformation, which grew in three distinct phases: an initial phase of 6 months with gradual subsidence of 1 to 2.8 mm across the Tohoku region, followed by a transition phase of 6 months with further gradual subsidence of 3.3 mm on the east coast, and a final phase of 3 months with an upheaval growth of 1 to 3 mm across Tohoku until the March 2011 M9 EQ event, as summarized in section 7.4.

In January 2010, the initial phase began, initiating a westward restoring force of the west coast compressed eastwardly by the continental plate-driving force. In June 2010, the east coast began the transition phase with a further gradual subsidence of 3.3 mm, accompanied by less than 1 mm subsidence on the west coast and the Top in Fig. 4. The east coast began pulling the subducting plate westward by grasping the fault surface with the subsidence. The dotted arrow over the fault in Fig. 4b indicates the pulling action by the overriding eastern edge. The east coast's pulling was activated by the overriding plate-driving eastward force with the eastward displacement that countered by an enormous restoring force of the compressed west coast. In November 2010, the east coast began the final phase by experiencing an upheaval growth, gradually growing to 1.2 mm with a lifting force generation until the 2011 EQ event.

The daily displacement $d(h,j)$ has a background noise of about ± 20 mm. Thus, only the equations of motion with a smooth $D(h,\tau) - V(h,\tau)$ path with the parameters $w = 200$ and $s = 300$ in days can quantify the bulge-bending deformation with subsidence and upheaval by minimizing yearly and seasonal variations and environmental noises in $\{c\}$ ($c = h$). The path has a resolution of 0.1 mm, four orders of magnitude greater than the daily noise level of ± 20 mm. The $D(h,\tau) - V(h,\tau)$ path and $A(h,\tau)$ require displacement $d(h,j)$ ranging from $d(h, \tau - 350)$ to $d(h, \tau + 350)$ and from $d(h, \tau - 500)$ to $d(h, \tau + 500)$, as shown in Eqs. (1) – (3).

We reassign the three phases of initial, transitional, and final as; an initial phase (referred to as Phase 0), the first phase (Phase 1), and the second phase (Phase 2) as in Table 2 (section 7.4). Each phase showed an approximately linear path on the $D(h,\tau) - V(h,\tau)$ plane, where $V(h,\tau) \approx k1 \times D(h,\tau) + k2$, and $k1$ and $k2$ are constants, and $D(h,\tau)$ is a displacement from an offset origin. The time rate of $V(h,\tau)$ is $A(h,\tau) \approx k1 \times V(h,\tau) \approx k1^2 \times D(h,\tau) + k1 \times k2$. The observed constant $k1$ and $k2$ had the same sign in Phase 1 (abbreviated as Ph-1) and Phase 2 (abbreviated as Ph-2). The bulge-bending deformations in Ph-1 and Ph-2 showed that the $A(h,$ $\tau$) was always positive (as in Figs. 7-1b, 7-2c, 7-3c, and Table 2). Therefore, it follows that $k1^2 \times D(h, \tau) > -k1 \times k2$, indicating that the positive acceleration (a lifting force on the fault surface) can occur with subsidence of negative $D(h, \tau)$. The lifting force with negative $D(h, \tau)$ is a restoring force. In contrast, a positive $D(h, \tau)$ indicates an upheaval growth caused by bulging, which is always a non-restoring (non-elastic) lifting force.

The positive acceleration exerted on each GPS station is due to bulging that generated a lifting force on the fault surface. The relative force, $F(h, \tau)$, is approximately equal to $A(h, \tau) - k1 \times k2$, which is in turn approximately equal to $k1^2 \times D(h, \tau)$, and indicates a non-restoring upheaval force with positive $D(h, \tau)$ moved up by the bulge-bending deformation. The buildup of lifting force on the east coast gradually reduced the static frictional strength of the subduction interface, eventually causing the shear stress to exceed the frictional strength and leading to the decoupling of the overriding and subducting plates.

## 7.1 Bulge on the east coast (Onagawa station)

The east coast has an array of GPS stations that show co-seismic downward vertical displacements, as seen in Fig. 3a. Onagawa station (38.4492° N, 141.4412° E) is one such station.

### 7.1.1 Until March 10, 2011 (Figs. 7-1a and 7-1b), one day before the Tohoku M9 EQ

The $D(h, \tau)$ and the $D(h, \tau) - V(h, \tau)$ path in Figs. 7-1a and 7-1b show an upheaval growth of 1.2 mm over 115 days, starting at the dot-arrows labeled "Bulge starts" and S2. The S2 onset time is $\tau = 3615$, which was November 29, 2009, 350 days before November 14, 2010, in time $j$. This date precedes December 22, 2010, when the westward speed of the subducting northwestern Pacific Plate became abnormal, as discussed in section 6.

The $D(h, \tau) - V(h, \tau)$ path in Fig. 7-1b shows three approximately linear segments: S0 in Phase 0 (Ph-0) with a nearly zero slope, S1 in Phase 1 (Ph-1) with a negative slope in subsidence, and S2 in Phase 2 (Ph-2) with a positive slope in upheaval growth. All segments experience positive $A(h, \tau)$, which grew linearly in time $\tau$, suggesting that the bulge-bending deformation on the east coast GPS station was under a lifting force.

Segment S1 of the $D(h, \tau) - V(h, \tau)$ path has $\Delta D(h, \tau) = -3.3$mm (subsidence) and $\Delta V(h, \tau) = + 0.0142$ mm/day over 158 days in time $\tau$, from $\tau = 3457$ on June 24, 2009, to $\tau = 3615$ on November 29, 2009, and from $j = 3807$ on June 9, 2010, to $j = 3965$ on November 14, 2010, in time $j$. The average acceleration is $+ 8.99 \times 10^{-5}$ mm/day$^2$, exerting the upward acceleration (Force) on the east coast GPS station. The average force is indicative of the bulge-bending strength. The time rate unit conversion is mm/day$^2 = 1.33959 \times 10^{-11}$ cm/sec$^2$.

Segment S2 has $\Delta D(h, \tau) = 1.2$ mm (upheaval) and $\Delta V(h, \tau) = + 0.0054$ mm/day over 115 days, from $j = 3965$ on November 14, 2010, to $j = 4080$ on March 10, 2011, one day before the M9 EQ. The average acceleration is $+ 4.7 \times 10^{-5}$ mm/day$^2$.

The $D(h, \tau) - V(h, \tau)$ paths (S0) to S1 in Figs. 7-1a and 7-1b show a nearly constant downward speed of $V(h, \tau) \approx -0.018$ mm/day in the frequency region selected with $w = 200$ and $s = 300$, except for a significant co-seismic displacement. The absence of appreciable changes in $V(h, \tau)$ suggest that $A(h, \tau) \approx 0$. The $A(h, \tau)$ in Fig. 7-1b shows a slight fluctuation until it starts to increase steadily at S0. Thus, the regular subsidence deformation at the east coast in Fig. 4a had no significant vertical force component.

The positive average acceleration (lifting force) suggests that segments S0, S1, and S2 are part of a bulge-bending deformation under the negatively (westwardly) increasing $V(E, \tau)$ and $A(E, \tau)$ with the westward displacement, as shown in Fig. 7-1c. Thus, the well-known elastic-rebound theory cannot explain the non-restoring force acting on the east coast and the sudden appearance of the upward force component ($A$ ($h$, $\tau$) > 0) nearly normal to the fault line (surface) shown in Fig.4.

The downward (westward) bend on the $D$ ($E$, $\tau$) – $V$ ($E$, $\tau$) path in Fig. 7-1c started at $\tau$ = 3434 on June 1, 2009, which is May 17, 2010, in time $j = \tau + 350$. The westward bend-onset characterized by the directional change of $A$ ($E$, $\tau$) > 0 (eastward) to $A$ ($E$, $\tau$) < 0 is Ph-0 in Table 2, preceded S1-onset by 23 days. The westward movement from Ph-0 was $D$ ($E$, $\tau$) = − 13.2 mm by March 10, 2011, one day before the Tohoku M9 EQ.

Segment S1 of the $D$ ($h$, $\tau$) – $V$ ($h$, $\tau$) path is part of the bulge-bending deformation process at Onagawa station. However, we refer to the transition from S1 to S2 as the bulge-bending deformation onset (bulge-onset) or 'Bulge starts' for the detection purpose.

As summarized in section 7.4, segment S1 has a lifting force twice higher than the S2 force. The S1-onset was at time $\tau$ = 3457 (on June 14, 2009), including $d$ ($h$, $j$) at time $j$ = 3807 (on June 9, 2010). The date precedes July 11, 2010, when the westward motion's trend change began, as described in section 6. The S1 ends at the bulge-onset, $\tau$ = 3615, which was on November 29, 2009, 350 days behind November 14, 2010, in time $j$. This date precedes December 22, 2010, when the abnormal westward speed of the subducting Pacific Plate became about three times higher than the standard speed. Thus, the bulge-bending deformation is the geophysical origin of the abnormal westward motion of the subducting oceanic plate.

We note that each linearity of S1 and S2 holds for the respective $D$ ($h$, $\tau$) – $A$ ($h$, $\tau$) path. However, $A$ ($h$, $\tau$) at the S2-onset ($\tau$ = 3615) requires $d$ ($h$, $j$) at $j$ = 4115 (> $j$ = 4080, March 10, 2011). Thus, the second linear segment is not available.

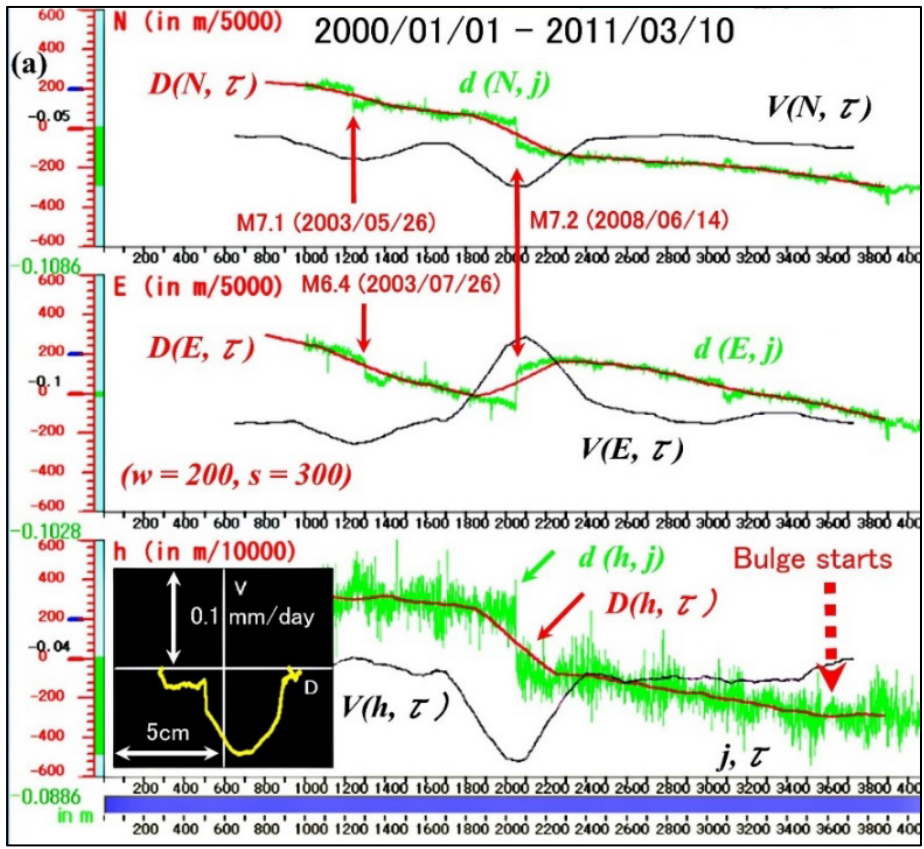

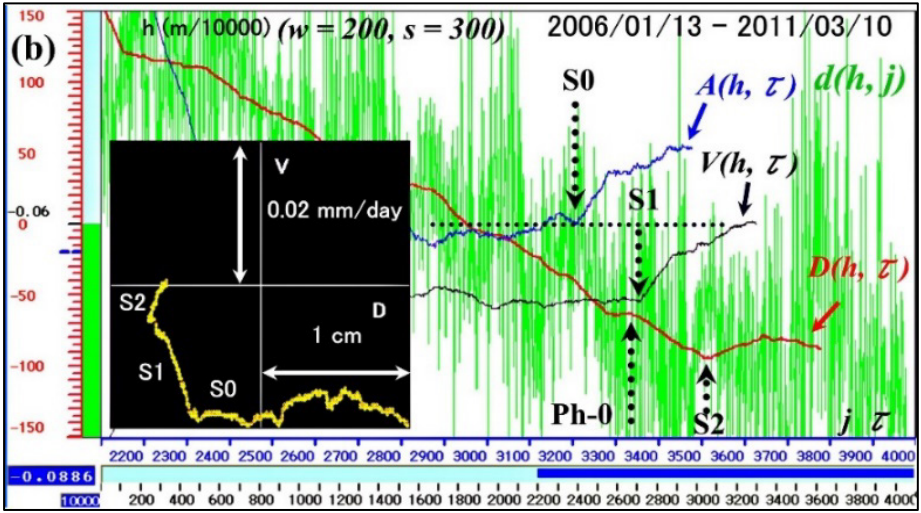

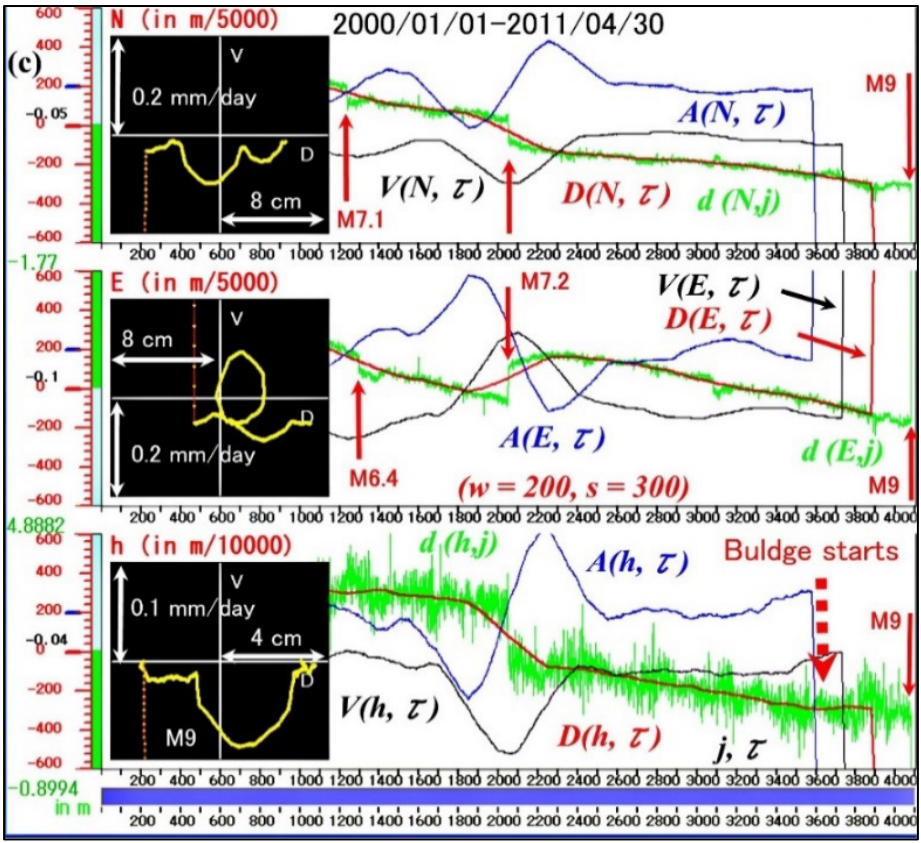

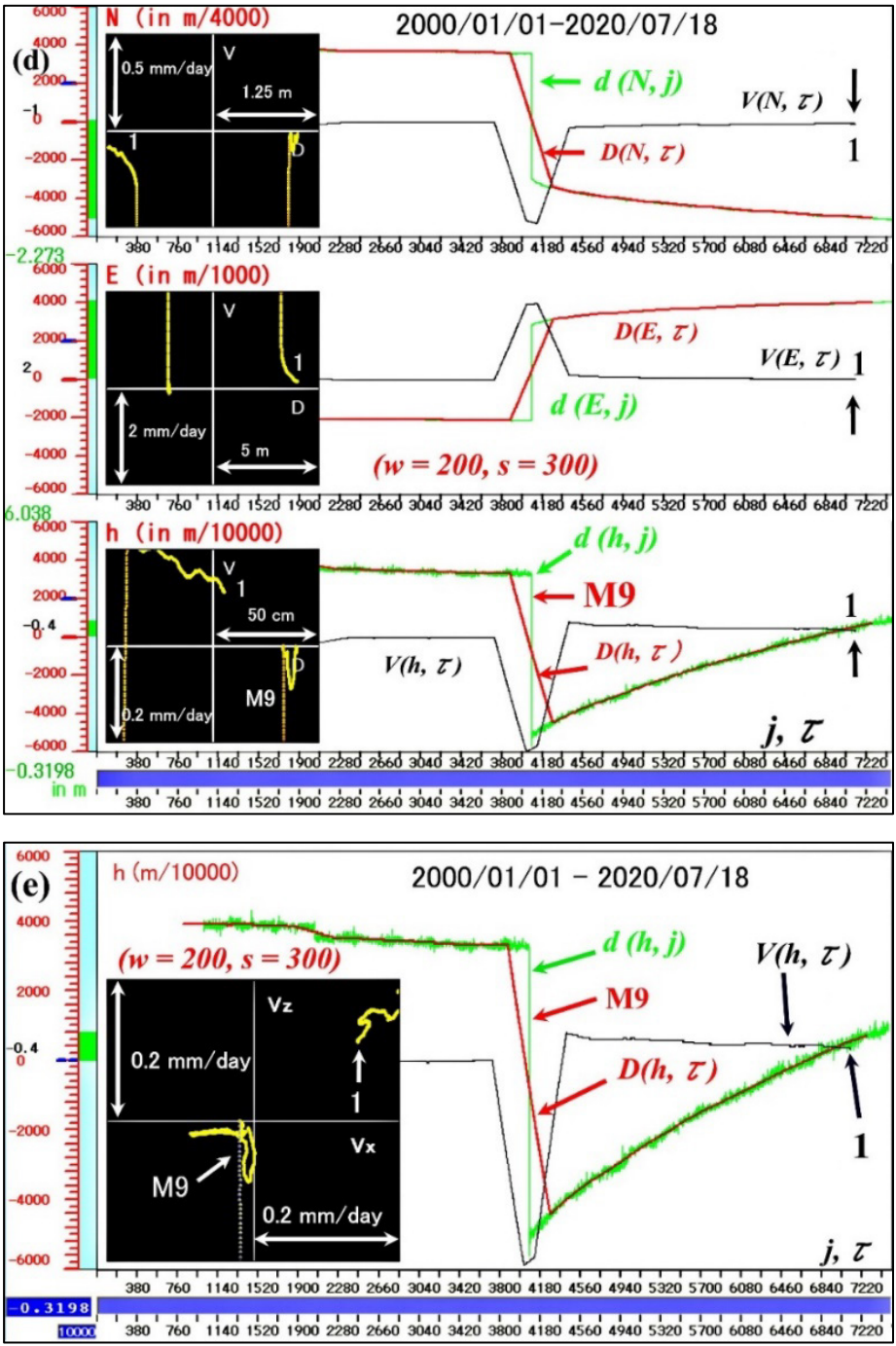

**Fig. 7-1. The bulge and M9 observations at Onagawa station.**

(a) The series {$c$} covers the period from January 1, 2000 ($j = 0$) to March 10, 2011 ($j = 4080$), with sudden large co-seismic shifts. The magnitudes and dates are indicated on $d(c, j)$ at arrows. Parameters $w = 200$ and $s = 300$ are used for $D(c, \tau)$, $V(c, \tau)$, and $A(c, \tau)$. The $D(h, \tau) – V(h, \tau)$ phase plane has a scale of (5 cm, 0.1 mm/day) with the offset origin $D(h, \tau) = -0.04$ m and $V(h, \tau) = 0$ mm/day. The bulge-onset is labeled at arrow 'Bulge starts' on the $D(h, \tau)$ at $\tau = 3615$, November 29, 2009 (November 14, 2010, in time $j$). (b) The expanded window is from January 13, 2006, to March 10, 2011. The magnified phase plane has a scale of (1 cm, 0.02 mm/day) with the origin $D(h, \tau) = -0.062$ m ($-0.06$ m $-0.002$ m). Approximately linear segments are labeled as S0, S1, and S2 on the phase path. Each segment's onset location is at a dot-arrow with its name. The S2 is the 'Bulge starts' in Fig. part a. Arrow Ph-0 between S0 and S1 is the onset location of an abnormal movement on the $D(E, \tau) - V(E, \tau)$ path in Fig. part c. A horizontal dot-line at the offset origin 0 ($-0.06$) is the abscissa for $V(h, \tau)$ and $A(h, \tau)$; namely, $V(h, \tau) = A(h, \tau) = 0$. The magnitudes of $V(h, \tau)$ and $A(h, \tau)$ are on relative scales. (c) The series $\{c\}$ covers the period from January 1, 2000 ($j = 0$) to April 30, 2011 ($j = 4107$). It includes the M9 EQ on March 11, 2011 ($j = 4081$), arrow M9 on $d(c, j)$. The $D(c, \tau) - V(c, \tau)$ planes have a scale of (8 cm, 0.2 mm/day) with the offset origins for $c = N$ and $E$, and (4 cm, 0.1 mm/day) for $c = h$. The $D(c, \tau) - V(c, \tau)$ path shows the doted path jumped by the M9 EQ. Every $d(c, j)$ and its column height have the co-seismic shifts saturated by the M9 EQ with its digital value from the position at $j = 0$. The southward shift is 1.90 m ($= -1.77 - 0.05 - 0.08$), the eastward shift is 4.7482 m ($= 4.8882 - 0.10 - 0.04$), and the downward shift is 0.9694 m ($= -0.8994 - 0.04 - 0.03$). (d) The series $\{c\}$ covers the period from January 1, 2000 ($j = 0$) to July 18, 2020 ($j = 7473$). The $D(c, \tau) - V(c, \tau)$ planes have the scale of (1.25 m, 0.5 mm/day) for $c = N$, and (5 m, 2 mm/day) $c = E$, and (50 cm, 0.2 mm/day) for $c = h$. The $D(c, \tau) - V(c, \tau)$ path ends at label 1. The location of the M9 EQ has arrow M9 on $d(h, j)$ that saturated the $D(c, \tau) - V(c, \tau)$ path. (e) The series $\{h\}$ covers the period from January 1, 2000, to July 18, 2020. The time-delayed $Vx = V(h, \tau - 600)$ and $Vz = V(h, \tau)$ path is on the (0.2 mm/day, 0.2 mm/day) plane. The M 9 event saturated the path. The last $V(h, \tau)$ and the corresponding point in the $Vx$ - $Vz$ plane have label 1.

### 7.1.2 On and After the Tohoku M9 EQ (Figs. 7-1c - 7-1e)

Figure 7-1c shows that the M9 EQ saturated $d(c, j)$ with 1.90 m southward, 4.7482 m eastward, and 0.9694 m downward shifts, respectively, as summarized in Table 3.

The co-seismic shifts due to the Tohoku M9 earthquake in Fig. 7-1d represent changes in $d(c, j)$ at $j = 4081$. As of July 18, 2020, the crustal motion is at arrow 1 on the $V(c, \tau)$ and label 1 on the phase path. The path at label 1 shows the positive (eastward) $V(E, \tau)$, which is the opposite of the westward trend observed in Fig. 7-1c. This suggests that the crustal state had not returned before the 2011 Tohoku M9, which is consistent with the GSI's eastward observation in Fig. 13-1-1 (Appendix B).

One way to analyze the crustal state is to use a time-delayed $Vx$ -$Vz$ path, where $Vx$ is a time-delayed velocity and $Vz$ is the present velocity [7]. An example with $Vx = V(h, \tau - 600)$ and $Vz = V(h, \tau)$ is in Fig. 7-1e, where the path is on the (0.2 mm/day, 0.2 mm/day) plane. The $V(h, \tau)$ at label 1 represents the crustal state in the 2020s. The 2020s ($Vx, Vz$) state is away from the state before the 2011 Tohoku M9, suggesting that the current east coast's vertical motion is still affected by the M9 EQ.

### 7.2 Bulge on the west coast (Ryoutsu2 station)

The west coast is the shore facing the Japan Sea in the schematics of Fig. 4 with the co-seismic upward displacement in Fig. 3a. Although Ryoutsu2 station (38.0633° N, 138.4717° E) is on an island off the west coast, the island is geologically a part of the main island, as seen in Fig. 3b.

### 7.2.1 Until March 10, 2011 (Figs. 7-2a - 7-2d), one day before the Tohoku M9 EQ

Figures 7-2a to 7-2d show the path of the upheaval growth of 2.3 mm in $\{h\}$ at Ryoutsu2 station on the west coast until one day before the Tohoku M9 EQ. The growth starts at the dot-arrow "Bulge starts" on the $D(h, \tau)$ and $D(E, \tau)$ paths in Figs. 7-2a and 7-2b, which corresponds to time $\tau = 4982$ on November 13, 2009 (or October 29, 2010, in time $j = 5332$).

The $D(h, \tau) - V(h, \tau)$ path in Fig. 7-2c shows three approximately linear segments, segment 0 (S0), segment 1 (S1), and segment 2 (S2), which respectively correspond to Ph-0, Ph-1, and Ph-2 of the bulge-bending deformation process. The onset location of each segment is at the corresponding dot-arrow, S0, S1, and S2. Segment S0 starts with an increasing $A (h, \tau)$ whose acting direction changed upward (positive above the horizontal dot-line in Fig. 7-2c), indicating a lifting force, like the east coast.

An anomalous $V (E, \tau)$ started at dot-arrow Ph-0, characterized by the change of $A (E, \tau) > 0$ (eastward) to $A (E, \tau) < 0$, as in section 7.4. The Ph-0 preceded S0, suggesting that an underlying bulge-bending over Tohoku began with the westward push exerted by a restoring force of the compressed west coast, counteracting the eastward displacement caused by the eastward plate-driving force. The bulge-onset is detected at the sharp corner from S1 to S2 on the $D (h, \tau) – V (h, \tau)$ path at dot-arrow S2 on the $D (h, \tau)$ path. The bulge-onset is at $\tau = 4838$ on June 22, 2009 (or June 7, 2010, in time $j = 5188$), and S2 has an upheaval growth of 2.3 mm from the bulge onset.

The bulge-onset detection by $PW (h, \tau) \geq 500$ is shown in Fig. 7-2d. Threshold 500 at the $PW$ red bar is about twice the maximum power at dot-arrow 1, observed during the regular deformation. Thus, the power level of 500 is high enough to detect any abnormal deformation. The bulge-onset detection was at dot-arrow 3 on October 28, 2010, which changed $V (h, j)$ and $A (h, j)$ in bold. They stay bold under $PW (h, j) \geq 500$. The onset detection by $PW (h, j)$ was one day before the qualitative detection in Figs. 7-2a - 7-2c. After the bulge-onset in Fig. 7-2d, the $PW (h, j)$ continued to rise to $PW = 2322$ until one day before the M9 EQ on March 11, 2011.

The second window in Fig. 7-2a shows the eastward motion started to slow down after reaching the $D (E, \tau) – V (E, \tau)$ path peak at $\tau = 4642$ on December 8, 2008 (or November 23, 2009, in time $j = 4992$). The $A (E, \tau)$ is zero at the peak and changed negative (westward) to slow down $V (E, \tau)$. The slowing action began at arrow Ph-0 in Fig. 7-2c, preceding the S1 onset by 196 days. The $D (E, \tau)$ net displacement from the slowing-onset (Ph-0) and the S1-onset to one day before the 2011 Tohoku M9 EQ was +18.4 mm and + 10.0 mm, respectively, as in Table 2.

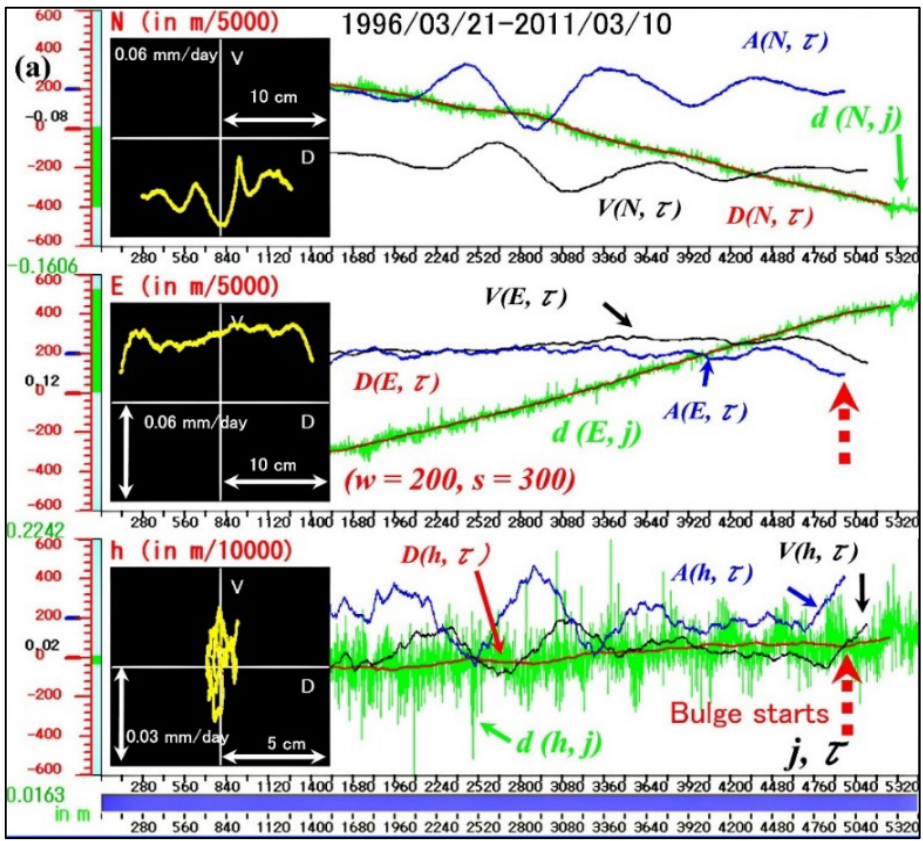

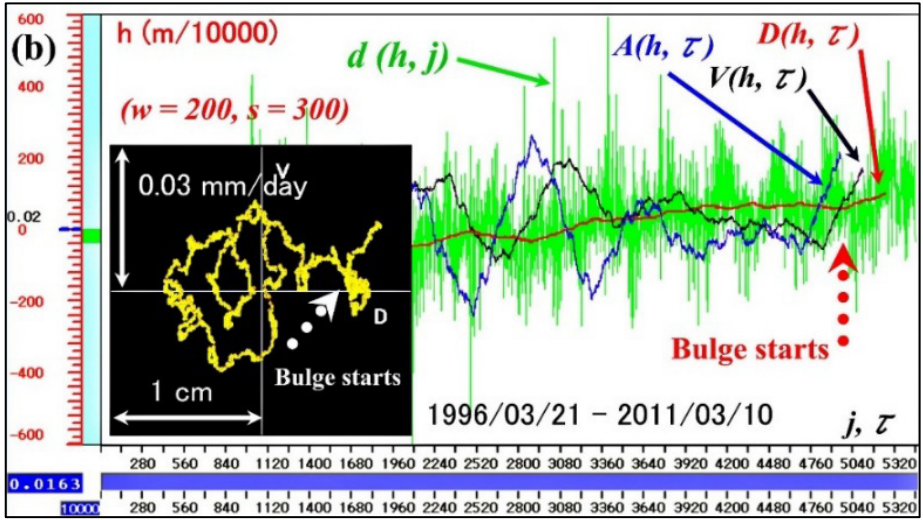

(b) h (m/10000)
(w = 200, s = 300)
d (h, j)    A(h, τ)    D(h, τ)
V(h, τ)
0.03 mm/day
1 cm    Bulge starts
Bulge starts
1996/03/21 − 2011/03/10    j, τ

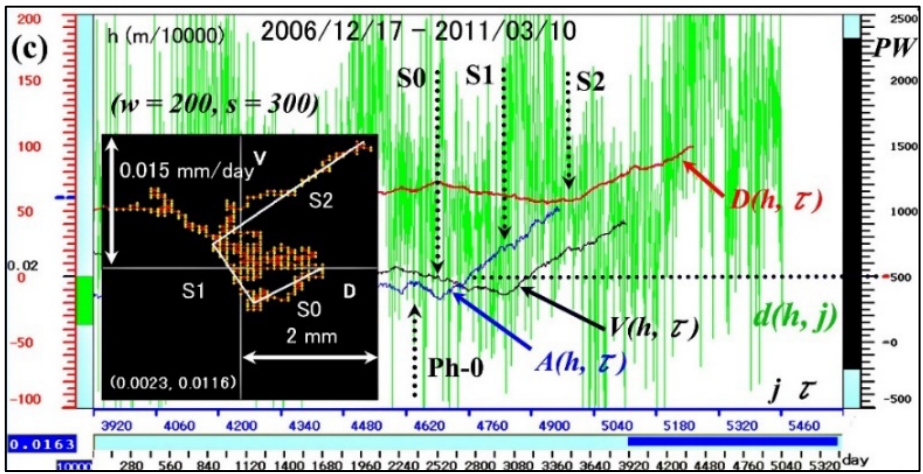

(c) h (m/10000)    2006/12/17 − 2011/03/10    PW
(w = 200, s = 300)
S0  S1  S2
0.015 mm/day
S2
S1    D
S0
2 mm
(0.0023, 0.0116)    Ph-0
D(h, τ)
V(h, τ)
d(h, j)
A(h, τ)    j, τ

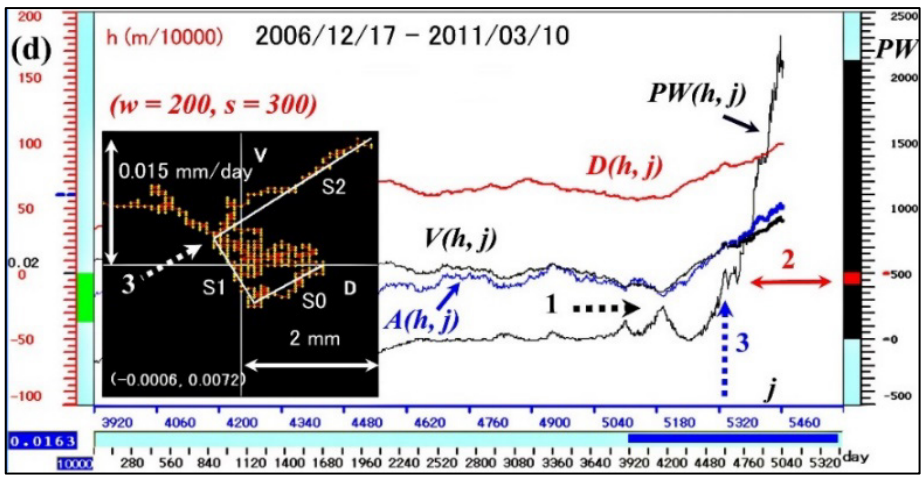

(d) h (m/10000)    2006/12/17 − 2011/03/10    PW
(w = 200, s = 300)    PW(h, j)
0.015 mm/day
S2    D(h, j)
V(h, j)
S1    D    2
S0
3    1
A(h, j)    3    j
2 mm
(−0.0006, 0.0072)

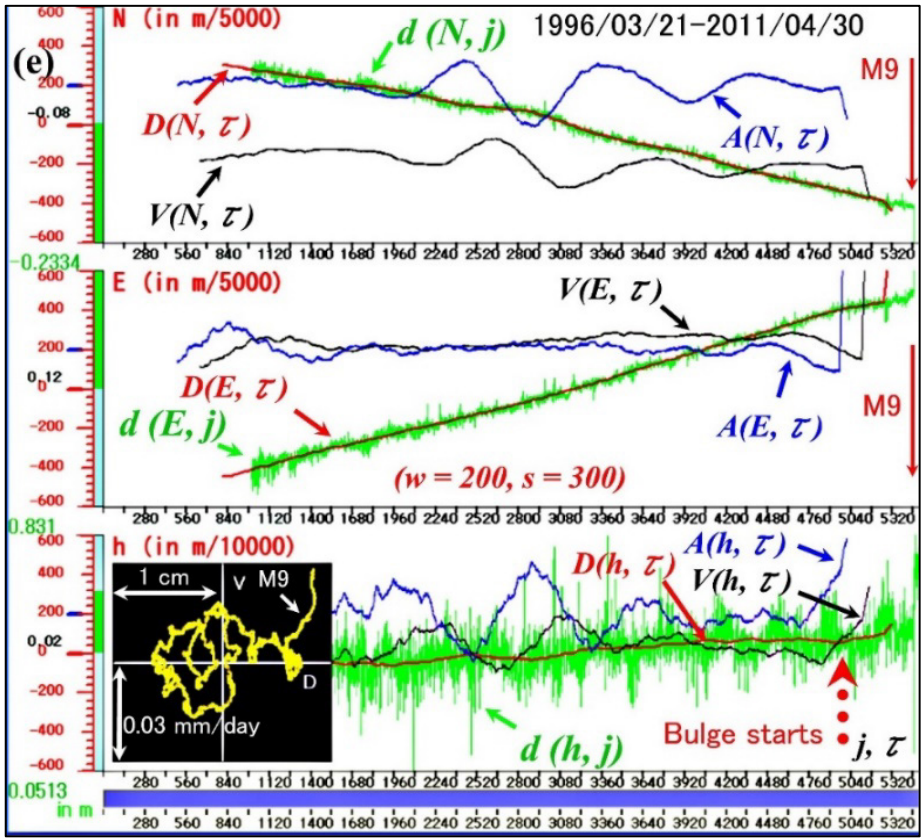

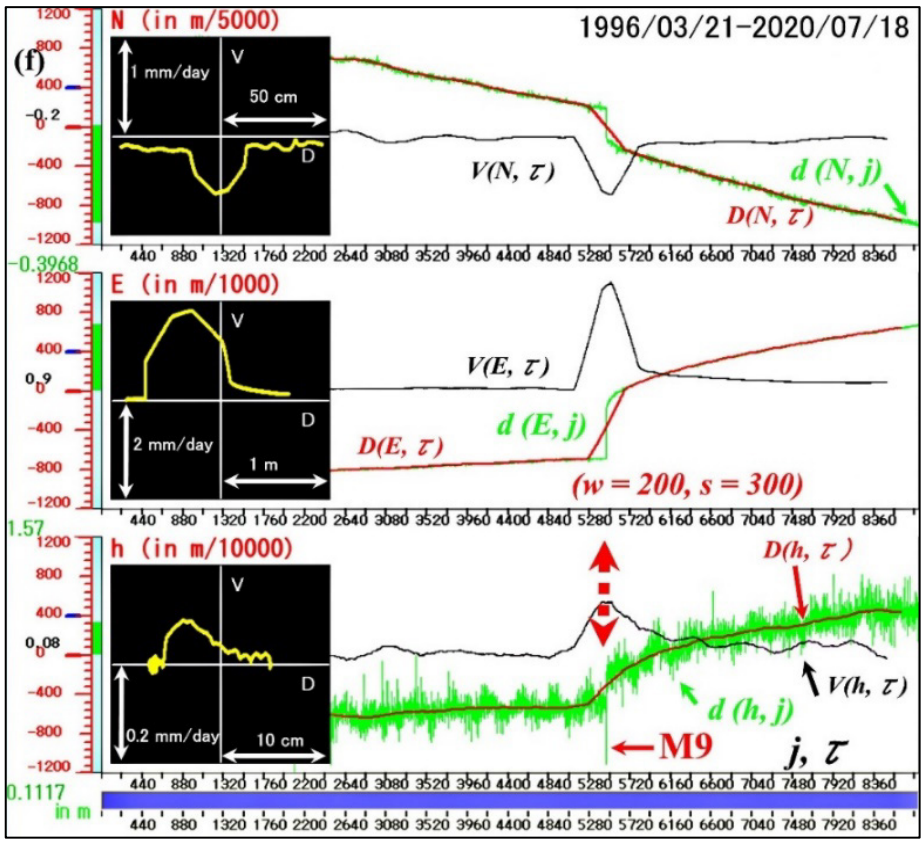

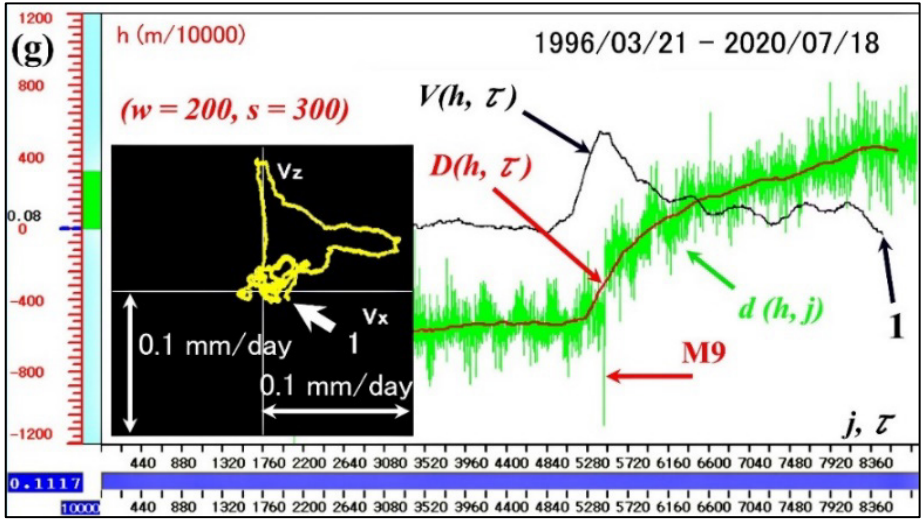

**Fig. 7-2. The bulge and M9 observations at Ryoutsu2 station on an island off the west coast.**
(a) The series {$c$} covers the period from March 21, 1996 ($j = 0$) to March 10, 2011 ($j = 5464$), with parameters $w = 200$ and $s = 300$ used for $D(c, \tau)$, $V(c, \tau)$, and $A(c, \tau)$. The $D(c, \tau) – V(c, \tau)$ planes have a scale of (10 cm, 0.06 mm/day) for $c = N$ and $E$, and (5 cm, 0.03 mm/day) for $c = h$. $V(c, \tau)$ and $A(c, \tau)$ are in relative scales, and their origins are at scale 0 for $V(c, \tau)$ and 200 (blue line) for $A(c, \tau)$. (b) The series {$h$} covers the period from March 21, 1996, to March 10, 2011. The $D(h, \tau) – V(h, \tau)$ plane has a scale of (1 cm, 0.03 mm/day). The bulge-onset at dot-arrow 'Bulge starts' has time $\tau = 4982$ on November 14, 2009 (in time $j = \tau + 350$ on October 10, 2010). (c) The expanded time window is from December 17, 2006, to March 10, 2011. The $D(h, \tau) – V(h, \tau)$ plane has a scale of (2 mm, 0.015 mm/day). The path has three approximately linear segments in white lines, labeled as S0, S1, and S2. The S2 has a scale of ($\Delta D(h, \tau)$ in m, $\Delta V(h, \tau)$ in mm/day) = (0.0023, 0.0116) displayed on the phase plane. The bulge-onset is at the sharp S1 – S2 bend. Each onset, S1, S2, and S3, is at each dot-arrow. The Ph-0 dot-arrow is the onset location for Phase 0 in Table 2. A horizontal dot-line is the abscissa for $V(h, \tau) = A(h, \tau) =$ zero at the level of the offset origin 0 (0.02). The $V(h, \tau)$ and $A(h, \tau)$ are on a relative scale. The "Bulge starts" in Fig. part b was at dot-arrow S2 of $\tau = 4982$ (on October 29, 2010, in time $j = 5332$). (d) The expanded time window is from December 17, 2006, to March 10, 2011. The bulge-onset detection was at $j = 5331$ (October 28, 2010) by the power monitoring $PW(h, j) \geq 500$, one day earlier than Fig. part c. Level 500 at the red scale bar adopted a standard level at arrow 1. The bulge-onset detection on $V(h, j)$ and $A(h, j)$ is at the vertical dot-arrow 3, which also points to the bulge-onset on the (2 mm, 0.015 mm/day) path. The same linear segment division in Fig. part c is on the path. The S1 has ($\Delta D(h, \tau)$ in m, $\Delta V(h, \tau)$ in mm/day) = (0.0006, 0.0072) displayed on the phase plane. Arrow 2 points to the power change from 420 (at $j = 5330$) to 506 (at $j = 5331$) on the displaced red column. The power on March 10, 2011 ($j = 5464$) is at the black column height. (e) The series {$c$} covers the period from March 21, 1996 ($j = 0$) to April 30, 2011 ($j = 5515$). The M9 EQ was at $j = 5465$ on March 11, 2011. The $d(N, j)$, $d(E, j)$, and their column heights show the co-seismic shifts saturated by the M9 EQ. Their digital values were from the position on March 21, 1996 ($j = 0$). The $D(h, \tau) – V(h, \tau)$ plane has a scale of (1 cm, 0.03 mm/day). The path shows the bulge-onset, and arrows point to the M9 EQ. (f) The series {$c$} covers the period from March 21, 1996 ($j = 0$) to July 18, 2020 ($j = 8858$). The $D(c, \tau) – V(c, \tau)$ planes have (50 cm, 1 mm/day) for $c = N$, and (1 m, 2 mm/day) $c = E$, and (10 cm, 0.2 mm/day) for $c = h$. The M9 downward spike on $d(h, j)$ at arrow M9 and a vertical dot-arrow are at $j = 5465$ (March 11, 2011). (g) The series {$h$} covers the period from March 21, 1996 ($j = 0$) to July 18, 2020 ($j = 8858$). The time-delayed $Vx = V(h, \tau – 600)$ and $Vz = V(h, \tau)$ plane has a scale of (0.1 mm/day, 0.1 mm/day). The last $V(h, \tau)$ and the corresponding point in the $Vx$ - $Vz$ plane have label 1.

### 7.2.2 On and After the Tohoku M9 EQ (Figs. 7-2e - 7-2g)

Figure 7-2e has the M9 EQ added on Figs. 7-2a and 7-2b. The M9 moved the west coast station to the south and the east by 0.0734 m and 0.6310 m, respectively. The event also lifted the station by 0.0313 m.

In Fig. 7-2f, the Tohoku M9 co-seismic shift is on the $\{c\}$ at $j = 5465$. As of July 18, 2020, the path appears freed from the Tohoku M9 EQ influence except for the $\{E\}$ motion. The $D(E, \tau) - V(E, \tau)$ path is not on the same $V(E, \tau)$ level as before the M9.

Figure 7-2g shows a time-delayed $Vx$ - $Vz$ path, where $Vx = V(h, \tau - 600)$ and $Vz = V(h, \tau)$. The path and $V(h, \tau)$ at arrow 1 is a 2020s state. The present state is similar to the one before the 2011 Tohoku M9 EQ, suggesting that the current $V(h, \tau)$ at arrow 1 is entirely free from the influence of the M9 EQ.

### 7.3 Bulge on the Top (Murakami station)

The Top is the peak position in the schematics of Fig. 4, a location of the no-displacement-ridge-line in Fig. 3. Murakami station (38.2307° N, 139.5069° E) is at the Top.

### 7.3.1 Until March 10, 2011 (Figs. 7-3a - 7-3d), one day before the Tohoku M9 EQ

Figures 7-3a and 7-3b show the bulge-onset location at the Bulge starts dot-arrow for Murakami station on $D(c, \tau)$ and $V(c, \tau)$ for $c = E$ and $h$, respectively.

Figure 7-3c shows the $D(h, \tau) - V(h, \tau)$ path divided into four approximately linear segments. The S0 into Ph-0 and S1 in Ph-1 are like those at the west coast (Ryoutsu2 station). Phase-2 has the segment divided into S2a and S2b. Each onset location is at each corresponding dot-arrow, S0, S1, S2a, and S2b. The S0 began with an increasing $A(h, \tau)$, whose action shortly changed to lifting (positive above the horizontal dot-line) the Top. The bulge-onset is the S2a-onset dot-arrow pointing to $D(h, \tau)$, $V(h, \tau)$, and $A(h, \tau)$. It is at a corner from S1 to S2a on the $D(h, \tau) - V(h, \tau)$ path. The S2a began at $\tau = 3538$ on September 12, 2009 (or August 28, 2010, in time $j = 3883$) and changed to S2b. The S2b started at $\tau = 3620$ on December 3, 2009 (or November 18, 2010, in time $j = 3970$). The net upheaval growth is 2.5 mm over 199 days before the Tohoku M9 EQ, as shown in Table 2.

The $D(E, \tau) - V(E, \tau)$ path in Fig. 7-3a at the first segment's onset (S1 onset) is at $V(E, \tau) = 0$ mm/day. The bulge grew with the westward motion, and the westward speed reached $V(E, \tau) = -0.0086$ mm/day at $\tau = 3732$ (on March 10, 2011, in $j = 4082$), one day before the M9 EQ. The station moved westward by 1.1 mm from $D(E, \tau)$ at $V(E, \tau) = 0$ mm/day.

Figure 7-3d shows the bulge-onset detection by the power monitoring $PW(h, j) \geq 500$. The threshold is 500 on a relative scale at the PW scale. The level is about 100 higher than the horizontal dotted level (arrow 1), which is the maximum power observed at the regular crustal deformation before the unexpected bulge-onset. The detection changed $V(h, j)$ and $A(h, j)$ bold, and the corresponding location on the phase path is at the vertical dot-arrow 3. The unexpected $PW(h, j)$ rose to 1664 and dropped to about 1300 one day before the M9 EQ.

### 7.3.2 On and After the Tohoku M9 EQ (Figs. 7-3e - 7-3g)

The co-seismic shift caused by the M9 EQ is shown in Figure 7-3e, with a southward shift of 0.1969 m and an eastward shift of 1.1607 m. There is no noticeable abrupt upward or downward change above the noise level of ±20 mm, which saturates the $\{N\}$ and $\{E\}$ displays.

In Figure 7-3f, the co-seismic shift at $j = 4083$ is shown on the $\{N\}$ and $\{E\}$. The $D(h, \tau) - V(h, \tau)$ path in Fig. 7-3f shows three large downward movements, with the first one coming from the M9 EQ and including a downward spike noise of M9 on $d(h, j)$ (label M9). However, the second and third movements have no clear geophysical origin. The downward trend suggests a recovering state from the bulge-bending deformation.

The time-delayed $Vx$ - $Vz$ path is shown in Fig. 7-3g, where $Vx = V(c, \tau - 600)$ and $Vz = V(c, \tau)$. Each time-delayed path and $V(c, \tau)$ at label 1 is the state as of July 18, 2020. The paths and $V(c, \tau)$ show a 2020s crustal state characterized by $V(c, \tau)$ that is still different from the state before the 2011 Tohoku M9 EQ, except for $V(h, \tau)$, suggesting that only the vertical motion, $V(h, \tau)$, is free from the M9 EQ influence.

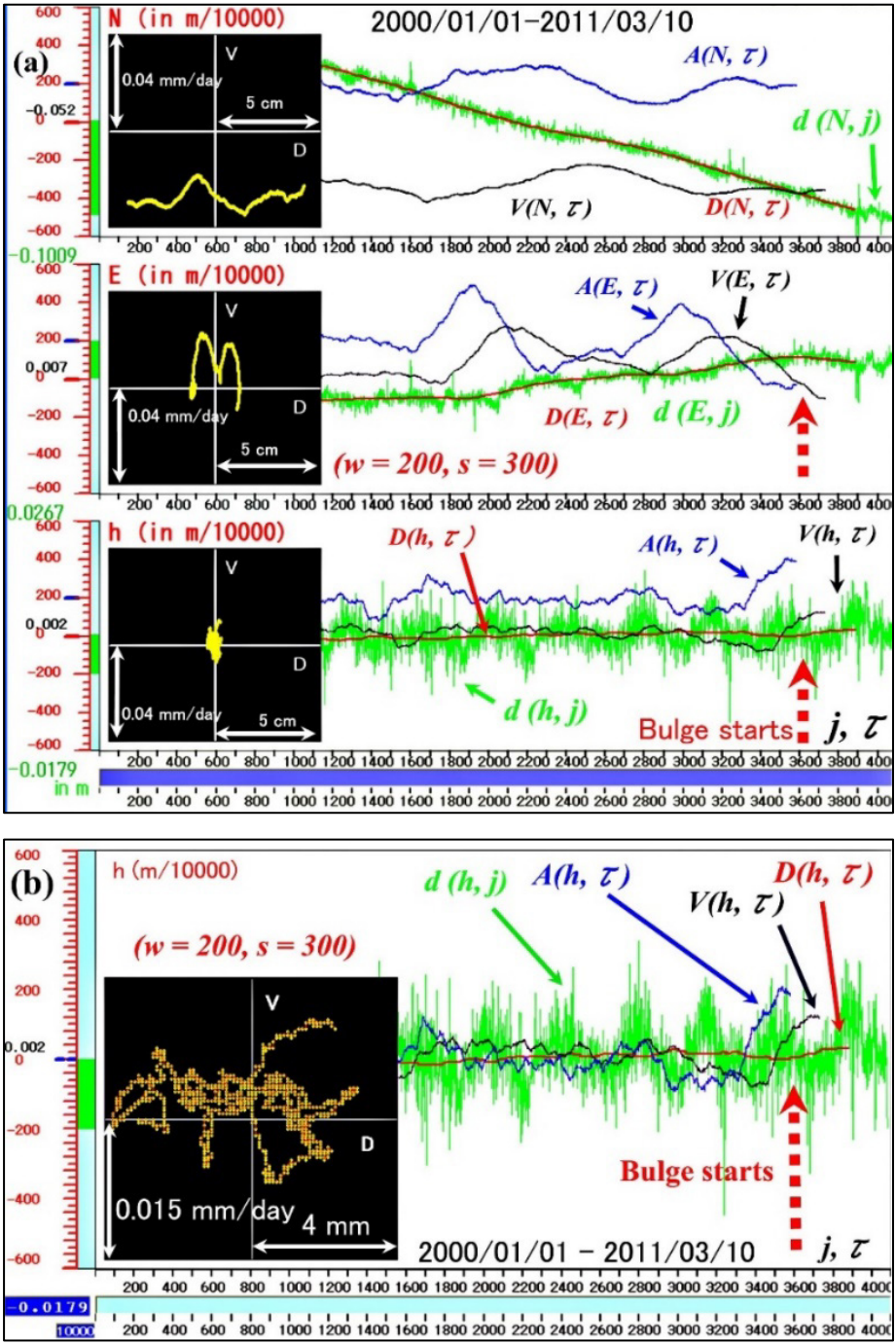

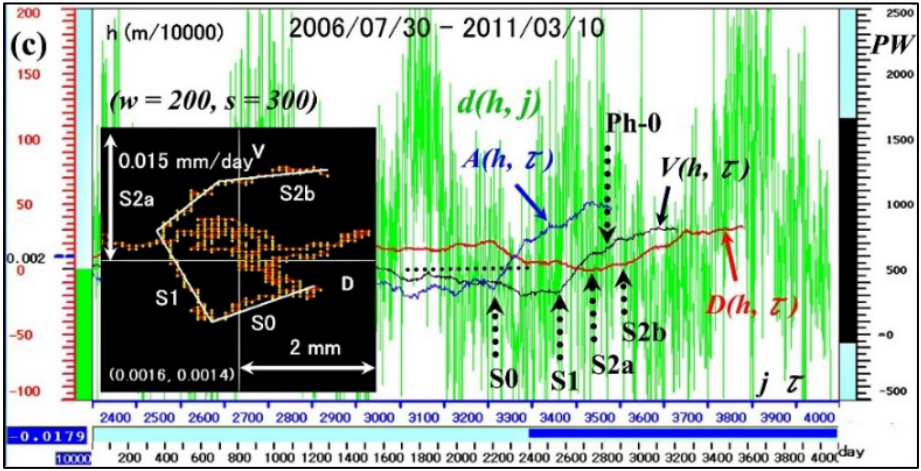

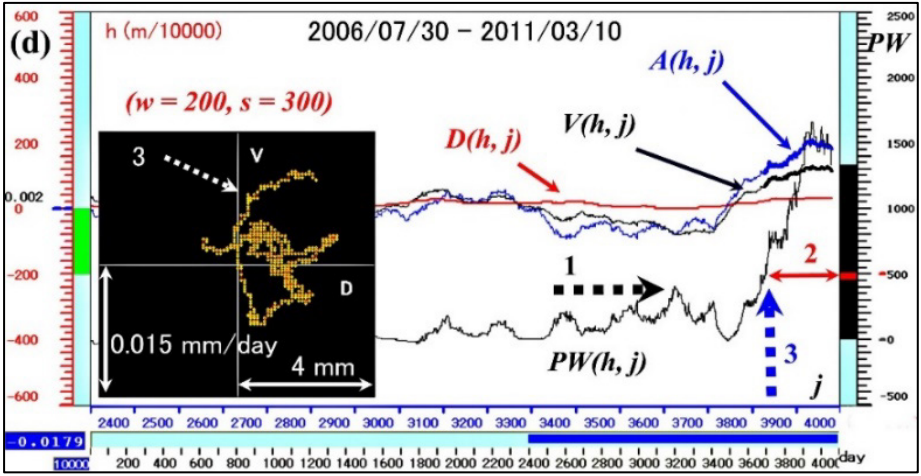

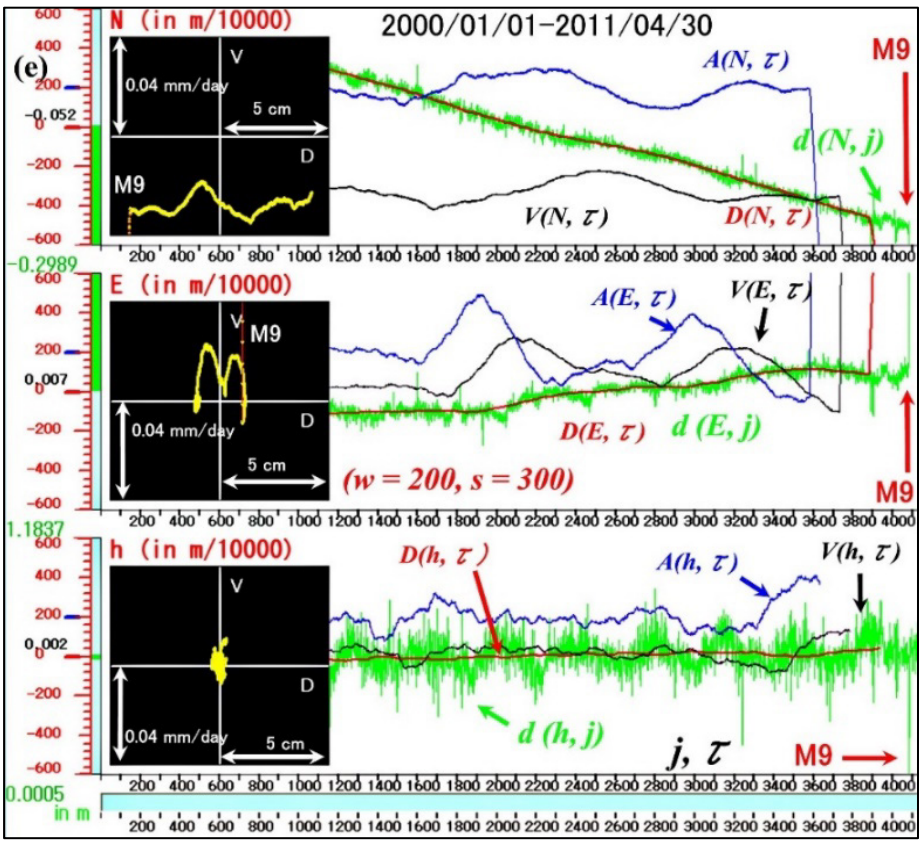

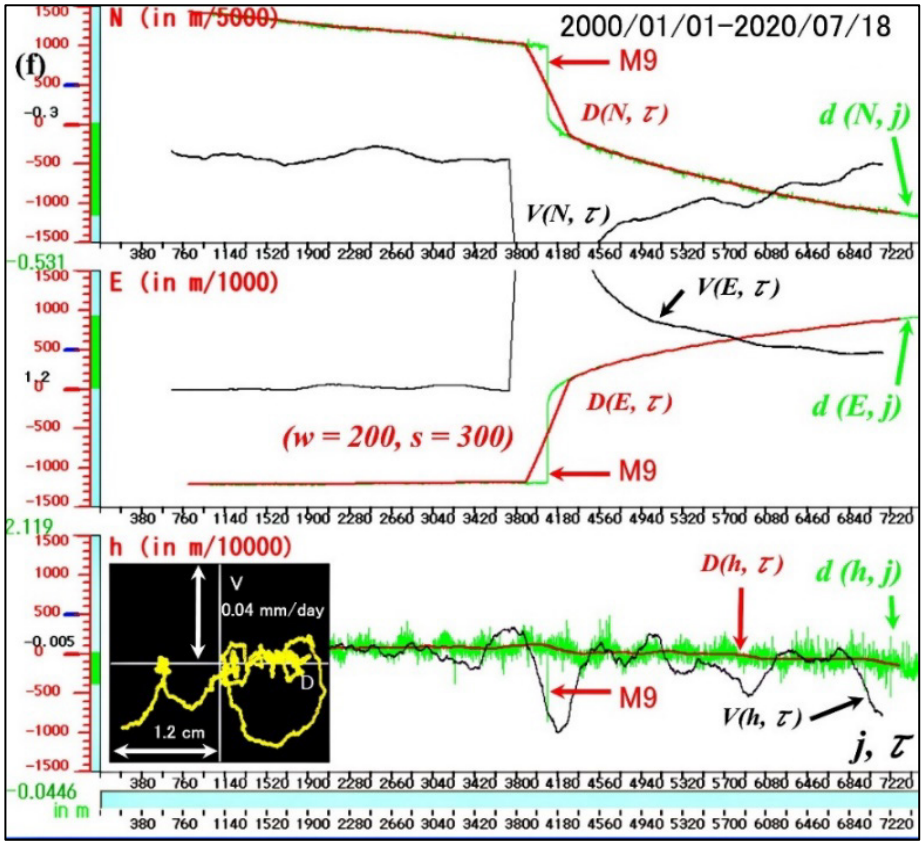

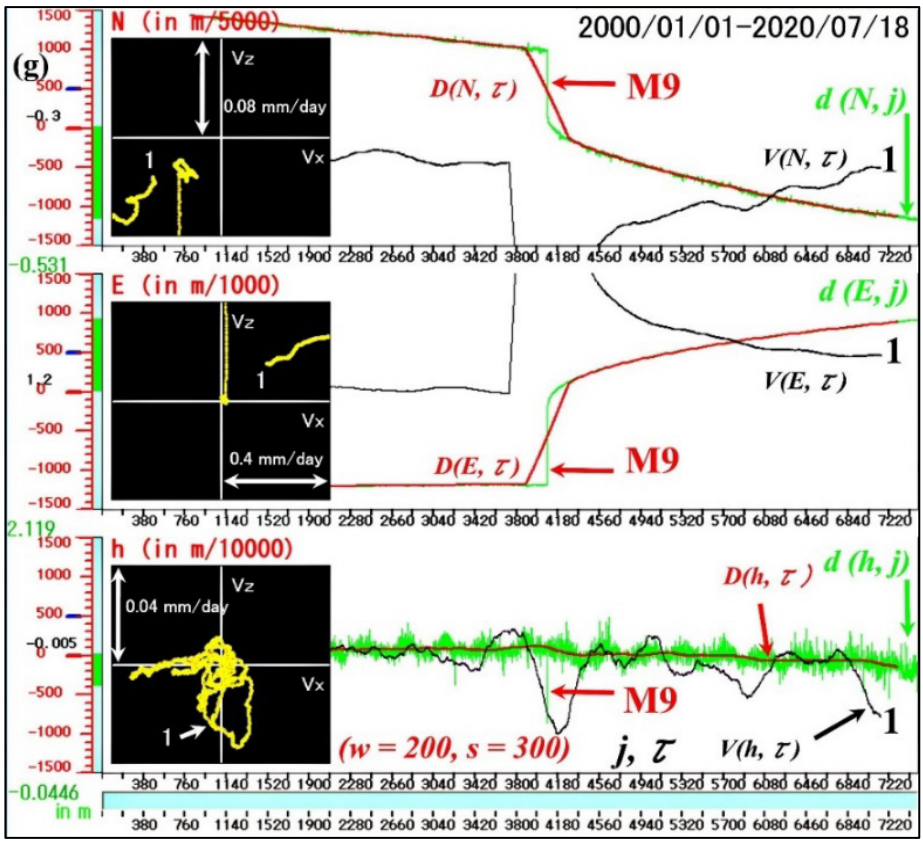

**Fig. 7-3. The bulge and M9 observations at Murakami station.**
(a) The series $\{c\}$ covers the period from January 1, 2000 ($j = 0$) to March 10, 2011 ($j = 4082$), with parameters $w = 200$ and $s = 300$ used for $D(c, \tau)$, $V(c, \tau)$, and $A(c, \tau)$. The $D(c, \tau) - V(c, \tau)$ plane has a scale of (5 cm, 0.04 mm/day) for $c = N, E,$ and $h$. The bulge-onset is at the dotted arrows with the label Bulge starts. (b) The series $\{h\}$ covers the period from January 1, 2000, to March 10, 2011. The $D(h, \tau) - V(h, \tau)$ plane has a scale of (4 mm, 0.015 mm/day). The bulge-onset is at the dotted arrows with the label Bulge starts. (c) The expanded time window is from July 30, 2006 ($j = 2400$) to March 10, 2011 ($j = 4082$). The $D(h, \tau) - V(h, \tau)$ plane has a scale of (2 mm, 0.015 mm/day). The path has four approximately linear segments in white lines, labeled as S0, S1, S2a, and S2b. The S2a has a scale of ($\Delta D(h, \tau)$ in m, $\Delta V(h, \tau)$ in mm/day) = (0.0009, 0.0055), and S2b has a scale of ($\Delta D(h, \tau)$ in m, $\Delta V(h, \tau)$ in mm/day) = (0.0016, 0.0014) displayed on the phase plane. The bulge-onset is at a corner from S1 to S2a. Each onset, S0, S1, S2a, and S2b, is at a corresponding dot-arrow. The onset location of an abnormal $V(E, \tau)$ change from eastward ($> 0$) to westward ($< 0$) is at dot-arrow Ph-0 (not as $A(E, \tau)$ change on the east and west coast). The directional change in $A(E, \tau)$ corresponding to the peak $V(E, \tau)$ at $\tau = 2041$ in Fig. part a was before S0, which was on September 27, 2009 ($j = 2391$). A horizontal dot-line is the abscissa for $V(h, \tau) = A(h, \tau) =$ zero at the level of the offset origin 0 (0.002). The $V(h, \tau)$ and $A(h, \tau)$ are on a relative scale. (d) The expanded time window is the same as in Fig. part c. The $PW(h, j) \geq 500$ monitoring detected a bulge-onset at $j = 3931$ on October 10, 2010, at arrow 3. The threshold of 500 is at the red bar. The detection bolded $V(h, j)$ and $A(h, j)$. The displaced red column height at arrow 2 is the power change from 459 (at $j = 3930$) to 504 (at $j = 3931$). The power on March 10, 2011 ($j = 4082$) is the column height in black. (e) The original $\{c\}$ is from January 1, 2000 ($j = 0$) to April 30, 2011 ($j = 4133$). The Tohoku M9 EQ was at $j = 4083$ on March 11, 2011. The $D(c, \tau) - V(c, \tau)$ plane has (5 cm, 0.04 mm/day) for $c = N, E,$ and $h$. (f) The series $\{c\}$ covers from January 1, 2000 ($j = 0$) to July 18, 2020 ($j = 7494$). The M9 downward spike on $d(h, j)$ was at $j = 4083$ on March 11, 2011. The $D(h, \tau) - V(h, \tau)$ plane has (12 cm, 0.04 mm/day). (g) The series $\{c\}$ covers from January 1, 2000 ($j = 0$) to July 18, 2020 ($j = 7494$). The time-delayed $Vx = V(c, \tau - 600)$ and $Vz = V(c, \tau)$ planes have a scale of (0.08 mm/day, 0.08 mm/day) for $c = N$, and (0.04 mm/day, 0.04 mm/day) for $c = E$ and $h$. The last $V(c, \tau)$ and the corresponding point in the $Vx$ - $Vz$ planes are labeled 1.

## 7.4 Summary of the bulge-bending deformation and co-seismic shifts

Table 2 summarizes the bulge deformation observations, where the East coast (Onagawa station), West coast (Ryoutsu2 station), and Top (Murakami station) are denoted as E, W, and T, respectively. The bulge-bending deformation process is divided into three phases: Phase 0 (Ph-0), Phase 1 (Ph-1), and Phase 2 (Ph-2). Each phase has a linear segment on the $D(h, \tau) - V(h, \tau)$ path, denoted as S0, S1, and S2, respectively. The S2 segment for the Top station is further divided into S2a and S2b. The data for S0, S2a, and S2b are in rows below each main row of segments S1 and S2. The time intervals in days for segments S1 (and S0), S2 (and S2a and S2b), and the westward movement in all phases are denoted as Int-1, Int-2, and Int-3.

The onset date for each segment is represented as time $j$ (= $\tau + 350$). Ph-0 also refers to the onset of an unusual motion that started before segment S0 (the west coast) and segment S1 (the east coast), characterized by a directional change in acceleration $A(E, \tau)$ from eastward ($> 0$) to westward ($< 0$) on the east and west coasts, marked by the onset of a significant change in velocity $V(E, \tau)$ on the $D(E, \tau) - V(E, \tau)$ path. The Top station experienced the directional change before S0. However, Ph-0 on the Top refers to the directional change in $V(E, \tau)$ from westward to eastward to emphasize the Top's westward movement after Ph-0 during S2a and S2b. The onset date for Ph-0 preceded Ph-1, which is a reference, by Δ day (minus) in days. If the onset follows Ph-1 as in Top (T), $\mathit{\Delta}$ day is positive.

$D$ ($c$) and $V$ ($c$) are the abbreviations for $D$ (c, τ) and $V$ (c, τ) in Table 2. If the change in westward movement is positive, such as in row W (west coast) with a value of 18.4 mm, it means eastward movement by 18.4 mm. The sign convention obeys the $D$ ($c$, τ) − $V$ ($c$, τ) plane for $c = E$ and $h$. The mm/day^2 with Int-1 is an average acceleration defined by $\Delta V$ ($h$) /Int-1. The time rate conversion from day to sec is 1mm/day = $1.1574^{-6}$ cm/sec, and 1mm/day$^2$ = $1.3396^{-11}$ cm/sec$^2$.

**Table 2 (Three phases of the bulge-bending deformation over Tohoku)**

| | Three phases of the bulge deformation, Phase 0 (Ph-0, S0), Phase 1 (Ph-1), and Phase 2 (Ph-2, S2a, S2b) | | | | | | | | | | | | | | | | |
| | Ph-1 | | | | | | Ph-2 | | | | | | Ph-0 + Ph-1 + Ph-2 | | | | Ph-0 |
| | Segment S1 (in mm) | | | | | | Bulge-Onset | Segment S2 (in mm) | | | | | Westward(−) Movement (in mm) | | | | |
| | τ | j in (y/m/d) | ΔD (h) | ΔV (h) | mm/day^2 | Int-1 | j in (y/m/d) | τ | ΔD (h) | ΔV (h) | mm/day^2 | Int-2 | ΔD (E) | ΔV (E) | Int-3 | mm/day^2 | Δday |
|---|---|---|---|---|---|---|---|---|---|---|---|---|---|---|---|---|---|
| E | 3457 | 2010/06/09 | -3.3 | 0.0142 | 8.99E-05 | 158 | 2010/11/14 | 3615 | 1.2 | 0.0054 | 4.70E-05 | 115 | -13.2 | -0.018 | 296 | -5.95E-05 | -23 |
| S0 | 3308 | 2010/01/11 | -2.8 | 0.0001 | 6.71E-07 | 149 | | | | | | | | | | | |
| | | | | | | | | | | | | | | | | | |
| W | 4838 | 2010/06/07 | -0.6 | 0.0072 | 5.00E-05 | 144 | 2010/10/29 | 4982 | 2.3 | 0.0116 | 8.79E-05 | 132 | 18.4 | -0.022 | 472 | -4.58E-05 | -196 |
| S0 | 4694 | 2010/01/14 | -1 | -0.0044 | -3.06E-06 | 144 | | | | | | | 10.0 | -0.018 | 299 | -5.85E-05 | -23 |
| | | | | | | | | | | | | | | | | | |
| T | 3461 | 2010/06/12 | -0.8 | 0.0104 | 1.44E-04 | 72 | 2010/08/26 | 3533 | 2.5 | 0.0069 | 3.47E-05 | 199 | -1.2 | -0.008 | 187 | -4.33E-05 | 116 |
| S0 | 3313 | 2010/01/15 | -1.5 | -0.0038 | -2.57E-06 | 148 | | | | | | | | | | | |
| S2a | | | | | | | 2010/08/26 | 3533 | 0.9 | 0.0055 | 6.32E-05 | 87 | | | | | |
| S2b | | | | | | | 2010/11/18 | 3620 | 1.6 | 0.0014 | 1.25E-05 | 112 | | | | | |

Row S0 below E in column Phase 1 (Ph-1) shows that the east coast (E) subsided by $\Delta D$ ($h$) = − 2.8 mm over 149 days (Int-1) is segment S0, with subsidence of $\Delta D$ ($h$) = − 1 mm and $\Delta D$ ($h$) = − 1.5 mm on the west coast (W) and the Top (T), respectively. This subsidence leads to increased static friction between the eastern edge of the overriding continental plate and the fault surface, effectively griping significant barriers (on fault) [10] 2.8 mm deep.

Column Phase 1 (Ph-1) shows that the east coast edge further subsided by $\Delta D$ ($h$) = − 3.3 mm over 158 days with the lifting force of 8.99 ×10$^{−5}$ mm/day$^2$ (average acceleration defined by $\Delta V$ ($h$) / Int-1), during which the east coast pulled down the subducting oceanic plate along with the bulging upward (lifting) force on the fault surface. The pulling in Ph-1 (S1) on the east coast began on June 9, 2010, while the west coast began S1 on June 7, 2010.

Column Phase 2 (Ph-2) shows the east coast gradually moved up by $\Delta D$ ($h$)= 1.2 mm over 115 days, an upheaval growth of 1.2 mm, the west coast by $\Delta D$ ($h$) = 2.3 mm, and the Top by $\Delta D$ ($h$) = 2.5 mm.

The $\Delta D$ ($h$) in Ph-1 and Ph-2 and $\Delta D$ ($E$) over the entire deformation phases suggest the bulge-bending schematic with uplift force (lifting force) components shown in Fig. 4-b. The gradual accumulation of lifting force weakened the static frictional strength of the subduction interface, eventually causing the shear stress to exceed the frictional strength and resulting in the decoupling of the overriding and subducting plates.

Column Ph-0 + Ph-1 + Ph-2 shows that the onset of an unusual motion, Ph-0, on the $D$ ($E$, τ) − $V$ ($E$, τ) path was 173 days earlier on the west coast compared to the east coast (as shown by Δ days in column Ph-0), suggesting that the bulge-bending over Tohoku began generating the westward restoring force of the west coast against the over-riding continental plate-driving force. One row below W in the same column shows an expected state of the west coast on the $D$ ($E$, τ) − $V$ ($E$, τ) path at the time preceding the east coast's Ph-0 onset by Δ day = − 23.

As indicated by the averaged accelerations (forces) in column Ph-0 + Ph-1 + Ph-2, the westward force exerted by the compressed west coast is a restoring force because it opposes the eastward displacement moved by the over-riding continental plate-driving eastward force. In contrast, the westward forces acting on the east coast and the Top are not restoring forces because their displacements are westward. The over-riding continental plate-driving force compressed the west coast by $\Delta D$ ($E$) = +10.0 mm (eastward), initiating the underlying bulge-bending deformation during Ph-1. During Ph-1 + Ph-2, the same eastward compression generated the east coast's pulling action with a westward-pulling force that moved the east coast coupled with the subducting oceanic plate and the Top by $\Delta D$ ($E$) = −13.2 mm and $\Delta D$ ($E$) = −1.2 mm, respectively, as shown in column $\Delta D$ ($E$).

Thus, the over-riding continental plate-driving eastward force compressed the west coast during Ph-1, which initiated the underlying bulge-bending deformation. The same eastward compression during Ph-1 + Ph-2 generated the east coast's pulling action.

As a result, the enormous restoring force of the compressed west coast stored the elastic potential energy of the Tohoku megathrust EQ during Ph-0 + Ph-1 + Ph-2. The final $\Delta D$ ($h$) in Ph-2 on the east coast triggered the decoupling of the overriding and the subducting plates and released the megathrust EQ energy stored on the compressed west coast, as illustrated in Fig. 4c.

Table 3 summarizes the co-seismic shifts and current states on the Tohoku M9 EQ. The sign convention for $d$($N$) and $d$($h$) is minus (−) for southward and downward shifts, respectively. The Tohoku M9 EQ of 2011 continues to impact the current crustal state at each GPS station in 2021. The column "Under the Tohoku M9" shows the $d$($c$) components under the M9 EQ. Table 3 also reveals the significant eastward restoring movement of the bulge-bent east coast with a slower restoring time compared to the bulge-bent west coast.

At the Onagawa GPS station, the co-seismic shift $d$($E$) of 4.75 m, $d$($N$) of − 1.90 m, and $d$($h$) of − 0.97 m are assumed proportional to the slip displacement components of the Tohoku EQ fault along the respective geographic axis $E$ (eastward), $N$ (northward), and $h$ (upward). An estimate of the strain accumulation period in years at the Onagawa station using Fig. 7-1a can be obtained from the average $V$ ($E$, $\tau$) of − 0.01559 m/year, $V$ ($N$, $\tau$) of − 0.00636 m/year, and $V$ ($h$, $\tau$) of − 0.0060 m/year, which give 305, 299, and 162 years, respectively. Therefore, the strain accumulation periods estimated from the horizontal slip components are approximately 300 years. The significant difference in the accumulation periods between the horizontal and vertical components suggests that the co-seismic downward shift $d$($h$) is primarily due to a rapid restoration of the bulge-bent deformation (a convex shape) to the regular deformation (a concave shape) on the east coast, as illustrated in Fig. 4c. Thus, we can expect that the period of regular deformation had lasted for the last three hundred years.

After the Tohoku EQ event, the vertical displacement $d$ ($h$, $\tau$) has been moving up to restore the accumulated deformation, as shown in Fig. 7-1e. The co-seismic shift of $d$($E$) = 1.1607 m at the Murakami station (Top) with no vertical displacement $d$($h$) = 0 suggests that the co-seismic uplift of $d$($h$) = 0.0313 m at the Ryoutsu 2 station (the west coast) is resulted from restoring the bulge-bent deformation to the regular-bent deformation, as illustrated in Fig. 4c.

**Table 3 (Co-seismic shifts and current states on the Tohoku M9 EQ)**

| Co-seismic shift (in m) | | | | Under the Tohoku M9 |
|---|---|---|---|---|
| GPS station | $d$($N$) | $d$($E$) | $d$($h$) | $d$($c$) |
| E-coast (Onagawa) | -1.9000 | 4.7482 | -0.9694 | $c = E, N, h$ |
| W-coast (Ryoutsu2) | -0.0734 | 0.6310 | 0.0313 | $c = E$ |
| Top (Murakami) | -0.1969 | 1.1607 | 0.0000 | $c = E, N$ |

## 8 The current GPS observations on the northwestern Pacific Plate and Tohoku

Chichijima-A, Hahajima, and Minamitorishima stations, located on the subducting western edge of the Pacific Plate and the northwestern Pacific Plate, respectively, have not shown any abnormal westward motion since the 2011 Tohoku M9 EQ, as depicted in Figs. 6-2a, 6-3b, and Fig. 6-4b.

Onagawa station, situated on the east coast of Fig. 4, has not returned to its pre-Tohoku M9 EQ state, as seen in Figs. 7-1d and 7-1e. Murakami station, on the Top of Fig. 4, is affected by the Tohoku M9 EQ in $V(E, \tau)$ and $V(N, \tau)$, as depicted in Fig. 7-3g. However, $V(h, \tau)$ is free from the M9 EQ except for the downward trends, as shown in Fig. 7-3f. Ryoutsu2 station, located on the west coast of Fig. 4, has $V(E, \tau)$ influenced by the Tohoku M9 EQ, as seen in Fig. 7-2e. The current observations suggest that the eastern edge of Tohoku, which overrides the subducting northwestern Pacific Plate, is sliding eastward without significant constraint. However, recent observations, as shown in Figs. 13-2-3d and 13-2-3e (Appendix B), suggest that the Chichijima-A station exhibited an abnormal westward motion of the subducting oceanic plate, coupling the M7.3 and M6.9 earthquakes that occurred on February 13, 2021, and March 20, 2021, respectively. The M7.3 ruptured within the subducting plate's slab, while for the M6.9, the observation suggests that the overriding eastern edge caught hold of a barrier on the subducting plate that was unbroken by the 2011 Tohoku M9 and ruptured the M6.9 barrier, as shown in Fig. 13-2-3e. The rupturing process suggests that it was a miniature of the Tohoku 2011 M9 EQ event.

## 9 Deterministic prediction of imminent megathrust EQs

The cross-correlation of P-Ws with the observed daily displacement time series $\{c\}$ defines the relations between the $D(c, \tau)$, $V(c, \tau)$, and $A(c, \tau)$ for the subducting oceanic plate motion coupled with the overriding crustal bulge-bending deformation. The $D(c, \tau) – V(c, \tau)$ path of the oceanic plate and the over-riding crustal bulge-bending motion quantified the genesis process of the 2011 Tohoku M9 EQ and subsequent tsunami, as schematically illustrated in Fig. 4. The process allows for real-time prediction of imminent megathrust EQs in subduction zones up to three months in advance [1].

The GPS prediction begins with detecting the oceanic plate's abnormal motion. The power monitoring of $PW(c, j) \geq$ Th (predetermined thresholds) detects the anomaly onset. The current GPS observations to watch for the unexpected motion of the subducting northwestern Pacific Plate and the Philippine Sea plate are detailed in Appendix B. As of March 11, 2023, the observations suggest no imminent megathrust ruptures in the subduction zones of the northwestern Pacific Plate and the Philippine Sea Plate [section 13.4 in Appendix B]. Additionally, detecting any bulge-onset that divides the $D(h, \tau) – V(h, \tau)$ path into two linear segments, each of which obeys linear elasticity, is another powerful tool (section 7.1).

Similarly, seismicity observations are a time series, $\{c\} = \{d(c, 1), d(c, 2), …, d(c, j), …\}$, where $c = LAT, LON, DEP, INT,$ and $MAG$ [1, 7, 10]. Each component $c$ represents the epicenter location (latitude $LAT$, longitude $LON$, and focal depth $DEP$), inter-event interval ($INT$) between the consecutive events, and magnitude $MAG$. Time $j$ is a chronological event index having a unique relationship with the origin time (event time). The time interval between consecutive events ($INT$) reflects changes in the regional stress state [11]. The cross-correlation of P-Ws with the observed seismicity time series $\{c\}$ determines the relations between $D(c, \tau)$ and $A(c, \tau)$ for the large EQ genesis process over several months [1, 7, 10]. This process allows predicting an imminent large earthquake's fault size and motion, magnitude, and rupture time in the event index a few months in advance. Index conversion to the rupture time in date is necessary [1, 10]. The successful applications of this method to significant hindsight EQs are available for anticipated large EQs [1].

A cumulative (moving) sum of $d$ ($INT$, $j$) and $d$ ($DEP$, $j$) over a predetermined parameter $w$ is proportional to the strain energy density stored in the regional crust [1, 10]. Each sum shows a smooth stress accumulation and a rapid release to an imminent significant EQ rupture in a selected region [1]. The observation led to the successful hindsight prediction of many significant EQs in Japan within a few days, including the 2011 Tohoku M9 EQ [1, 10]. Thus, the time series analyses of GPS displacement and EQ source parameters with P-Ws are available for predicting an imminent megathrust EQ in subduction zones [2], as shown in Fig. 9 [1].

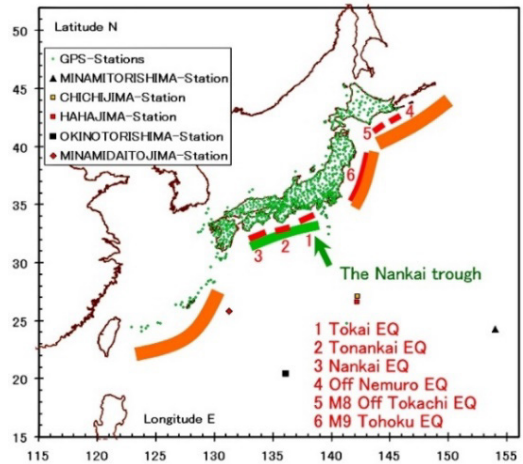

**Fig. 9. Schematics of repeating large EQs (faults), trenches, troughs, and GPS stations.**
A dense network of 1,240 GPS stations laid by GSI is in green dots. The schematic locations of trenches are in brown, the Nankai trough is in green [2], and the faults of repeating large EQs [2] are in red. The trench off northeastern Japan's east coast was the M9 Tohoku EQ on March 11, 2011, whose fault is label 6.

## 10 Acknowledgments

The patent (130 pages and 85 figures) has two observations for the claims: seismicity and GPS observations. This article updated the GPS observations. Physical Wavelets analyses of displacement time series $\{c\}$ and their analyzed-data drawing of Fig. 5-1 - Fig.7-3 used an EQ prediction software written for the two patents, [1] and [B1] (Appendix B).

The contents of Figs. 3 and 4 with GPS observations were in four presentations at Japanese and AGU fall meetings. They are A32-11 (Seismological Society of Japan, 2011 Fall Meeting Oct. 14), NH23A-1543 (AGU 2011 Fall Meeting Dec. 6), G23B-0794 (AGU 2013 Fall Meeting Dec. 10), and Japan Geoscience Union Meeting 2020 [SSS07-02], and 2023 [HDS06-08].

## 12 Appendix A (Mathematical operators)

Suppose we have a Brownian particle [A1]. We then have the well-known Brownian motion's stochastic differential equations: the Langevin equation and another derived using Ito's calculus. A phenomenal description of a large number of such particles is the diffusion equation. However, in deriving the equation, we have ignored the non-differentiability of the particle-probability-distribution function with respect to the particle's probabilistic position [A1]. In contrast, Physical Wavelets (P-Ws) are operators that define equations of stochastic motion, producing a smooth path in phase space [A2]. Thus, P-Ws can find the genesis process of the 2011 Tohoku M9 EQ from the non-differentiable daily displacement time series observed at GPS stations.

### 12.1 Physical Wavelets (P-Ws)

#### 12.1.1 Introduction

Some examples of P-Ws in time $t$ and $\tau$ are illustrated in Figs. 12-1a and 12-1b. The inverted $D1W(t)$ of Fig. 12-1a with the narrower width $\Delta t$ is the well-known Haar wavelet, and the $DDW(t)$ is then the scaling function for the Haar wavelet. The pair forms a basis of a complete orthonormal set for the most straightforward multiresolution analysis of time series [A3]. In contrast, the cross-correlation between P-Ws and any non-differentiable path of particle motion defines its position ($D$), velocity ($V$), and acceleration ($A$). Since 1985, the $D1W(t-\tau)$ of Fig. 12-1b has been an imperative tool to extract the real-time rate change ($V$) from noisy pressure fluctuations for an oscillometric on-line-digital-blood-pressure unit design [A4-A6]. Thus, the usages of P-Ws differ from those in the well-known wavelet analysis.

P-Ws can detect faint anomalies from noisy signals and assign physical laws to them. The detected anomaly is a deterministic and physics-based precursor to an imminent disaster under critical observation. For example, a product of $V$ and $A$ defined with P-Ws is proportional to the kinetic energy ($KE$) rate change, which is the power. Monitoring the power by comparing it with predetermined threshold levels automates detecting any abnormally increased rate change leading to the disaster. The automatic or manual threshold setting is adaptive to the maximum rate change in a routine operation. Such precursor detections have prevented the sudden material fractures of rotating heavy manufacturing machinery a few hundred milliseconds before disasters [A7-A9]. The detected precursor has physical laws that help the manufacturing system improve.

The displacements observed at the GPS stations have the crustal responses to the lunar tidal force loading [A2, A10-A12]. The displacement time series show the fortnightly and the synodic period. The automated power monitoring of the time series detects the abnormally increased power of their crustal responses. The anomalies are the sudden deviations from the periodic and regular lunar tidal force loading. Unusual responses near imminent large earthquake (EQ) epicenters were the precursors seen weeks before the EQs [A10-A11].

In studying nonlinear dynamics, the phase space constructed from the observed time series with P-Ws is physically more tractable than the state space reconstructed by so-called time delay embedding, as in Figs. 13-2-3d and 13-2-3e [A8, A13-A14]. An array of $DDW(t)$s can estimate the number of independent variables or dynamical degrees of freedom, creating a chaotic time series [A2]. The detection algorithm is the same as finding false nearest neighbors for estimating the minimum embedding dimension of an attractor constructed by a delay-embedding theorem. The P-Ws may claim that the minimum embedding dimension is the number of independent variables creating time series.

Statistical analyses of time series with P-Ws give us some physical intuition on their extracting quantities. For example, the Allan variance is the two-sample variance taken with the $D1W(t)$ of Fig. A1a, giving us a statistical $KE$. The variance for 1/f fluctuation is finite regardless of the time width of $D1W(t)$, which suggests the system with 1/f has an endless *KE* to stay active [A13, A14].

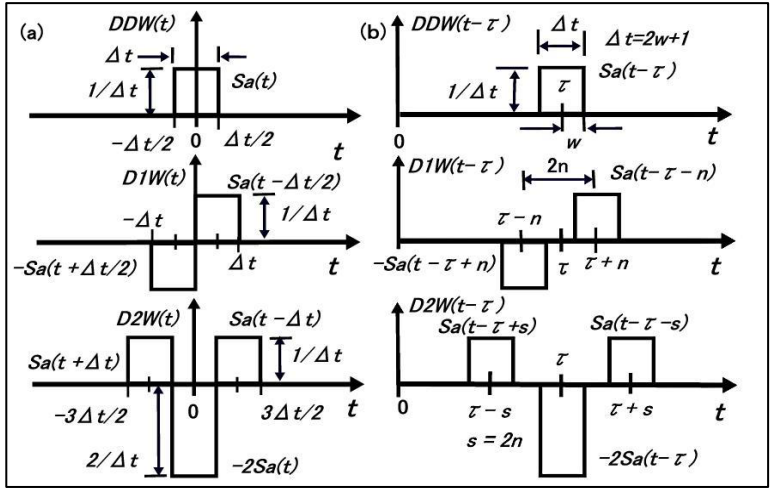

**FIG. 12-1. The layouts of square waves to construct Physical Wavelets.**
(a) The interval to take differences for $D1W(t)$ and $D2W(t)$ is integer $\Delta t$ ($= 2w+1$) that is the width of $Sa(t)$ or $DDW(t)$. (b) The interval can be any integer different from the width $\Delta t$ of $Sa(t-\tau)$. The layouts are for $D1W(t-\tau)$ and $D2W(t-\tau)$ with $s = 2n$ where integer $s > \Delta t$. A few other layouts with different conditions among $s$, $2n$, and $\Delta t$ are not shown.

### 12.1.2 Physical Wavelets (P-Ws)

Consider the motion of a virtual particle of unit mass in coordinate space. We first assume its *c*-component position, denoted by $D(c, t)$, is time-differentiable. Denoting a small interval of $t$ by $\Delta t$ ($\geq 0$), the conventional definition of the *c*-component of the particle velocity is,

$$V(c,t) = \lim_{\Delta t \to 0}[D(c, t + \Delta t) - D(c, t)] / \Delta t = dD(c,t) / dt. \quad (A1)$$

The differential operator $d/dt$ has the time reversal property of $d/d(-t) = -d/dt$. This time reversal of $-t$, while keeping the interval $\Delta t$ positive, changes the forward difference in Eq. (A1) into, $[D(c, -(t - \Delta t)) - D(c, -t)] = -[D(c, t) - D(c, t - \Delta t)]$, for which $D(c, -t) = D(c, t)$. The difference does not obey the time reversal of $d/dt$. The correct representation is then the central difference,

$$V(c,t) = \lim_{\Delta t \to 0}[D(c, t + \Delta t / 2) - D(c, t - \Delta t / 2)] / \Delta t = dD(c,t) / dt. \quad (A2)$$

Using the Dirac delta function $\delta(\tau)$, Eq. (A2) is

$$V(c,t) = \lim_{\Delta t \to 0}\{\int_{-\infty}^{+\infty} D(c, \tau)[\delta(\tau - t - \Delta t / 2) - \delta(\tau - t + \Delta t / 2)]d\tau\} / \Delta t. \quad (A3)$$

The $\delta(\tau)$ is an even function of time $\tau$ with the property of

$$D(c,t) = \int_{-\infty}^{+\infty} D(c, \tau)\delta(\tau - t)d\tau. \quad (A4)$$

The $\delta(t)$ may be replaced with a square wave of $Sa(t)$, as in Fig. A1, whose height and width are $1/\Delta t$ and $\Delta t$, respectively. As $\Delta t \to 0$, $Sa(t)$ has the same property as $\delta(t)$. Replacing $\delta(t)$ with $Sa(t)$, Eq. (A4) is

$$D(c, \tau) = \lim_{\Delta t \to 0} \int_{-\infty}^{+\infty} D(c, t) Sa(t - \tau) dt. \tag{A5}$$

Similarly, Eq. (A3) is

$$V(c, \tau) = \lim_{\Delta t \to 0} \{ \int_{-\infty}^{+\infty} D(c, t) [Sa(t - \tau - \Delta t / 2) - Sa(t - \tau + \Delta t / 2)] dt \} / \Delta t. \tag{A6}$$

Assuming $V(c, \tau)$ is differentiable, acceleration $A(c, \tau)$ is then

$$A(c, \tau) = \lim_{\Delta t \to 0} [V(c, \tau + \Delta t / 2) - V(c, \tau - \Delta t / 2)] / \Delta t = dV(c, \tau) / d\tau$$
$$= \lim_{\Delta t \to 0} \{ \int_{-\infty}^{+\infty} D(c, t) [Sa(t - \tau - \Delta t) - 2Sa(t - \tau) + Sa(t - \tau + \Delta t)] dt \} / \Delta t^2 = d^2 D(c, \tau) / d\tau^2. \tag{A7}$$

By removing the limiting process, the differentiability of $D(c, t)$ is not the prerequisite to defining $V(c, t)$ and $A(c, t)$. Integrals of Eqs. (A5) – (A7) are the cross-correlation functions between $D(c, t)$ and a set of square waves. The square waves form the observational windows (operators) with which to detect the particle's motion at time $\tau$. The operator in Eq. (A5) detects the position (or displacement) of the particle exposed over interval $\Delta t$, so it is the displacement detector, $Sa(t - \tau) = DDW(t - \tau)$. Similarly, the operator in Eq. (A6) is the first-order difference detector, $Sa(t - \tau - \Delta t/2) - Sa(t - \tau + \Delta t/2) = D1W(t - \tau)$. The operator in Eq. (A7) is the second-order difference detector, $Sa(t - \tau - \Delta t) - 2Sa(t - \tau) + Sa(t - \tau + \Delta t) = D2W(t - \tau)$. The layouts of these detection windows are in Fig. 12-1 at time $t = 0$. The $D1W(t)$ and $D2W(t)$ are odd and even functions of time $t$ to obey each time reversal property of $d/dt$ and $d^2/dt^2$. We denote $D1W(t - \tau)/\Delta t$ and $D2W(t - \tau)/(\Delta t)^2$ by $VDW(t - \tau)$ and $ADW(t - \tau)$, respectively. The $Sa(t)$ in the definitions may be other representations for the $\delta(t)$.

The $DDW(t)$ is even, $VDW(t)$ is odd, and $ADW(t)$ is even with respect to $t$. Therefore, $DDW(t - \tau)$ and $VDW(t - \tau)$ are orthogonal to each other at time $t = \tau$, and so are $VDW(t - \tau)$ and $ADW(t - \tau)$. However, $DDW(t - \tau)$ and $ADW(t - \tau)$ are not. The orthogonality between $DDW(t - \tau)$ and $VDW(t - \tau)$ guarantees that $D(c, \tau)$ and $V(c, \tau)$ are independent of one another. They completely define the state of the moving particle in the $D(c, t) - V(c, t)$ plane (space) at time $\tau$ and then uniquely define $A(c, \tau)$ [A15]. Its subsequent motion will draw the predicted path in space. We name these detection windows Physical Wavelets and specifically $DDW(t - \tau)$ displacement-defining operator, $VDW(t - \tau)$ velocity-defining operator, and $ADW(t - \tau)$ acceleration-defining operator.

### 12.1.3 Equations of stochastic motion

We now assume the particle motion changes its direction and speed discontinuously. We denote its $c$-component position at time $t$ by $d(c, t)$ and its non-differentiable path by the observed data $\{c\} = \{d(c, 0), d(c, 1), d(c, 2), \ldots, d(c, j), \ldots\}$ where integer $j$ is the chronological event index at the observation time $t$. We have the $c$-component position of the particle motion smoothed over $\Delta t = 2w+1$, for which integer $w \geq 1$. It is

$$D(c, \tau) = \int_{-\infty}^{+\infty} \{c\} DDW(t - \tau) dt = [1 / (2w+1)] \sum_{j=-w}^{w} d(c, \tau + j). \tag{A8}$$

The interval to take differences in $D1W(t - \tau)$ and $D2W(t - \tau)$ is $\Delta t = 2w+1$. Let the interval be an integer of $2n$ for $D1W(t-\tau)$ and another integer $s$ for $D2W(t-\tau)$. Their layouts are in Fig. 12-1b. The P-Ws find $V(c, \tau)$ and $A(c, \tau)$ for $s = 2n$ as,

$$V(c,\tau)=\int_{-\infty}^{+\infty}\{c\}\,VDW(t-\tau)dt=[D(c,\tau+s/2)-D(c,\tau-s/2)]/s \qquad \text{(A9)}$$

and

$$A(c,\tau)=\int_{-\infty}^{+\infty}\{c\}\,ADW(t-\tau)dt=[D(c,\tau+s)-2D(c,\tau)+D(c,\tau-s)]/s^2. \qquad \text{(A10)}$$

In these definitions, each square wave, $Sa(t)$ of Fig. 12-1b, collects $d(c, j)$ as in Eq. (A8).

The relations between Eq. (A8), Eq. (A9), and Eq. (A10) are the equations of motion [A15] for the particle motion changing its direction and speed discontinuously, which may carry the periodically fluctuating components embedded in $\{c\}$.

The extraction of specific periodicity from $\{c\}$ is significant if the mutual correlation between P-Ws and $\{c\}$ is strong. In the fluctuation (frequency) domain, the Fourier transform of P-Ws is the extracting function by the correlation theorem. The respective Fourier transforms of $DDW(t)$, $VDW(t)$, and $ADW(t)$ are then;

$$DDW(f)=\frac{\sin(\pi f\Delta t)}{\pi f\Delta t}\quad, \qquad \text{(A11)}$$

$$VDW(f)=\frac{2}{i}\frac{\sin(\pi f\Delta t)}{\pi f\Delta t\,s}\sin(\pi f s) \qquad \text{(A12)}$$

and

$$ADW(f)=-4\frac{\sin(\pi f\Delta t)}{\pi f\Delta t\,s^2}\sin^2(\pi f s)\quad. \qquad \text{(A13)}$$

The frequency $f$ is in 1/day. The symbol $i$ in Eq. (A12) is a complex number, $i \times i = -1$. Equation (A11) works as a low pass filter, Eq. (A12) as a bandpass filter, and Eq. (A13) as another bandpass filter. A functional alternative to $Sa(t)$ improves these filtering functions. Therefore, Eqs. (A8) – (A10) show the respective filtered physical quantities. The $w$ and $s$ can be any integer by which to filter out the selected frequency components of $D(c, \tau)$, $V(c, \tau)$, and $A(c, \tau)$. These extracted components draw the $D$-$A$, $D$-$V$, $V$-$V$, and $A$-$A$ paths [A8, A9, A11, A13].

The Physical Wavelets operations stated above can be represented by bra and ket vectors in analogy to quantum mechanics. In the matrix representation of Physical Wavelets, displacement-detecting (or defining) wavelet (operator) $DDW(t-\tau)$, velocity-defining operator $VDW(t-\tau)$, and acceleration-defining operator $ADW(t-\tau)$ are in row vectors localized at time $\tau$ and a non-differentiable time series $\{c\}$ is a column vector representing the monitored physical state. Then defining each physical quantity takes the inner product of two vectors, which is the same as taking the cross-correlation of each wavelet (operator) with time series $\{c\}$ in continuous function representation. In the vector representation, $DDW(t-\tau)$ is $< DDW(t-\tau)\,|$ in the bra vector, and $\{c\}$ is $|\,c >$ in the ket vector. We may write Eq. A8 – A10 as:

$< DDW(t-\tau)\,|\,c > =$ Eq. A8, $< VDW(t-\tau)\,|\,c > =$ Eq. A9, and $< ADW(t-\tau)\,|\,c > =$ Eq. A10.

### 12.1.4 Automated power monitoring

If an anomaly is a change comparable to the amplitudes of the observed fluctuations in $\{c\}$, it will be indistinguishable from the background fluctuations. It may appear as the changes in their phases and magnitudes. The following method can detect such anomalies in real time [A7-A10].

We define a time-rate change of the kinetic energy by the product of $V(c, \tau)$ and $A(c, \tau)$ in Eq. (A9) and Eq. (A10), which is the power, $PW(c, \tau)$. We then reassign the $V(c, \tau)$ and $A(c, \tau)$ at time $\tau$ to $V(c, j)$ and $A(c, j)$ at the current time $j = \tau + w + s$, defined for $A(c, \tau)$. The power at time $j$ is given by

$$PW(c, j) = V(c, j) \times A(c, j) = \frac{1}{s} KE(c, j) \times \left(1 - \frac{V(c, j-s)}{V(c, j)}\right) \quad . \tag{A14}$$

The $KE(c, j)$ is the kinetic energy, defined as the $V(c, j)$ squared. Velocity $V(c, j)$ extracts the fluctuations of period $2s$ for which $D(c, j)$ and $D(c, j-s)$ will have the opposite sign to each other. The $KE(c, j)$ magnifies the relative velocity change in Eq. (A14) parentheses. Thus, power $PW(c, j)$ becomes maximum near either troughs or peaks of the periodic fluctuations of $2s$ in $A(c, j)$. The larger the amplitude of $A(c, j)$ localized within $2s$ becomes, the larger $PW(c, j)$ becomes. Thus, any anomaly becomes the corresponding large $PW(c, j)$. A predetermined threshold detects the rising $PW(c, j)$, which becomes higher than the threshold level and gives the anomaly-onset time. The threshold level may automatically adopt the $PW(c, j)$'s maximum amplitude during the standard condition.

### 12.2 References on Appendix A

## 13 Appendix B (Current oceanic plate observations as of March 11, 2023)

The daily crustal displacement time series, $\{c\} = \{d\,(c, 1), d\,(c, 2), …, d\,(c, j), …\}$ represents the relative change from a graphical reference, and the chronological event index $j$ starts from $j = 1$ unless specified. The geographic axis $c$ is $E$ (west to east), $N$ (south to north), and $h$ (down to up) in right-handed coordinates $(E, N, h)$, as shown in Fig. 5-1.

The observation at Minamitorishima station in the Northwest Pacific Ocean (Figs. 3 and 9) is imperative in detecting the imminent megathrust EQ genesis process off Nemuro (label 4, in Fig. 9). The real-time monitoring requires the station's stable operation and increased number of GPS stations around and within the Russian border to detect a bulge-bending deformation, as in section 7.1. The current Minamitorishima operation appears unstable, and thus, we only update the observed $\{c\}$ at Chichijima-A station on the western edge of the subducting northwestern Pacific Plate and Minamidaito-Jima station on the Philippine Sea Plate using earthquake prediction software [B1, B2].

As of March 11, 2023, the motion of the northwestern Pacific Plate and the Philippine Sea Plate in the subduction zone of Tohoku and western Japan is normal for imminent megathrust EQs [section 13.4]. However, the motion of Chichijima-A (and Hahajima) was abnormal from January 2, 2021, until April 10, 2021. During the abnormal period, the M7.3 and the M6.9 ruptured off the Tohoku east coast on February 13, 2021, and March 20, 2021, respectively. The abnormal motion was a shortened and miniature version of the 2011 Tohoku M9 EQ, as in Figs. 13-2-3d and 13-2-3e.

We define the $D\,(c, \tau)$, $V\,(c, \tau)$, and $A\,(c, \tau)$ with P-Ws of $w \approx 6$ and $s \approx 15$ ($w = 6$ and $s = 15$, and $w = 7$ and $s = 20$) to extract the lunar synodic fluctuations of period 29.5. The updates follow those graphics in Figs. 5-1 and 6-1. The oceanic plate's $\{E\}$ and $\{N\}$ observations reveal that:

1.   The horizontal displacement follows the lunar synodic tidal loading (29.5-day period).
2.   The westward-moving speed triggering significant EQ events becomes approximately twice higher than the standard one. As of May 22, 2021, the observed significant EQ events are M6.9, M6.4, M7.9, M8.1, M7.3, and M6.9. The M6.9 and M7.3 were the coupled events stated above. The M8.1 was west off Ogasawara, 681.7 km deep, one of Wadati-Benioff zone EQs (Figs. 46 - 49, in [B2]).

3. The bulge-bending deformation (section 7.1) pulled the subducting Pacific Plate westward, as in Fig. 4b. The westward speed increased approximately three times higher than usual, triggering the M7.9 EQ on December 22, 2010. The pulling action ultimately led to the 2011 Tohoku M9 EQ and tsunami, as depicted in Fig. 4c. Thus, the M7.9 EQ was a precursory event to the M9 EQ, as in sections 6.1 and Figs. 13-2-3f - 13-2-3i.

4. The geophysical origin for the increased speed, triggering the other large EQs, is unclear. It may have coupled with so-called slow slip events in subduction zones [B3], as discussed in sections 13.2.4, 13.2.5, and 13.3.5.

5. However, the GPS observations on the 2011 Tohoku M9 EQ genesis process suggest that slow slip events are not in the megathrust EQ genesis process in subduction zones.

6. Power monitoring shows that the M7.9 and M8.1 events ruptured after a few consecutive abnormal responses to the synodic loading (in sections 13.2.3 and 13.3.4). The two cases suggest that the first anomalous response is the precursor to the upcoming event larger than about M8. A power monitor is an automated tool to detect unusual motion in time *j*.

## 13.1 The oceanic plates and the overriding continental plates

### 13.1.1 The subducting northwestern Pacific Plate, the Philippine Sea Plate, and their plate boundaries

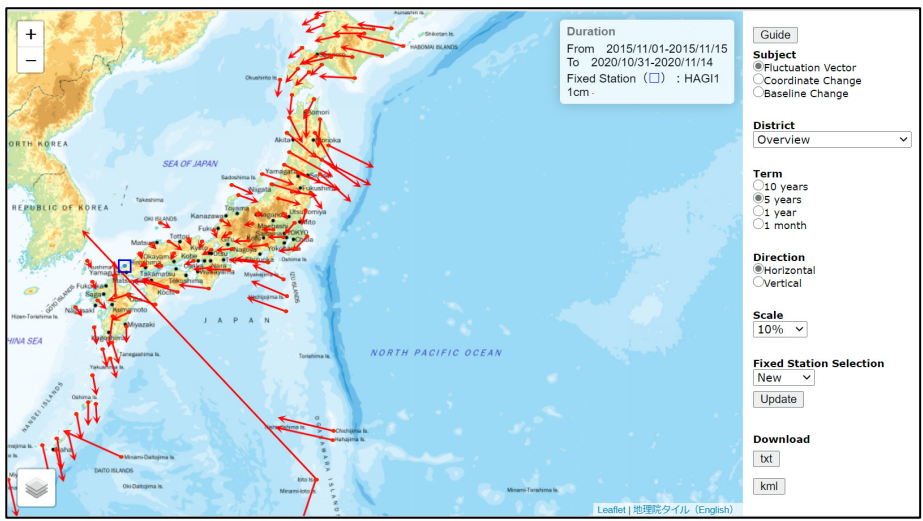

**Fig. 13-1-1. The horizontal displacements at the website https://mekira.gsi.go.jp/index.en.html**
Each displacement over five years at the GPS stations is from a reference at the station labeled in the square (Hagi1). The abnormal movements at Ioto Island (below Chichijima Island) are due to volcano activities [B4].

Before the 2011 Tohoku M9 EQ, the eastward movements along Tohoku's east coast were opposite. The M9 EQ has reversed the movement direction. Thus, Tohoku's east coast is still under the 2011 M9 EQ.

The Chichijima-A station (Chichijima Is) is on the subducting northwestern Pacific Plate, and Minamidaito-Jima station (Minami Daitojima Is) is on the Philippine Sea plate.

### 13.1.2 Geological configurations of overriding plate boundaries and oceanic plate motions

The Pacific Plate moves roughly perpendicular to the northern main-island (Tohoku) eastern shoreline of about 500 km (as in Fig. 13-1-1). Due to the enormous fault coupling, the entire shoreline had been pulled down, as illustrated in Fig. 4a. As the crustal deformation along the shoreline turned into a bulge-bending pulling (Phase 1 in section 7.4), the further bending started accelerating the pull, as in Fig. 4b. In Phase 2 (section 7.4), the pull was decelerated and stopped by the bulge force with a gradual upheaval growth of 1.2 mm on the east coast. The upheaval growth triggered the enormous force recoiled by the restoring force of the compressed west coast against the overriding plate, leading to the megathrust earthquake and tsunami illustrated in Fig. 4c. Based on an empirical magnitude relation of M = 2 × (Log₁₀ (500 Km) + 1.8) = 9 [B5], we can estimate 500 km fault length's rupture to be the M9 EQ.

In contrast, the Philippine Sea Plate moves roughly parallel to the 600 km east coastline of the southern main island and Shikoku, as in Fig. 13-1-1. The coastline's subsidence is local, appearing only in Tokai, Tonankai, and Nankai, as in Fig. 13-1-2. The entire coastline's subsidence to couple with the significant fault on the subducting Philippine Sea Plate and the subsequent bulge-bending deformation to pull down the oceanic plate is necessary for the Nankai-trough Mw 9.1 EQ with 34-meter-high tsunami [B6].

However, the subsidence is missing along the entire coastline, suggesting that the occurrence of the Nankai-trough Mw 9.1 events is implausible. Instead, the subsidence scattered along the coastline suggests that M7 to M8 EQs are more likely.

We may also estimate the unrealistic EQ of 600 km fault length to be M9.2 by the empirical relation of M = 2 × (Log₁₀ (600 Km) + 1.8) = 9.2 [B5].

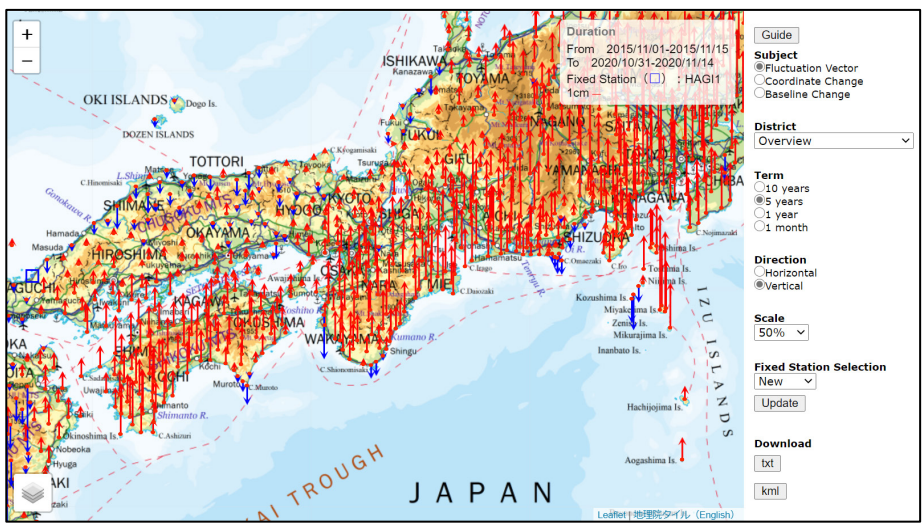

**Fig. 13-1-2. The vertical displacements at the website https://mekira.gsi.go.jp/index.en.html**
Each vertical displacement over five years at the GPS stations is from a reference at the station labeled in the square (Hagi1).

### 13.1.3 Chichijima-A station on the subducting northwestern Pacific Plate

The GPS station had a 39-day-deficit out of the total 4730-day-observation. The deficits include the removed spike data of Tohoku M9 EQ (March 11, 2011).

As in Fig. 5-1, the $d(N, j)$, $d(E, j)$, and $d(h, j)$ are in the first, second, and third windows from the top, respectively. Each abscissa is time $j$ and $\tau$ from December 4, 2007 ($j = 1$). Each ordinate is each displacement $d(c, j)$ (green) and $D(c, \tau)$ (red) in meters. Each graphical origin has its offset assigned for the entire display of $\{c\}$. Their values are 0.03 m, − 0.11 m, and 0 m from the top window. The northward, eastward, and upward displacements from the origin are positive. The magnifications on each offset scale are 10000, 2500, and 5000 for the entire drawing of $d(c, j)$ in green. Each mercury manometer-like column has the last offset displacement at its bottom, which is 0.0872 m, − 0.3460 m, and 0.0186 m.

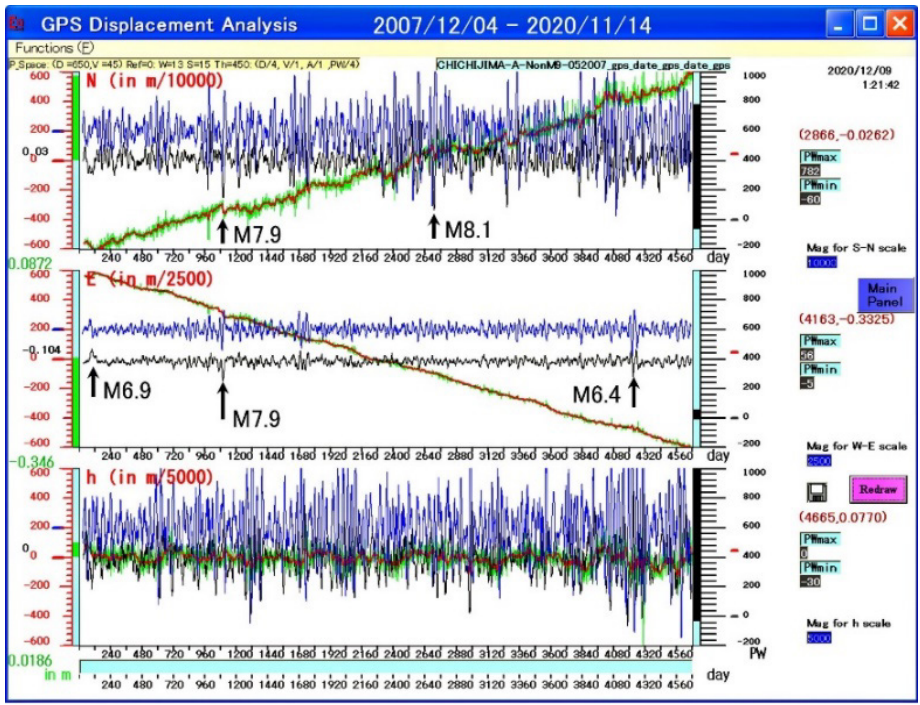

**Fig. 13-1-3a. Chichijima-A from December 4, 2007 ($j = 1$) to November 14, 2020 ($j = 4691$).**

Parameters $w$ and $s$ to define $D(c, \tau)$, $V(c, \tau)$ and $A(c, \tau)$ are $w = 6$ (W $= 13 = 2w + 1$) and $s = 15$ (S $= s$). The black $V(c, \tau)$ and blue $A(c, \tau)$ are in relative scales with their origins at scales 0 (red) and 200 (blue), respectively. $D(c, \tau)$ is in red. Time $\tau$ lags behind time $j$, $j = \tau + w$ for $D(c, \tau)$, $j = \tau + w + s/2$ for $V(c, \tau)$, and $j = \tau + w + s$ for $A(c, \tau)$.

The four-to-one magnification ratio for $\{N\}$ and $\{E\}$ shows that the average horizontal direction of motion is approximately 14 degrees from the west to the north. We note that the two-to-one ratio for $\{N\}$ and $\{h\}$ makes their environmental-noise fluctuation amplitudes similar. Thus, the $\{N\}$ and $\{h\}$ paths in the $D(c, \tau) - V(c, \tau)$ plane become similar without the northward moving trend in Fig. 13-1-3b. The large EQs (M6.9, M7.9, and M6.4 in Fig. 13-1-3a) are in <u>section 13.2</u>.

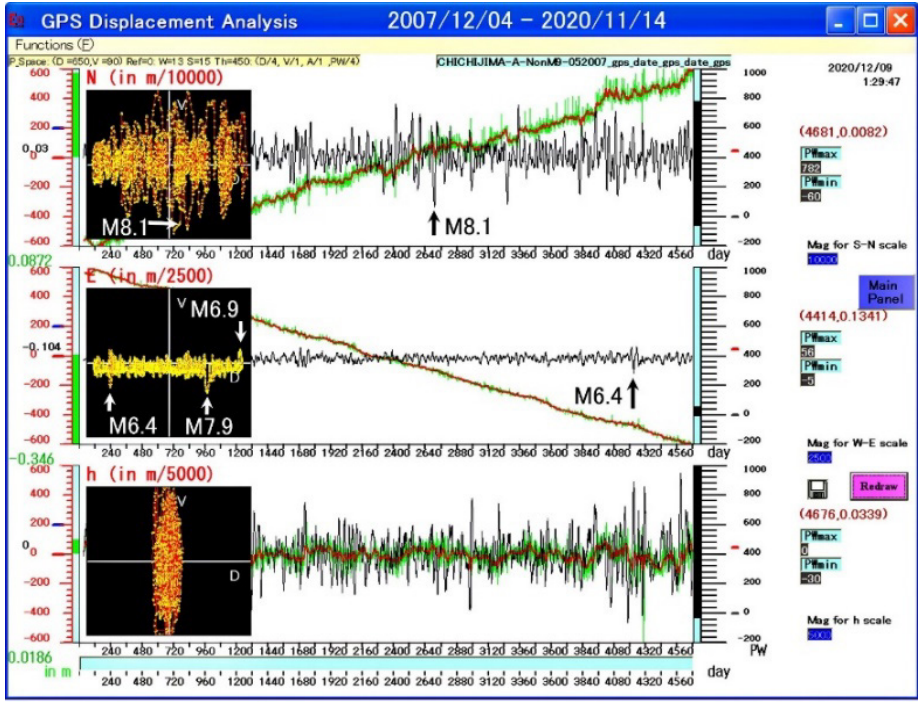

**Fig. 13-1-3b. Chichijima-A's $D\,(c,\tau) - V\,(c,\tau)$ path.**

As for the graphical parameters on Fig. 13-1-3b, the parameters are W= 13 = $2w + 1$ ($w = 6$) and S = $s =$ 15. The (D = 650, V = 90) has the respective scale conversion;

for $c = N$, (6.5 cm, 0.6 mm/day) by the magnification 10000 for $d\,(N, j)$,

for $c = E$, (26 cm, 2.4 mm/day) by the magnification 2500 for $d\,(E, j)$,

for $c = h$, (13 cm, 1.2 mm/day) by the magnification 5000 for $d\,(h, j)$.

The origin of the $D\,(c,\tau) - V\,(c,\tau)$ plane has each offset values; 0.03 m for $D\,(N, \tau)$, $-$ 0.104 m for $D\,(E, \tau)$, 0 m for $D\,(h, \tau)$. The (D / 4, V / 1, A / 1, PW / 4) is the reduced magnification for the respective drawing. The threshold (Th) explanation is in <u>section 13.2.3</u>. The right-side scales are for $PW\,(c,\tau)$ and $PW\,(c, j)$ as in <u>Fig. 6-1d</u>.

### 13.1.4 Minamidaito-Jima station on the Philippine Sea Plate

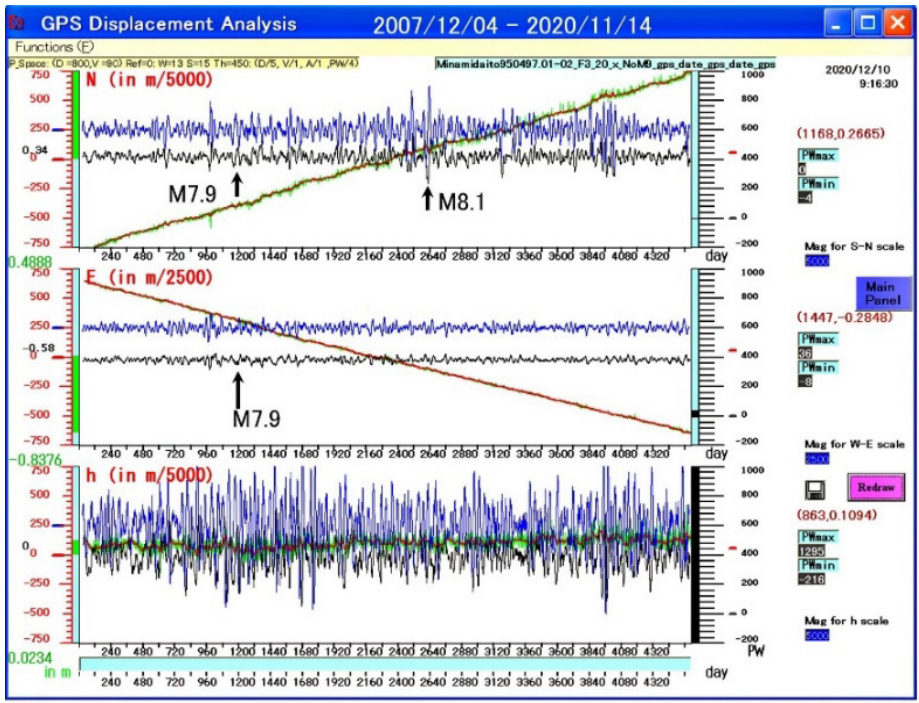

**Fig. 13-1-4a. Minamidaito-Jima from December 4, 2007 ($j = 1$) to November 14, 2020 ($j = 4612$).**

The GPS station has a scattered 118-day deficit, mainly during 2011 and 2012, out of the total 4730-day observation. Their deficits include the spike of March 11, 2011. The $d(N, j)$, $d(E, j)$, and $d(h, j)$ are in the first, second, and third windows from the top, respectively. Each abscissa is time $j$ and $\tau$ from December 4, 2007 ($j = 1$). Each ordinate is each displacement $d(c, j)$ (green) and $D(c, \tau)$ (red) in meters. Each graphical origin zero has an offset assigned for the entire display of $\{c\}$. Their values are 0.34 m, $-0.58$ m, and 0 m from the top window. The northward, eastward, and upward displacements from the origin are positive. Each offset scale has 5000, 2500, and 5000 magnifications for the entire graphing of $d(c, j)$. Each mercury manometer-like column has the last offset displacement at its bottom, which is 0.4888 m, $-0.8376$ m, and 0.0234 m.

Parameters $w$ and $s$ to define $D(c, \tau)$, $V(c, \tau)$ and $A(c, \tau)$ are $w = 6$ (W $= 13 = 2w + 1$) and $s = 15$ (S $= s$). The black $V(c, \tau)$ and blue $A(c, \tau)$ have the same drawings as Fig. 13-1-3a. They are in relative scales with their origins at scales 0 (red) for $V(c, \tau)$ and 200 (blue) for $A(c, \tau)$. Time $\tau$ lags behind time $j$, $j = \tau + w$ for $D(c, \tau)$, $j = \tau + w + s/2$ for $V(c, \tau)$, and $j = \tau + w + s$ for $A(c, \tau)$. The two to one magnification ratio for $\{N\}$ and $\{E\}$ shows that the average horizontal direction of motion is approximately 27 degrees from the west to the north. The magnifications of $\{N\}$ and $\{h\}$ are the same; however, the environmental-noise fluctuation amplitudes of $\{h\}$ become larger, and so do those of $V(h, \tau)$ and $A(h, \tau)$. The M7.9 is the M7.9 EQ in sections 6.1 and 13.2.

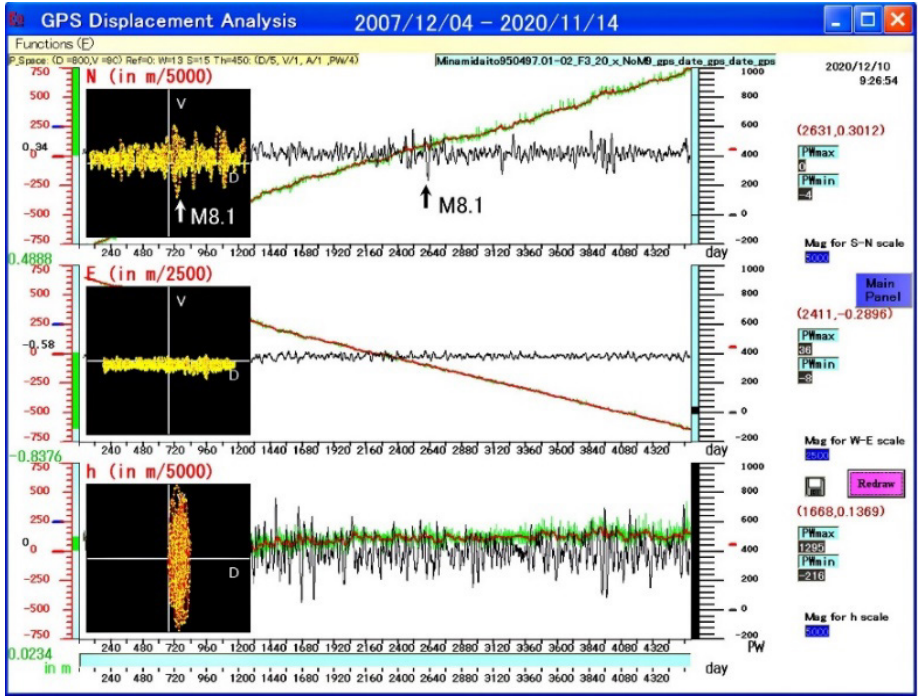

**Fig. 13-1-4b.** Minamidaito-Jima's $D(c, \tau) - V(c, \tau)$ path.

The parameters are W = 13 = 2$w$ + 1 ($w$ = 6) and S = 15 = $s$. Phase Space (D = 800, V = 90) for the respective $D(c, \tau) - V(c, \tau)$ is;

for $c = N$, (16 cm, 1.2 mm/day) by the magnification 5000 for $d(N, j)$,

for $c = E$, (32 cm, 2.4 mm/day) by the magnification 2500 for $d(E, j)$,

for $c = h$, (16 cm, 1.2 mm/day) by the magnification 5000 for $d(h, j)$.

The origin of the $D(c, \tau) - V(c, \tau)$ plane has each offset values: 0.34 m for $D(N, \tau)$, $-$ 0.58 m for $D(E, \tau)$, 0 m for $D(h, \tau)$. The (D / 5, V / 1, A / 1, PW / 4) is the reduced magnification for the respective drawing. The threshold (Th) is in section <u>13.2.3</u>. The right-side scales are for $PW(c, \tau)$ and $PW(c, j)$.

## 13.2 The subducting northwestern Pacific Plate

### 13.2.1 The overriding continental plate (the Tohoku area)

Before the 2011 Tohoku M9 EQ, Tohoku's east coast vertical motion was downward as Figs. 4 and 7-1a. The M9 EQ has reversed the vertical direction, as in section <u>7.1.2</u>. Figure 13-2-1 shows that the displacement along the east coastline is still upward over 500 km, namely under the 2011 Tohoku M9 EQ event. The displacements are over five years at the GPS stations, which are relative to the fixed station labeled in the square, Ryoutsu1 station, just above Ryoutsu2 station in <u>section 7.2</u>. The Tohoku area is still under the 2011 M9 EQ event.

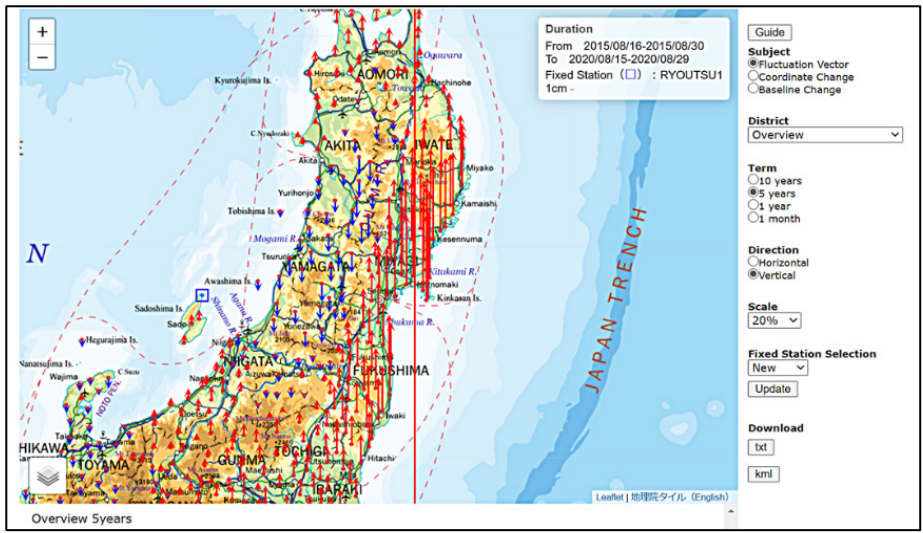

**Fig. 13-2-1. The vertical displacements at the website https://mekira.gsi.go.jp/index.en.html**

## 13.2.2 The westward motion of the subducting Pacific Plate (Chichijima-A station)

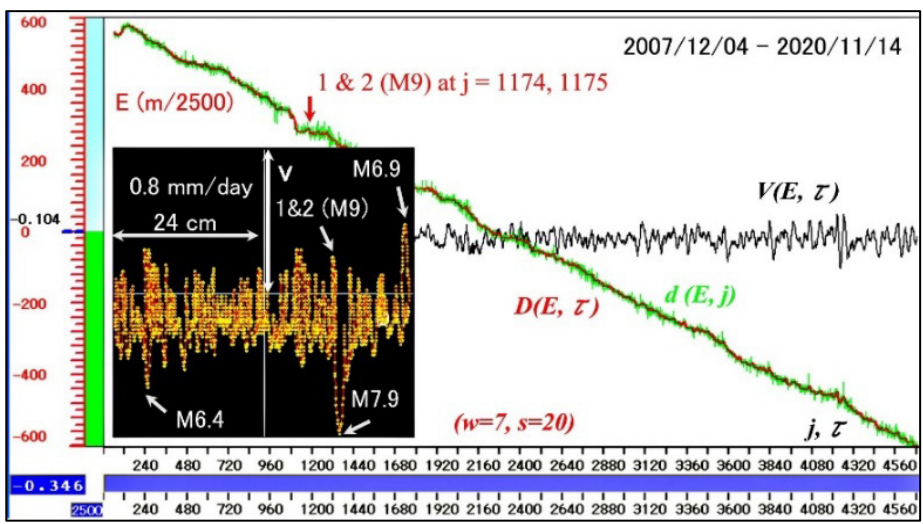

**Fig. 13-2-2a. Chichijima-A from December 4, 2007, to November 14, 2020.**

Figure 13-2-2a is the update of Fig. 6-2a. Label 1 & 2 (M9) points to the deficit of the M9 EQ on March 11, 2011, between $j = 1174$ and 1175. The unexpected westward and eastward motions on the path with $w = 7$ and $s = 20$ triggered EQs.

The eastward speed $V(E, \tau) = + 0.39$ mm/day at $\tau = 91$ (March 6, 2008) triggered the M6.9 EQ of the reverse faulting (STR 31º, DIP 69 º, SLIP 74 º) in the Kurile island region [B7].

The westward speed $V(E, \tau) = - 0.77$ mm/day at $\tau = 1096$ (December 22, 2010) triggered the M7.9 EQ of the normal faulting (STR 340º, DIP 57 º, SLIP − 56 º) near Chichijima island [B7].

The westward speed $V(E, \tau) = - 0.534$ mm/day at $\tau = 4228$ (August 4, 2019) triggered the M6.4 EQ of the reverse faulting (STR 198º, DIP 21 º, SLIP 87 º) off the Fukushima coast [B7].

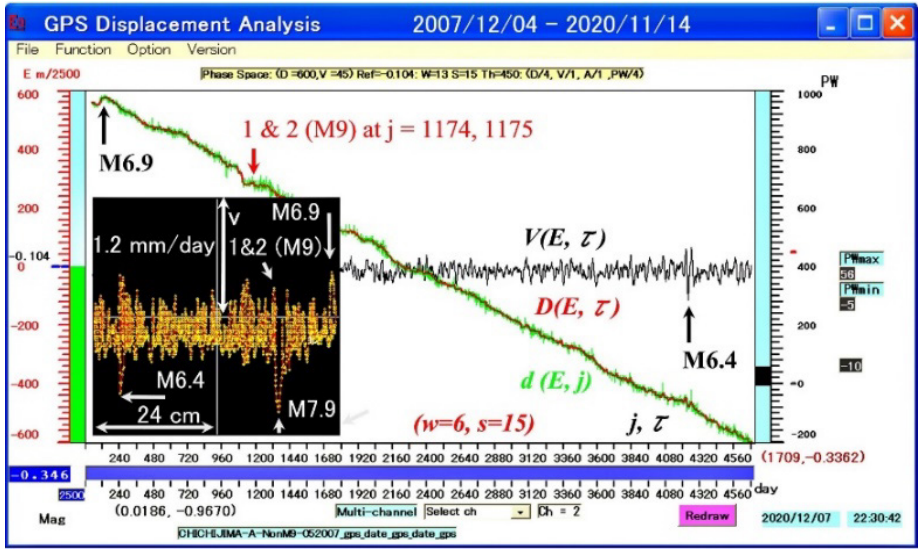

**Fig. 13-2-2b. Chichijima-A's $D(E, \tau) - V(E, \tau)$ path, from December 4, 2007, to November 14, 2020.**

The scale of phase space (D = 600, V = 45) is (24 cm, 1.2 mm/day) for the $D(E, \tau) - V(E, \tau)$ plane by the magnification 2500 for $d(E, j)$. The Ref = − 0.104 (= − 0.104 m + 0 / 2500 m) is the reading of $D(E, \tau)$ at the blue line on the left scale 0. It becomes the offset origin for the $D(E, \tau) - V(E, \tau)$ plane. The W and S parameters are W = 13 = 2$w$ + 1 ($w$ = 6) and S = 15 = $s$. The (D / 4, V / 1, A / 1, PW / 4) is the reduced magnification for the respective drawing.

Label 1 & 2 (M9) points to the deficit of the M9 EQ on March 11, 2011.

The unexpected motions on the path show the large $V(E, \tau)$ amplitudes triggering M6.9, M7.9, and M6.4. Their large amplitudes create the corresponding large $PW(E, \tau)$, as detailed in Appendix A 12-1-4.

### 13.2.3 Power monitoring $PW(E, \tau)$ on the westward motion at Chichijima-A

The observation at Chichijima station of Fig. 6-1b has the $D(E, \tau) - A(E, \tau)$ paths in Figs. 13-2-3f - 13-2-3i to show the equations. We begin updating the Chichijima-A observations.

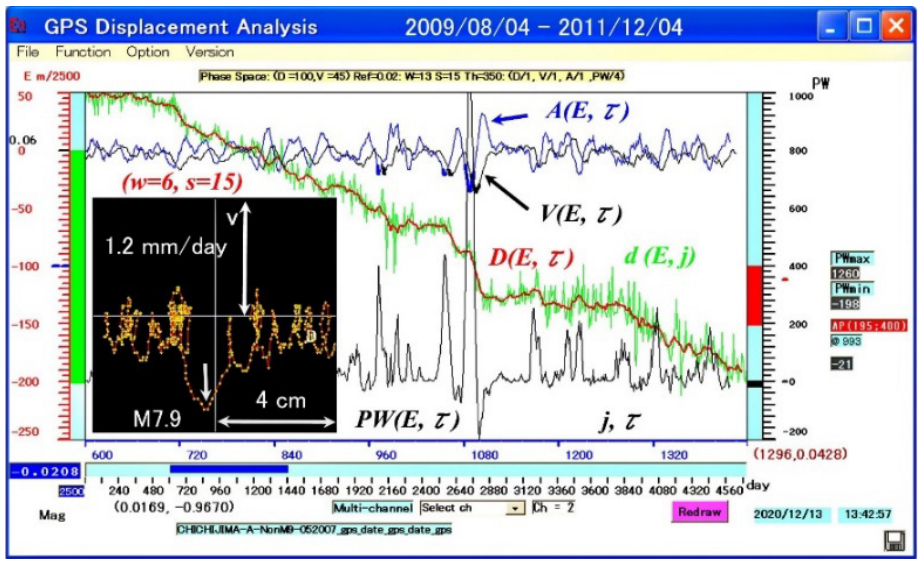

**Fig. 13-2-3a. $PW(E, \tau)$ from August 4, 2009, to December 4, 2011.**

Phase Space (D = 100, V = 45) is (4 cm, 1.2 mm/day) for the $D(E, \tau) - V(E, \tau)$ plane by the magnification 2500 for $d(E, j)$, whose offset origin is Ref = 0.02 (= 0.06 m − 0.04 m) read at the blue line on the left scale −100 (× m / 2500). Parameters W and S are W = 13 = 2$w$ + 1 ($w$ = 6) and S = 15 = $s$.

The path shows that the M7.9 event raptured at $V(E, \tau) = -0.9670$ mm/day at $\tau = 1096$ (on December 22, 2010). Its path-point is (0.0169, −0.9670), where 0.0169 m is the westward location from the offset origin 0.02m. The abnormal $V(E, \tau)$ triggered the normal faulting M7.9 EQ (STR 340º, DIP 57 º, SLIP −56 º) near Chichijima [B7].

The threshold (Th) for the abnormal power (AP) detection is Th = 350. The AP (195; 400) and @ 993 on the right is that $PW(E, j) \geq 350$ detected the AP changing from 195 to 400 at $j = 993$ ($\tau = 971$). There are three APs; the third corresponds with the abnormal westward $V(E, \tau)$, triggering M7.9 EQ. The power monitoring $PW(E, j)$ detected the unusual first response to the lunar synodic loading at $j = 993$ on September 10, 2010, about three months before the M7.9 EQ event.

In Fig. 13-2-3a, we shifted time $j$ back to time $\tau$ to show $D(E, \tau)$, $V(E, \tau)$, $A(E, \tau)$, and $PW(E, \tau)$, as in Fig. 6-1e. The $V(E, \tau)$ and $A(E, \tau)$ are bold, while $PW(E, \tau) \geq 350$.

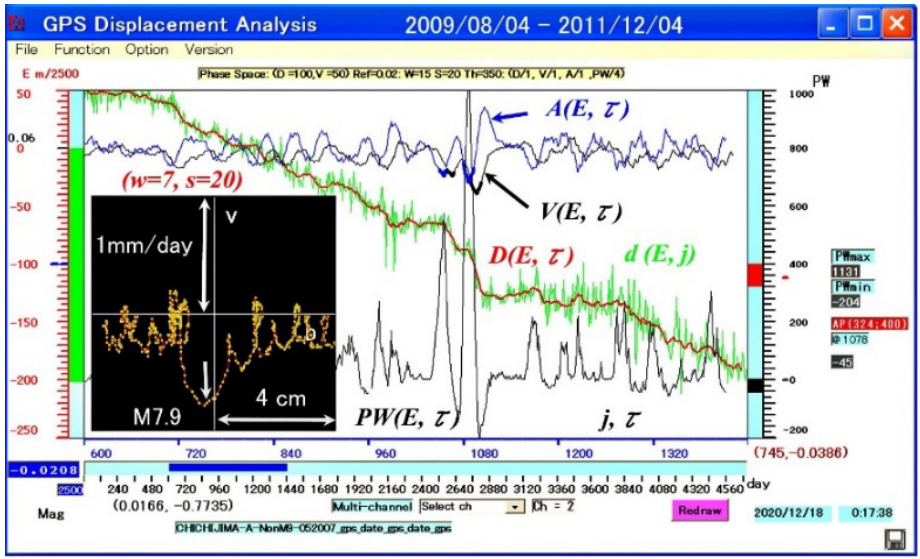

**Fig. 13-2-3b. $PW(E, \tau)$ with $w = 7$ and $s = 20$, from August 4, 2009, to December 4, 2011.**

Phase Space (D = 100, V = 50) is (4 cm, 1 mm/day) with the offset origin of Ref = 0.02 m (= 0.06 m − 100/2500 m). Parameters W and S are W = 15 = 2w + 1 ($w = 7$) and S = 20 = $s$.

The path shows that the M7.9 event raptured at $V(E, \tau) = -0.7735$ mm/day at $\tau = 1096$ on December 22, 2010. Its path-point reading is (0.0166, −0.7735), where 0.0166 m is the westward location from the offset origin 0.02m and $V(E, \tau) = -0.7735$ mm/day.

The threshold is Th = 350. The AP (324; 400) and @ 1078 is that $PW(E, j) \geq 350$ detected the first AP changing from 324 to 400 at $j = 1078$. The $V(E, \tau)$ and $A(E, \tau)$ are bold, while $PW(E, \tau) \geq 350$.

The $PW(E, \tau)$ shows that the M7.9 ruptured at the second anomalous synodic loading. It suggests the first loading at $j = 1078$ (on December 4, 2010) is a precursor to the normal faulting M7.9 EQ (STR 340º, DIP 57 º, SLIP −56 º) at $\tau = 1096$ (on December 22, 2010) near Chichijima [B7].

We may set the other $w$, $s$, and adoptive threshold (Th) for the $PW(E, j) \geq$ Th monitoring to re-observe the 2011 Tohoku M9 EQ. We then apply them to detect any plate's motion precursory to the anticipated megathrust ruptures.

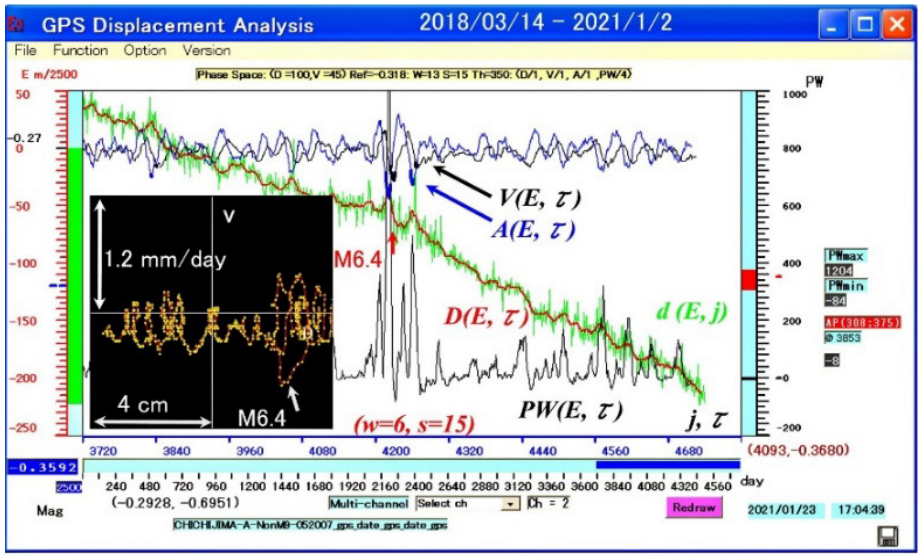

**Fig. 13-2-3c.** *PW* $(E, \tau)$ and M6.4 on August 4, 2019, as of January 2, 2021.

Phase Space (D = 100, V = 45) is (4 cm, 1.2 mm/day) with the offset origin of Ref = − 0.318 m (= − 0.27 m − 120/2500 m). The parameters to extract the lunar synodic loading are W = 13 = 2w + 1 (w = 6) and S = 15 = s.

The threshold is Th = 350. The AP (308; 375) and @ 3858 is that *PW* $(E, j) \geq 350$ detected the first AP changing from 308 to 375 at $j = 3858$. The $V (E, \tau)$ and $A (E, \tau)$ are bold, while *PW* $(E, \tau) \geq 350$.

The westward speed $V (E, \tau) = − 0.6951$ mm/day at $\tau = 4228$ (August 4, 2019) triggered the M6.4 EQ of reverse faulting (STR 198º, DIP 21 º, SLIP 87 º) off the Fukushima coast [B7].

As of January 2, 2021, the subducting northwestern Pacific Plate's {E} is normal, namely, no anomaly rupturing any megathrust EQ.

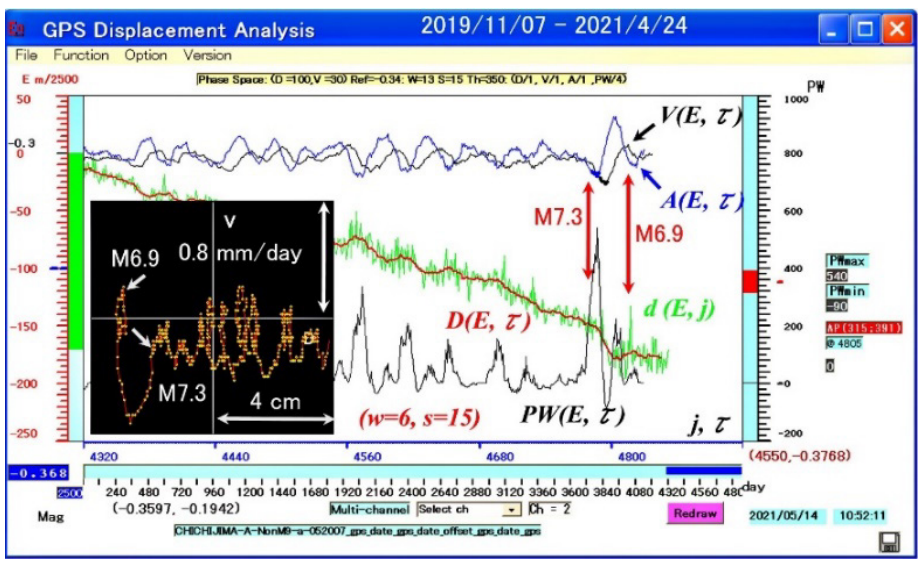

**Fig. 13-2-3d.** *PW* $(E, \tau)$, $D (E, \tau) − V (E, \tau)$ path, M7.3, and M6.9, as of April 24, 2021.

Figure 13-2-3d updates Fig. 13-2-3c, which shows M7.3 at $j = 4782$ (on February 13, 2021), and M6.9 at $j = 4817$ (on March 20, 2021) in the subduction zone. The M7.3 was the reverse faulting of (STR 191º, DIP 55º, SLIP 78º) with the epicenter in the subducted slab off Fukushima [B9]. The M6.9 was the reverse faulting of (STR 183º, DIP 20º, SLIP 72º) off Miyagi [B10].

A four-day segment in {c}, starting from February 20, 2021 (j = 4789 ~ 4793), had a uniform unrealistic displacement shift for which we ignored the four-day segment and made the offset smoothing in Figs. 13-2-3d, 13-2-3e and 13-2-5c. Hahajima station also had the same displacement shift.

The $D(E, \tau) - V(E, \tau)$ path shows an unusual westward motion, which coupled the M7.3 with the M6.9 event, as seen in <u>Fig. 13-2-3e</u>. The highest westward speed of $V(E, \tau) = -0.7223$ mm/day ($w = 6$ and $s = 15$) was at $\tau = 4795$ on February 26, 2021 (on March 11, 2021, in time $j = 4808 = 4795 + 6 + 15 / 2$). The westward speed with w = 7 and $s = 20$ was − 0.5670 mm/day, nearly reaching the maximum of − 0.78 mm/day in <u>Table 1</u>.

Chichijima-A and Hahajima stations are on the western edge of the Ogasawara Plateau (a western edge of the subducting Pacific Plate), as seen in <u>Fig. 13-1-1</u>. On the other hand, Hahajima showed the highest westward speed of $V(E, \tau) = -0.3738$ mm/day with w = 7 and $s = 20$ for the abnormal motion, about half the maximum of − 0.84 mm/day as in <u>Table 1</u>. Two stations responded differently to the fault (barrier) coupling with the Tohoku eastern edge near the M6.9 epicenter. The fault length estimation is L = 44.7 km by an empirical M = 2 × (Log$_{10}$ (L Km) + 1.8) <u>[B5]</u>.

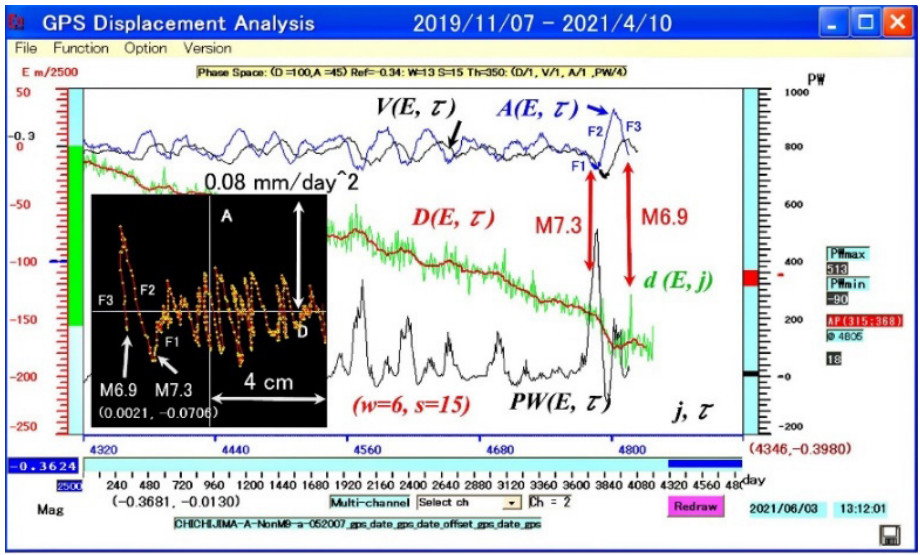

**Fig. 13-2-3e. $D(E, \tau) - A(E, \tau)$ path, M7.3, and M6.9, as of April 10, 2021.**
April 10, 2021, corresponds with time $j = 4838$ ($j = \tau + w + s$).

The $D(E, \tau) - A(E, \tau)$ path terminates at the M6.9 event ($\tau = 4817$, March 20, 2021). The $D(E, \tau)$ and $A(E, \tau)$ relation is on the (4cm, 0.08 mm/day$^2$) plane, which uses the same coordinate system as the $D(E, \tau) - V(E, \tau)$ plane. The $A(E, \tau)$ values in the upper half-plane is positive and eastward. The $A(E, \tau)$ at the GPS station is proportional to the external force $F(E, \tau)$, which is the sum of the subducting plate-driving force coupled with the overriding eastern edge and the lunar synodic tidal loading.

The $D(E, \tau) - A(E, \tau)$ path exhibits an oscillatory motion of the synodic tidal loading under the subducting plate's uniform westward movement. The lunar tidal force loading is significant when the path has uniformly dispersed data points. The densely packed points on the path represent the plate's constraint movement. Each oscillation exhibits a segment equation, $F(E, \tau) \approx A(E, \tau) \approx K \times D(E, \tau)$, with a time and synodic-cycle dependent constant $K$ (positive or negative). The observed three oscillations reveal three segment forces: F1, F2, and F3, which explain how M7.3 and M6.9 ruptured.

**Force 1 (F1) on segment 1 (a westward movement with $K = 0.0102$ /day$^2$)**

The $D(E, \tau) - A(E, \tau)$ path to the M7.3 event at $\tau = 4782$ (on February 13, 2021) shows $A(E, \tau) \approx K \times D(E, \tau)$ with a weak lunar synodic tidal modulation like a trend change on $D(E, \tau)$ for the Tohoku M9 EQ in section 6.1. The segment over the modulation has a linear change by $(\Delta D(E, \tau), \Delta A(E, \tau)) = (-5.7$ mm, $-0.0582$ mm/day$^2$) from December 31, 2020 ($A(E, \tau) = 0.0229$ mm /day$^2$ at $\tau = 4738$). The minus in the change is westward, which finds positive $K = 0.0102$ /day$^2$.

The $K$ is positive due to two constrained responses of the westward movement to synodic loadings. The westward force (negative $A(E, \tau)$) suppressed the lunar synodic tidal force loading by some constraint and ruptured the M7.3. Force 1 was abnormal, as shown with $PW(E, \tau) \geq 350$. The onset detection time is $j = 4805$ ($\tau = 4783$) two days after the M7.3 event.

The time rate unit conversion is mm/day$^2$ = 1.33959 $\times 10^{-11}$ cm/sec$^2$.

**Force 2 (F2) on segment 2 (a westward movement with $K = -0.0084$ / day$^2$)**

The $D(E, \tau) - A(E, \tau)$ path segment after the M7.3 shows a good linear relation, $A(E, \tau) = K \times D(E, \tau)$. The change $(\Delta D(E, \tau), \Delta A(E, \tau)) = (-10.9$ mm, $0.0914$ mm/day$^2$) finds $K = -0.0084$ /day$^2$. The observation on $V(E, \tau) = -0.7223$ mm/day at $\tau = 4795$ (February 26, 2021) is very similar to the abnormal motion on the Tohoku M9 EQ in section 6, suggesting that Force 2 is a pulling action by the Tohoku eastern edge. The edge might have caught hold of an unbroken barrier left on the subducting oceanic plate and pulled it westward.

**Force 3 (F3) on segment 3 (an eastward movement with $K = -0.0336$ /day$^2$)**

The $D(E, \tau) - A(E, \tau)$ path to the last phase point of M6.9 at $\tau = 4817$ (March 20, 2021) shows $A(E, \tau) \approx K \times D(E, \tau)$. The linear change $(\Delta D(E, \tau), \Delta A(E, \tau)) = (2.1$ mm, $-0.0706$ mm/day$^2$) finds $K = -0.0336$ /day$^2$. Force 3 ruptured the M6.9 with the large $K$.

**Summary on F1, F2, and F3:**

We summarize $F(E, \tau) \approx A(E, \tau) \approx K \times D(E, \tau)$ for each linear segment. Time $\tau$ for each force is the date starting the segment.

| Segment | $\tau$ | Date | $(\Delta D(E, \tau)$ mm, $\Delta A(E, \tau)$ mm/day$^2$) | $K$ (1 /day$^2$). |
|---|---|---|---|---|
| F1 | 4738 | Dec 31, 2020 | $(-5.7$ mm, $-0.0582$ mm/day$^2$) | $K = 0.0102$ /day$^2$ |
| F2 | 4784 | Feb 14, 2021 | $(-10.9$ mm, $0.0914$ mm/day$^2$) | $K = -0.0084$ /day$^2$. |
| F3 | 4801 | Mar 3, 2021 | $(2.1$ mm, $-0.0706$ mm/day$^2$) | $K = -0.0336$ /day$^2$. |

Force 1 has $F(E, \tau) \approx A(E, \tau) \approx K$ (positive) $\times D(E, \tau)$ relation on the M7.3 (reverse faulting within the subducted slab). Force 2 shows an effective lunar synodic tidal force loading, significantly reducing constrain with negative $K$ (dispersed points along the path). It pulled the subducting plate by $-10.9$ mm (westward), similar to the 2011 Tohoku M9 EQ in segment 2 of Fig. 13-2-3i. Force 3 is the reaction to Force 2 under the lunar tidal force loading. The overriding continental plate's eastern edge began to push back the barrier on the subducting plate by 2.1 mm, opposite the expected oceanic plate moving direction. The reaction (Force 3) moved the subducting plate eastward with the barrier, triggering the M6.9 EQ on March 20, 2021, as seen in segment 4-2 of Fig. 13-2-3i.

One probable reason for the establishment of the $A(E, \tau) \approx K$ (positive) $\times D(E, \tau)$ relation in Force 1 is a non-effective synodic loading due to constraint, such as catching hold of a barrier on the subducting plate by the overriding eastern edge. This is similar to the trend change observed in Fig. 6-1b, which had a preceding $A(E, \tau) \approx K$ (positive) $\times D(E, \tau)$, as seen in Figs. 13-2-3f and 13-2-3g. Thus, the Tohoku M9 discussed in

 suggests a miniature coupling of the Tohoku eastern edge with the subducting oceanic plate. The overriding edge pulled the subducting oceanic plate at the abnormal westward speed of $V(E, \tau) = -0.7223$ mm/day, as in .

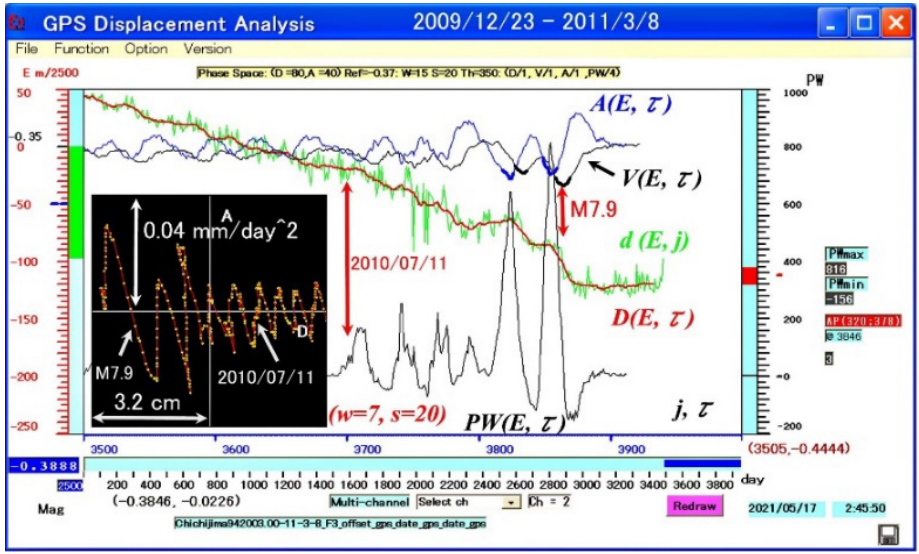

**Fig. 13-2-f.** $D(E, \tau) - A(E, \tau)$ path with $w = 7$ and $s = 20$ at Chichijima station for Fig. 6-1b.

Figure 13-2-3f is  with the $D(E, \tau) - A(E, \tau)$ path.

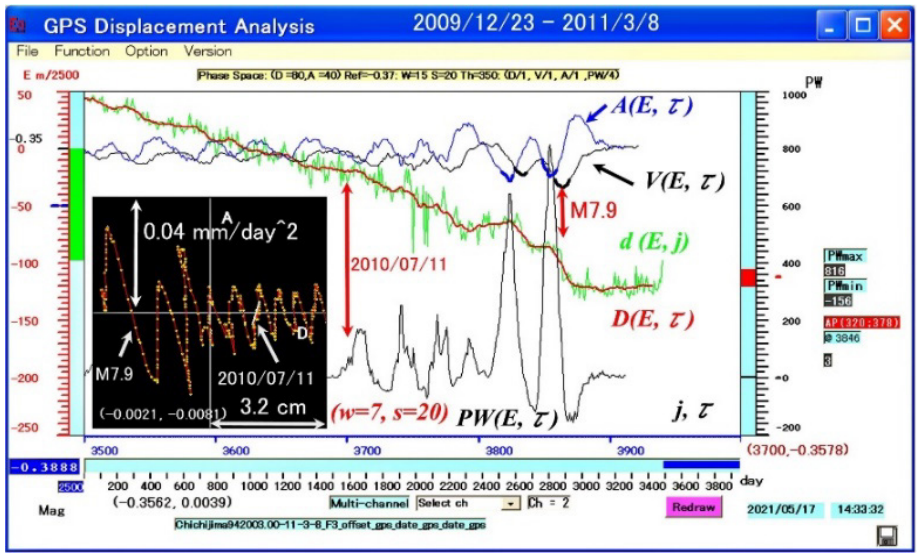

**Fig. 13-2-3g.** $D(E, \tau) - A(E, \tau)$ path with $A(E, \tau) \approx K \times D(E, \tau)$ for Fig. 6-1b.

Figure 13-2-3g shows a linear segment of the $D(E, \tau) - A(E, \tau)$ path at arrow 2010/07/11 as a white line. The linear change is $(\Delta D(E, \tau), \Delta A(E, \tau)) = (-0.0021, -0.0081)$ in (m, mm/day$^2$) from $A(E, \tau) = 0.0039$ mm /day$^2$ at $\tau = 3681$ on June 22, 2010, as shown below the time scale $(-0.3562, 0.0039)$ in (m, mm/day$^2$). Thus, the segment equation $F(E, \tau) \approx A(E, \tau) \approx K \times D(E, \tau)$ has $K = 0.0039$ /day$^2$. The $(3700, -0.3578)$ in $(\tau,$ m) below the $PW$ scale shows $D(E, \tau)$ at the trend change.

The segment in a white line in Fig. 13-2-3g preceded the trend change at label 2010/07/11. It had a positive $K$ in $A(E, \tau) \approx K \times D(E, \tau)$. The positive $K$ shows a weak response of the subducting plate to the lunar synodic tidal force loading, as in .

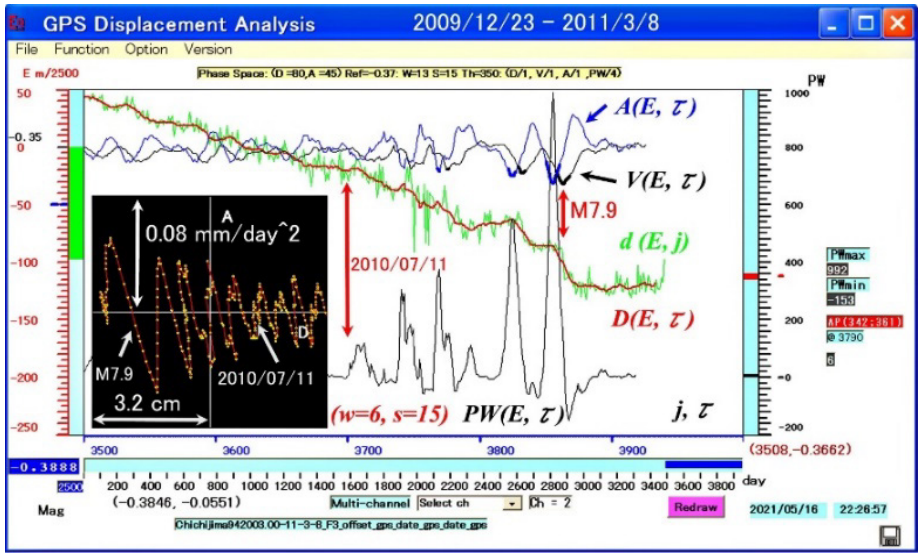

**Fig. 13-2-3h.** *D* (*E*, *τ*) – *A* (*E*, *τ*) path with *w* = 6 and *s* = 15 for Fig. 6-1b.

An insufficient synodic tidal loading on the segment appears on the *D* (*E*, *τ*) – *A* (*E*, *τ*) path with *w* = 6 and *s* = 15, as in Fig. 13-2-3h. Thus, the positive *K* in Fig. 13-2-3g is due to the tidal response constrained by coupling the subducting plate motion with the bulge-bending deformation on the overriding eastern edge (the Tohoku east coast).

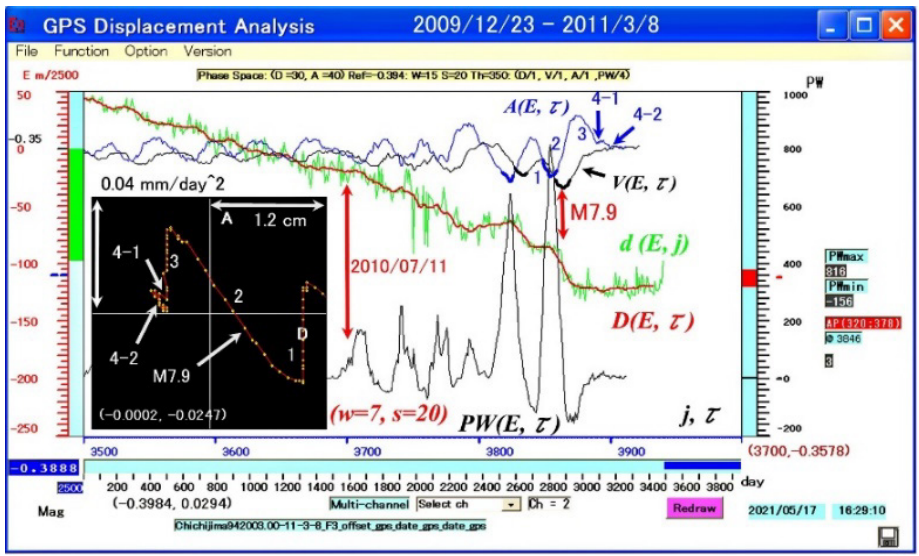

**Fig. 13-2-3i.** A *D* (*E*, *τ*) – *A* (*E*, *τ*) path for a last three-month Tohoku M9 EQ genesis process.

Figure 13-2-3i shows four linear segments of the *D* (*E*, *τ*) – *A* (*E*, *τ*) path for Fig. 6-1b, illustrating the last three-month megathrust EQ genesis process.

**Last three-month M9 EQ genesis process:**

The *A* (*E*, *τ*) has the corresponding indexes with the magnified *D* (*E*, *τ*) – *A* (*E*, *τ*) path. The path segments show the last three-month part of the fifteen-month Tohoku M9 EQ genesis process.

Each segment has the segment index number, starting time $\tau$, ($\Delta D$ ($E$, $\tau$) mm, $\Delta A$ ($E$, $\tau$) mm/day$^2$), and $K$ in $F$ ($E$, $\tau$) ≈ $A$ ($E$, $\tau$) ≈ $K \times D$ ($E$, $\tau$). Segment 4 has two sub-portions, 4-1 and 4-2. Last segment 4-2 ends at $\tau$ = 3913 on 9 Feb 2011 because $j$ = 3940 is Mar 8, 2011, ($\tau$ = 3940 − $w$ − $s$).

| Segment | $\tau$ | Date | ($\Delta D$ ($E$, $\tau$) mm, $\Delta A$ ($E$, $\tau$) mm/day$^2$) | $K$ (1 /day$^2$). |
|---------|--------|------|-------|------|
| 1 | 3844 | Dec 2, 2010 | (0. 1 mm, − 0.0330 mm/day$^2$) | $K$ = − 0.0330 /day$^2$ |
| 2 | 3854 | Dec 12, 2010 | (− 8.7 mm, 0.0434 mm/day$^2$) | $K$ = − 0.0050 /day$^2$. |
| 3 | 3878 | Jan 5, 2011 | (− 0.2 mm, −0.0247 mm/day$^2$) | $K$ = 0.1235 /day$^2$. |
| 4-1 | 3890 | Jan 17, 2011 | (− 1.6 mm, 0.0034 mm/day$^2$) | $K$ = − 0.0213 /day$^2$. |
| 4-2 | 3894 | Jan 21, 2011 | (1.5 mm, − 0.0068 mm/day$^2$) | $K$ = − 0.0453 /day$^2$. |

The oceanic plate has the unusual tidal force coupling of 33 days in segments 1 and 2. Constant $K$ in segment 2 is minimal, indicating no effective tidal force loading on the plate motion. The bulging eastern edge pulled the plate westward with the highest velocity that triggered the M7.9 at $\tau$ = 3864 on Dec 22, 2010, the arrow on the $D(E, \tau) − A(E, \tau)$ path at $A(E, \tau)$ ≈ 0. Segment 3 then shows the abnormally high external force loading with $K$ = 0.1235 /day$^2$ on the standard tidal loading, indicating that the bulge pulling has a sudden deceleration. Segment 4-1 recovers the standard tidal loading for only five days. Segment 4-2 shows the east coast moving eastward by 1.5 mm to prepare the rupturing process of the megathrust EQ, as in Fig. 4c.

### 13.2.4 The northward motion of the subducting Pacific Plate (Chichijima-A station)

Only 3 % of the average Pacific Plate horizontal motion has the northward component {$N$}. The {$N$} has its environmental-noise fluctuation amplitude about a half of {$h$} as shown in Figs. 13-1-3a and 13-1-3b. The observation of M7.9 at the downward (southward) $V$ ($N$, $\tau$) arrowed in Fig. 13-2-4a shows that the 3 % of the lunar synodic southward loading was a part of the normal faulting M7.9 EQ (STR 340º, DIP 57 º, SLIP −56 º) near Chichijima [B7].

The M8.1 (STR 32°, DIP 25°, SLIP −44°) on May 30, 2015, [B7] was on the lunar synodic loading, as arrowed on the $V$ ($N$, $\tau$) in Fig. 13-2-4a and the $V$ ($N$, $\tau$) and the path in Fig.13-2-4b. Its $PW$ ($N$, $\tau$) and $PW$ ($N$, $j$) detection are Fig. 13-2-5a.

However, the other large EQ events' identifications in the lunar synodic loading of $V$ ($N$, $\tau$) and $A$ ($N$, $\tau$) require noise separation path analyses. The paths are on the $V$ ($N$, $\tau$) − $V$ ($E$, $\tau$) and the $V$ ($N$, $\tau$) − $V$ ($h$, $\tau$), and the $A$ ($N$, $\tau$) − $A$ ($E$, $\tau$) and the $A$ ($N$, $\tau$) − $A$ ($h$, $\tau$) planes. Their observations and analyses will be arXiv [B8].

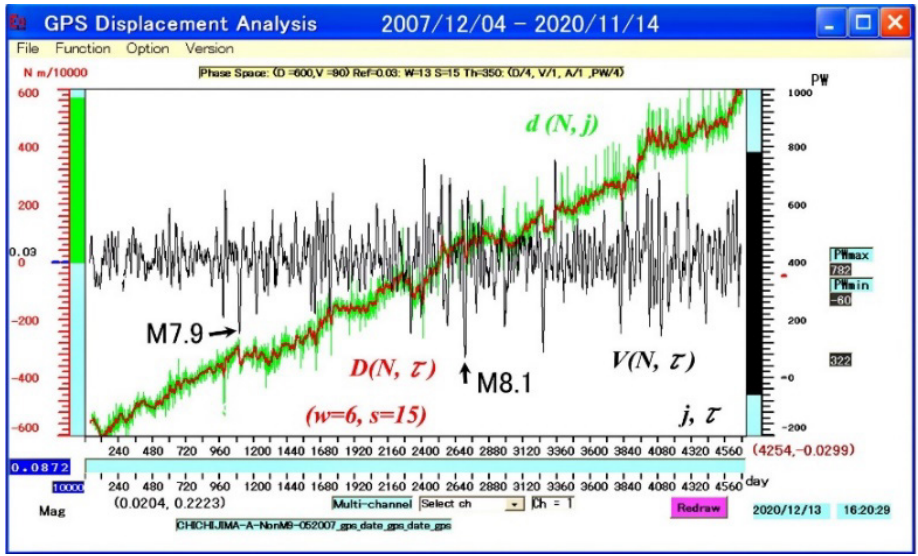

**Fig. 13-2-4a. Chichijima-A from December 4, 2007, to November 14, 2020.**

The $d(N, j)$ excludes the M9 EQ's spike data on March 11, 2011 (between $j = 1174$ and 1175). The drawing follows the eastward displacement $\{E\}$ in Figs. 13-2-2a and 13-2-2b.

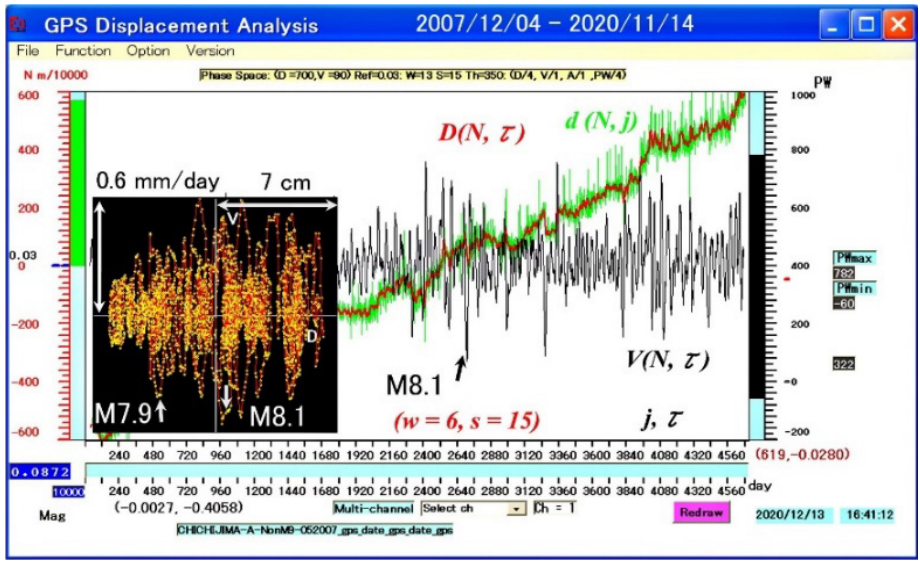

**Fig. 13-2-4b. Chichijima-A's $D(N, \tau) - V(N, \tau)$ path from December 4, 2007, to November 14, 2020.**

Phase Space (D = 700, V = 90) is (7 cm, 0.6 mm/day) by the magnification 10000 for $d(N, j)$. The Ref = 0.03 (in meters) is the reading at the blue line on the left scale of 0. It becomes the offset origin for the $D(N, \tau) - V(N, \tau)$ plane. Parameters W and S are W = 13 = $2w + 1$ ($w = 6$) and S = 15 = $s$. The (D / 4, V / 1, A / 1, PW / 4) is the reduced magnification for the respective drawing.

The path shows the localized northward and southward fluctuating motions whose amplitudes are approximately equal, suggesting the lunar synodic loading on $\{N\}$ dominates the fluctuations. They include sudden northward shifts of about 1 cm before and after the M8.1 event. They appear to be some co-seismic shifts; however, the geophysical origin is unclear. It may be a so-called slow slip event [B3]. Some details are in Figs. 13-2-5b and 13-2-5c for the Pacific Plate's subduction zone, and Figs. 13-3-5a and 13-3-5b for the Philippine Sea Plate's subduction zone.

### 13.2.5 Power monitoring *PW* (*N*, τ) on the northward motion at Chichijima-A

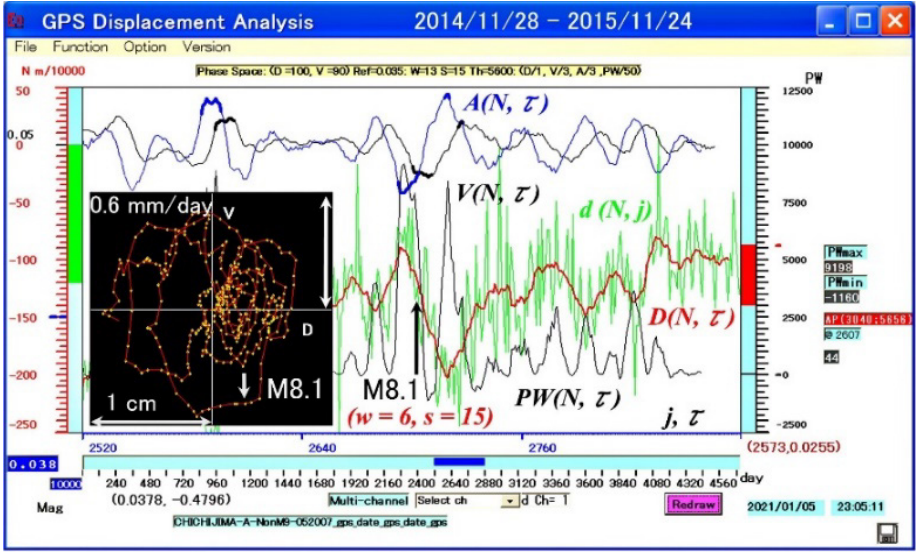

**Fig. 13-2-5a. *PW* (*N*, τ) and M8.1 on May 30, 2015 (*j* = 2703).**

The magnified date window is from November 28, 2014, to November 24, 2015. Phase Space (D = 100, V = 90) is (1 cm, 0.6 mm/day) by the magnification 10000 for *d* (*N*, *j*). The Ref = 0.035 (= 0.05 m − 150 /10000 m) is the reading at the blue line on the left scale − 150 (× m/10000). It becomes the offset origin for the *D* (*N*, τ) − *V* (*N*, τ) plane. The parameters to extract the synodic loading are W = 13 = 2*w* + 1 (*w* = 6) and S = 15 = *s*. The magnification is 10000 for *d* (*N*, *j*), four times *d* (*E*, *j*) so that *V* (*N*, τ) and *A* (*N*, τ) also become four times *V* (*E*, τ) and *A* (*E*, τ). Thus, the *PW* (*N*, τ) threshold to detect an abnormal power (AP) loading in time *j* is sixteen times that of *PW* (*E*, τ)'s 350, which is 5600. The (D / 1, V / 3, A / 3, PW /50) is the reduced magnification for the respective drawing.

The AP (3034; 5656) and @ 2607 is that *PW* (*N*, *j*) detected the first AP changing from 3034 (*j* = 2606) to 5656 (*j* = 2607) on February 23, 2015. The *V* (*N*, τ) and *A* (*N*, τ) are bold for *PW* (*N*, τ) ≥ 5600. The M8.1 event was on May 30, 2015 (at *j* = 2703), at a little off downward (southward) peak of *A* (*N*, τ) dotted-arrowed at τ = 2703. The location on the *D* (*N*, τ) − *V* (*N*, τ) plane was at (0.0378, − 0.4796); namely, the offset location is 0.0378 m, and the southward speed (negative) − 0.4796 mm/day. The triggering motion is in harmony with the M8.1 EQ's normal faulting (STR 32°, DIP 25°, SLIP −44°) [B7]. The first AP detection suggests it was a precursor three months before the M8.1 EQ.

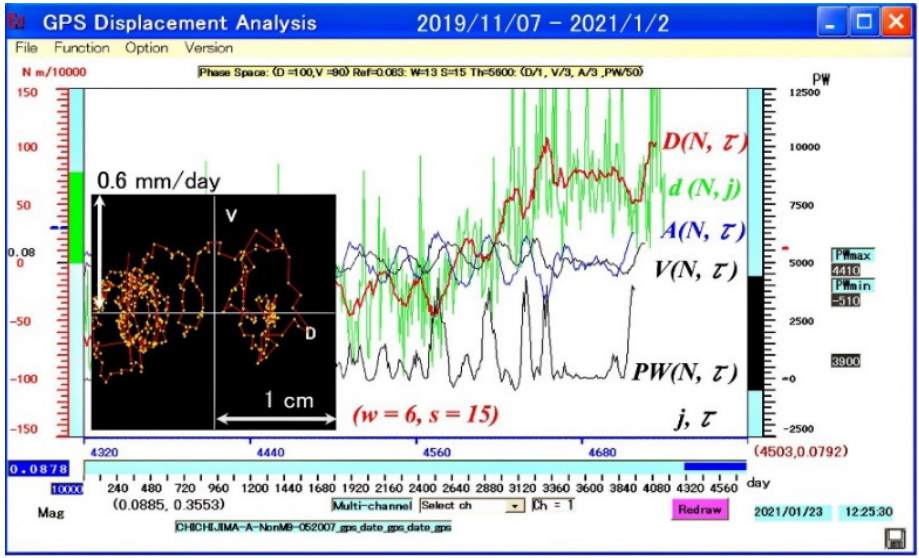

**Fig. 13-2-5b.** *PW* (*N*, τ) from November 7, 2019, to January 2, 2021.

Phase Space (D = 100, V = 90) is (1 cm, 0.6 mm/day) by the magnification 10000 for *d* (*N*, *j*). The Ref = 0.083 (= 0.08 m + 0.03 m) is the reading at the blue line on the left scale 30 (×m/10000). It becomes the offset origin for the *D* (*N*, τ) – *V* (*N*, τ) plane. The parameters to extract the synodic loading are W = 13 = 2*w* + 1 (*w* = 6) and S = 15 = *s*. The magnification is 10000 for *d* (*N*, *j*), four times *d* (*E*, *j*). Thus, the *PW* (*N*, *j*) threshold for detecting the abnormal power (AP) loading in time *j* is 5600, sixteen times *PW* (*E*, *j*)'s 350. The (D / 1, V / 3, A / 3, PW /50) is the reduced magnification for the respective drawing.

The *D* (*N*, τ) – *V* (*N*, τ) path shows the fluctuating motion under the lunar synodic loading on {*N*}. The sudden northward shift of about 1 cm accompanies one synodic oscillation during the jump so that the change is not co-seismic but maybe a so-called slow slip event in the subduction zone [B3]. The *D* (*N*, τ) – *V* (*N*, τ) path with *w* = 15 and *s* = 50 shows the same jumping motion without the synodic oscillation.

As of January 2, 2021, the subducting northwestern Pacific Plate's {*N*} is normal; namely, no anomaly to rupture any imminent megathrust in the subduction zone.

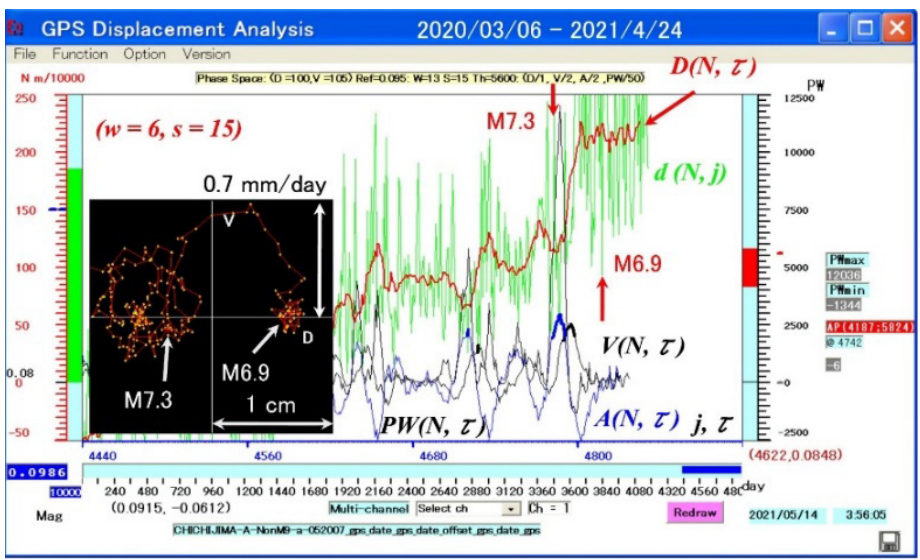

**Fig. 13-2-5c. *PW* (*N*, τ) and M7.3 and M6.9 on March 20, 2021 (*j* = 4817).**

Figure 13-2-5c updates Fig. 13-2-5b with the phase Space (D = 100, V = 105) of (1 cm, 0.7 mm/day) by the magnification 10000 for *d* (*N*, *j*). The Ref = 0.088 (= 0.08 m + 0.08 m) is the reading at the blue line on the left scale 80 (×m/10000). It is the offset origin for the *D* (*N*, τ) – *V* (*N*, τ) plane.

Figure 13-2-5c shows M7.3 at *j* = 4782 (on February 13, 2021), and M6.9 at *j* = 4817 (on March 20, 2021) in the subduction zone. The M7.3 was the reverse faulting of (STR 191º, DIP 55º, SLIP 78º) within the subducted slab off Fukushima [B9], suggesting the abnormal power detected at *j* = 4757 (on January 19, 2021) might have been the precursory stress loading to the slab. The second AP (abnormal power) implies the coupling of M7.3 and M6.9, as in Fig. 13-2-5e.

**13.2.6 Current Eastward and Northward motions at Chichijima-A, as of May 22, 2021**

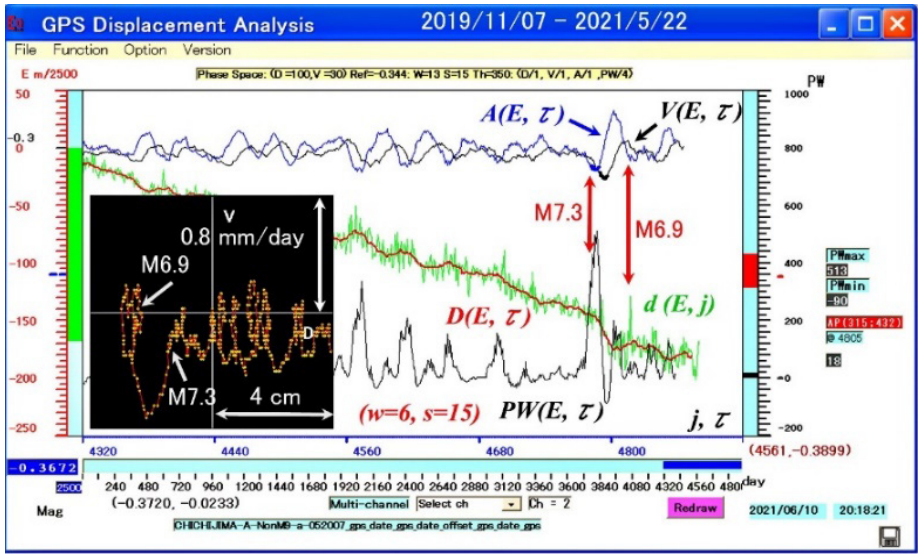

**Fig. 13-2-6a. *PW* (*E*, τ) on the eastward motion as of May 22, 2021.**

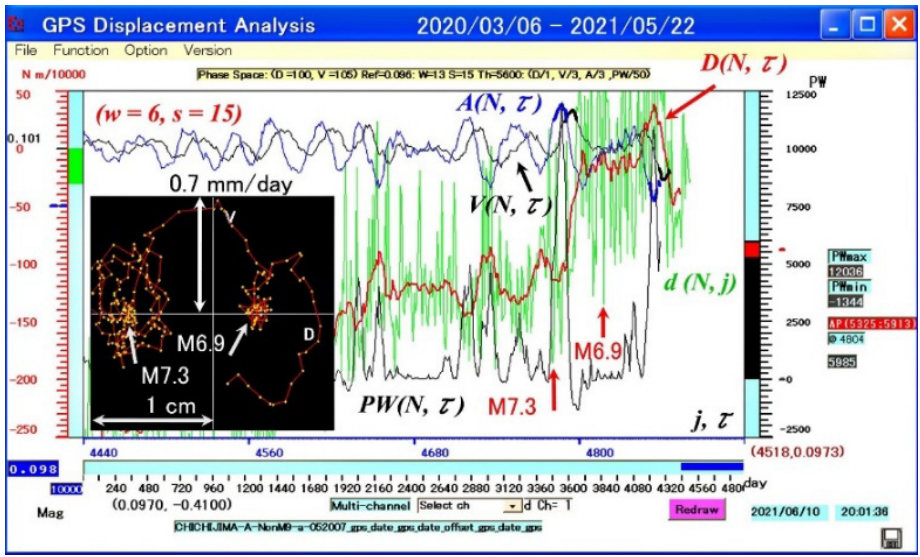

**Fig. 13-2-6b. *PW* (*N*, τ) on the northward motion as of May 22, 2021.**

Figures 13-2-6a and 13-2-6b updated Figs. 13-2-3d and 13-2-5c as of May 22, 2021. Some minor updates are due to the updates by GSI [B4].

As of May 22, 2021, the subducting northwestern Pacific Plate's {$E$} and {$N$} show no imminent megathrust ruptures. The abnormal power in {$N$} has no clear geophysical origin, as discussed in sections 13.2.4, 13.2.5, and 13.3.5.

## 13.3 The Philippine Sea plate

### 13.3.1 The overriding continental Plate (the main island's western part)

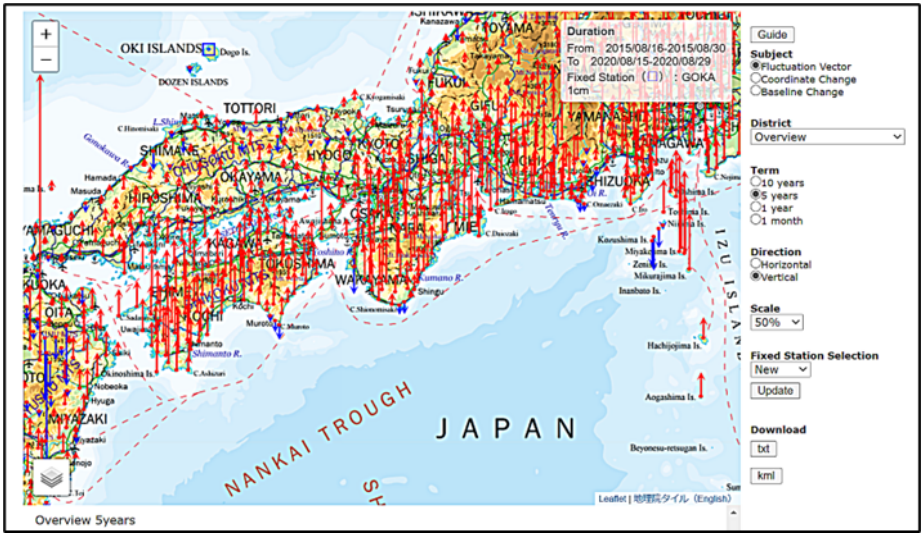

**Fig. 13-3-1. The vertical displacements at the website https://mekira.gsi.go.jp/index.en.html**

As discussed in section 13.1.2, Tohoku's entire eastern shoreline of about 500 km had a downward displacement by coupling the whole 500 km length fault with the subducting northwestern Pacific Plate motion. We may assume a similar fault coupling between the Eurasian Plate's eastern overriding edge (the Amurian Plate) and the Philippine Sea Plate [B2]. The fault coupling is the prerequisite for the anticipated Nankai-Trough megathrust EQ of Mw 9.1 and 34 m-height tsunamis [B6].

, the GPS observation along the eastern shoreline of about 600 km, as in Fig. 13-3-1, shows the scattered subsidence spots along the southern coast of the main island and Shikoku. They are in Tokai, Tonankai, Nankai, and the islands near Kozushima. Three separate independent faults are Tokai (label1), Tonankai (label 2), and Nankai (label 3) in Fig. 9. Thus, even with their simultaneous chain ruptures, without the fault coupling as in the 2011 Tohoku events, generating an enormous action to pull down the Philippine Sea Plate along the entire 600 km long coastline is unlikely. Namely, the anticipated Mw 9.1 EQ and 34 m-height tsunami [B6] will not occur. The non-scientific presumption may come from Japan's cabinet office's inability to prepare for the Tohoku M9 EQ, as in Appendix C.

We present the Philippine Sea Plate's motion. Each scattered fault coupling with the oceanic plate motion will be at arXiv [B8].

### 13.3.2 The westward motion of the Philippine Sea Plate (Minamidaito-Jima station).

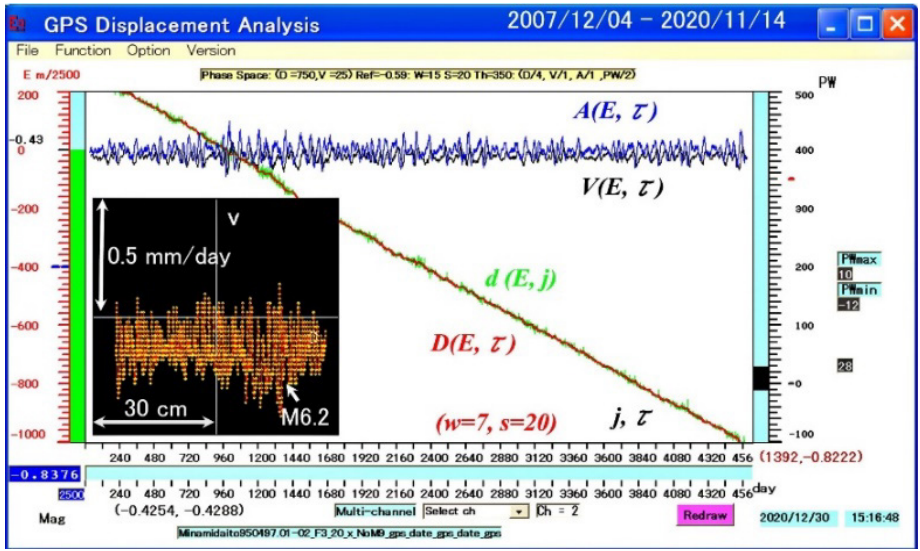

**Fig. 13-3-2. Minamidaito-Jima from December 4, 2007, to November 14, 2020.**

The $d(E, j)$ excludes the M9 EQ's spike data on March 11, 2011 (between $j = 1730$ and 1731). Phase Space (D = 750, V = 25) is (30 cm, 0.5 mm/day) by the magnification 2500 for $d(E, j)$. The Ref = − 0.59 (= − 0.43 m − 0.16 m) is the reading at the left scale − 400 (× m/2500). It is the offset origin, (− 0.59 m, 0 m/day), for the $D(c, \tau) – V(c, \tau)$ plane. The parameters to extract the lunar synodic loading are W = 15 = $2w + 1$ ($w$ = 7) and S = 20 = $s$. The (D / 4, V / 1, A / 1, PW / 2) is the reduced magnification for the respective drawing.

The path shows the unusual westward speed, −0.4288 mm/day, whose action triggered M6.2 EQ on August 22, 2010 ($j = 973$), as detailed in <u>Fig. 13-3-3a</u>.

### 13.3.3 Power monitoring $PW(E, \tau)$ on the westward motion at Minamidaito-Jima

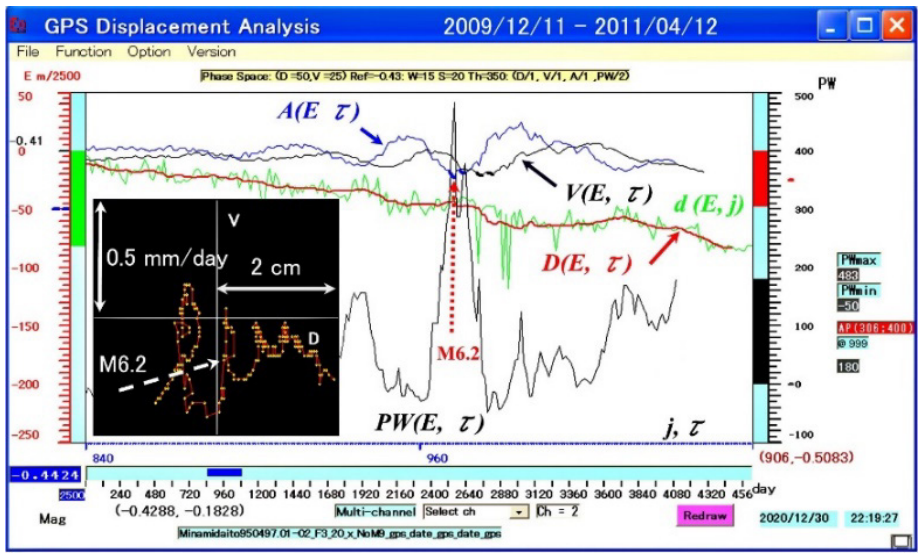

**Fig. 13-3-3a. $PW(E, \tau)$ and M6.2 on August 22, 2010 ($j = 973$).**

Phase Space (D = 50, V = 25) is (2 cm, 0.5 mm/day) by the magnification 2500 for $d(E, j)$. The Ref = − 0.43 (= − 0.41 m − 0.02 m) is the reading at the blue line on the left scale − 50 (× m/2500). It is the offset origin of the $D(E, \tau) - V(E, \tau)$ plane. The parameters to extract the synodic loading are W = 15= $2w + 1$ ($w = 7$) and S = 20 = $s$. The threshold for detecting the abnormal power (AP) loading in time $j$ is Th = 350. The (D / 1, V / 1, A / 1, PW / 2) is the reduced magnification for the respective drawing.

The AP (306; 400) and @ 999 on the right show the detected AP was from 306 (at $j$ = 998) to 400 (at $j$ = 999). The $V(E, \tau)$ and $A(E, \tau)$ become bold under $PW(E, \tau) \geq 350$. The M6.2 EQ ruptured at the abnormal $A(E, \tau)$ downward peak, as arrowed at $\tau = 973$. The M6.2 rupturing location on the path is ($-0.4288$, $-0.1828$), $-0.4288$ m eastward from the offset origin $-0.43$ m, and $V(E, \tau) = -0.1828$ mm/day at $\tau = 973$ (August 22, 2010). The AP shows that the abnormal downward force $A(E, \tau)$ triggered the normal faulting M6.2 EQ (STR 326°, DIP 63°, SLIP $-85°$) at depth 31km, 19°58.6' N, and 147°15.3' E [B7], far below Chichijima in Fig. 3.

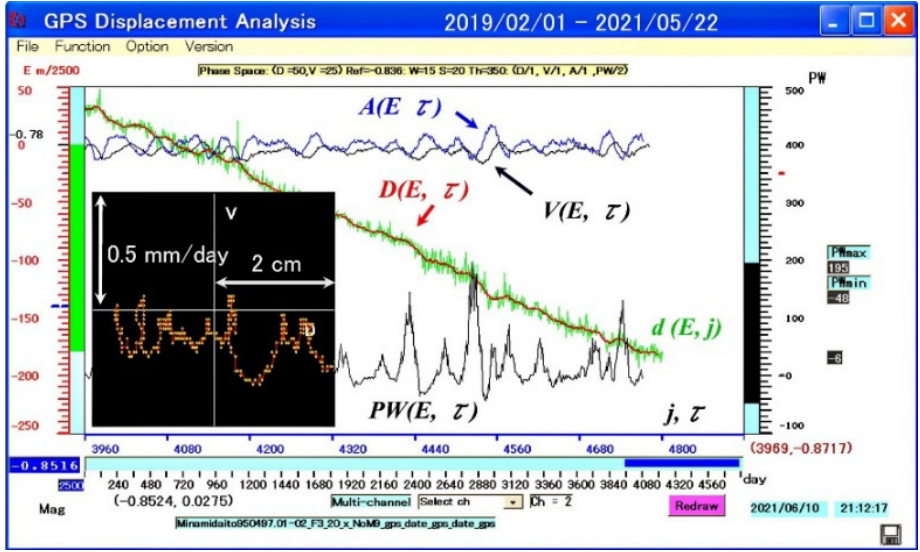

**Fig. 13-3-3b.** $PW(E, \tau)$ from February 1, 2019, to May 22, 2021.

Phase Space (D = 50, V = 25) is (2 cm, 0.5 mm/day) by the magnification 2500 for $d(E, j)$. The Ref = $-0.836$ (= $-0.78$ m $-0.056$ m) is the reading at the blue line on the left scale $-100$ ($\times$ m/2500). It is the offset origin for the $D(E, \tau) - V(E, \tau)$ plane.

The parameters to extract the synodic loading are W = 15 = $2w + 1$ ($w = 7$) and S = 20 = $s$. The threshold for detecting the AP is Th = 350. The (D / 1, V / 1, A / 1, PW / 2) is the reduced magnification for the respective drawing. The present Philippine Sea Plate's {E} is ordinary as of May 22, 2021.

### 13.3.4 The northward motion of the Philippine Sea Plate (Minamidaito-Jima station)

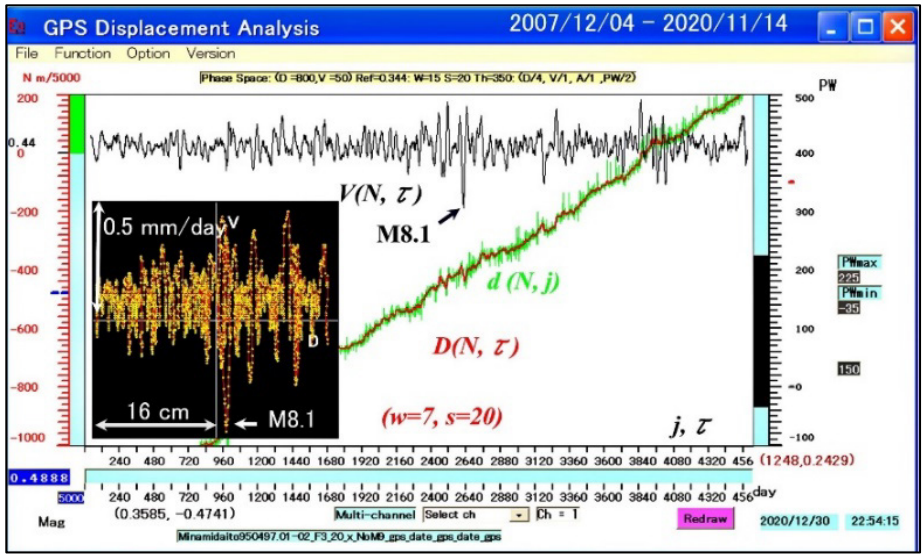

**Fig. 13-3-4. Minamidaito-Jima from December 4, 2007, to November 14, 2020.**

The $d(N, j)$ excludes the M9 EQ's spike data on March 11, 2011 (between $j = 1730$ and 1731). Phase Space (D = 800, V = 50) is (16 cm, 0.5 mm/day) by the magnification 5000 for $d(N, j)$. The Ref = 0.334 (= 0.44 m − 0.096 m) is the reading at the blue line on the left scale − 480 (× m/5000). It is the offset origin for the $D(N, \tau) - V(N, \tau)$ plane. The parameters to extract the synodic loading are W = 15 = 2w + 1 ($w = 7$) and S = 20 = $s$. The (D / 4, V / 1, A / 1, PW / 2) is the reduced magnification for the respective drawing.

The M8.1 EQ on May 30, 2015 ($j = 2618$) [B7] was at $V(N, \tau) = -0.4741$ mm/day, as arrowed M8.1 a little off the downward peak. It is the same M8.1 EQ in Figs. <u>13-2-4a</u>, <u>13-2-4b</u>, and <u>13-2-5a</u>.

The $D(N, \tau) - V(N, \tau)$ path shows the fluctuating motion with sudden northward shifts of about 1 cm, as in <u>Fig. 13-2-4b</u>. The detailed changes before the M8.1 and M6.2 events are in <u>Fig. 13-3-5a</u> and <u>Fig. 13-3-5b</u>, respectively.

### 13.3.5 Power monitoring $PW(N, \tau)$ on the northward motion at Minamidaito-Jima

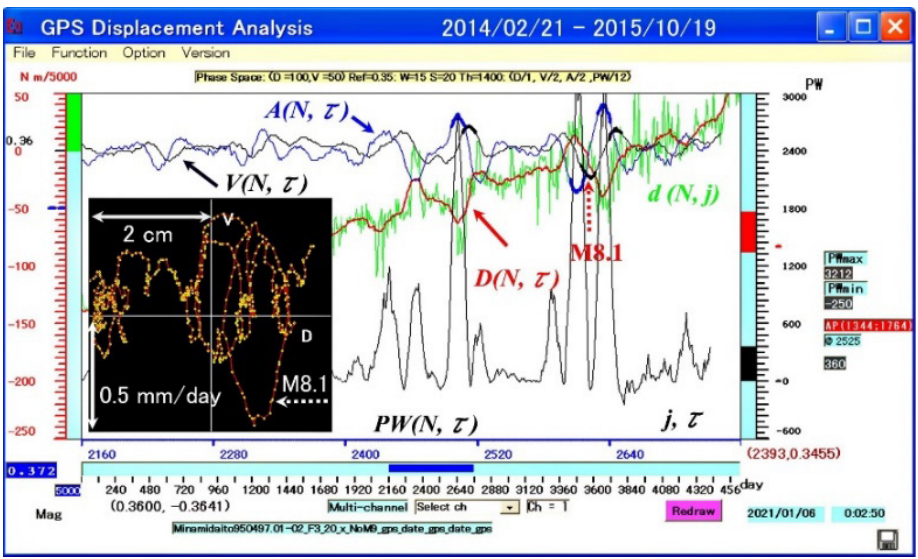

**Fig. 13-3-5a. $PW(N, \tau)$ and M8.1 on May 30, 2015 ($j = 2618$).**

The magnified date window is from February 21, 2014, to October 10, 2015. Phase Space (D = 100, V = 50) is (2 cm, 0.5 mm/day) by the magnification 5000 for $d(N, j)$. The Ref = 0.35 (= 0.36 m − 0.01 m) is the reading at the blue line on the left scale − 50 (× m/5000), which is the $D(N, \tau) - V(N, \tau)$ offset origin. The parameters to extract the synodic loading are W = 15 = 2w + 1 (w = 7) and S = 20 = s. The magnification is 5000 for $d(N, j)$, which is twice $d(E, j)$. Thus, $V(N, \tau)$ and $A(N, \tau)$ also become twice $V(E, \tau)$ and $A(E, \tau)$. The $PW(N, j)$ threshold for detecting the abnormal power (AP) loading in time $j$ is then four times $PW(E, j)$'s 350, which is Th = 1400. The (D / 1, V / 2, A / 2, PW /12) is the reduced magnification for the respective drawing.

The AP (1344; 1764) and @ 2525 is that $PW(N, j)$ detected the first AP changing from 1344 (at $j$ = 2524) to 1764 (at $j$ = 2525) on February 26, 2015. The $V(N, \tau)$ and $A(N, \tau)$ are bold under $PW(N, \tau) \geq 1400$. The M8.1 event was on May 30, 2015 ($j$ = 2618), at a little off the downward (southward) peak of $A(N, \tau)$ dot-arrowed at $\tau$ = 2618. The M8.1 EQ triggering motion is in harmony with that of the northwestern Pacific Plate, generating the normal faulting (STR 32°, DIP 25°, SLIP −44°) [B7].

The $D(N, \tau) - V(N, \tau)$ path shows the sudden northward shift of about 1 cm before the M8.1 event, accompanying a weak synodic oscillation. Thus, the non-co-seismic change may be a so-called slow slip event in the subduction zone [B3].

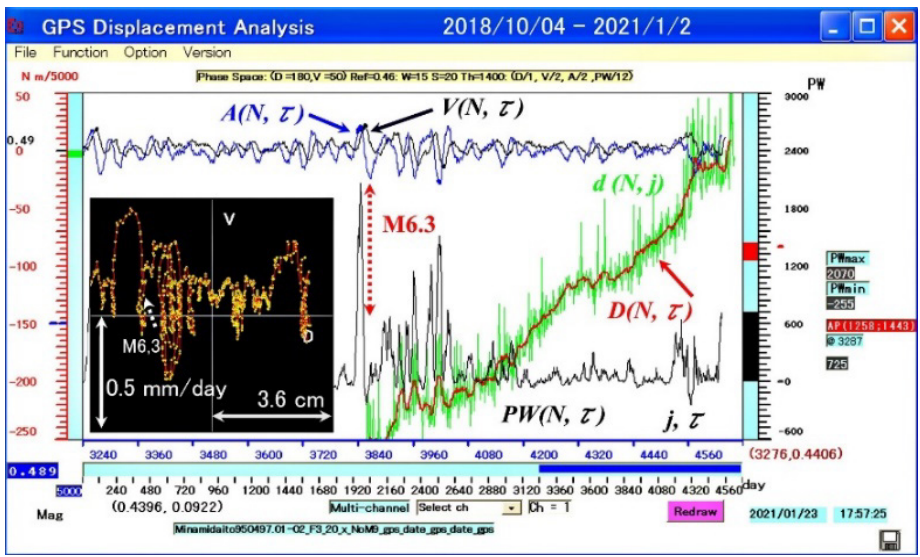

**Fig. 13-3-5b.** *PW* (*N*, τ) and M6.3 on October 24, 2018 (*j* = 3860).

The magnified date window is from October 4, 2018, to January 2, 2021. Phase Space (D = 180, V = 50) is (3.6 cm, 0.5 mm/day) by the magnification 5000 for $d(N, j)$. The Ref = 0.46 (= 0.48 m − 0.02 m) is the reading at the blue line on the left scale − 100 (× m/5000). It is the $D(N, \tau) - V(N, \tau)$ offset origin. Parameters W and S are W = 15 = 2w + 1 (w = 7) and S = 20 = s. The $PW(N, \tau)$ threshold is Th = 1400. The (D / 1, V / 2, A / 2, PW /12) is the reduced magnification for the respective drawing.

The AP (1258; 1520) and @ 3867 is that $PW(N, j)$ detected the abnormal power (AP) changing from 1258 (at $j$ = 3866) to 1520 (at $j$ = 3867) on October 31, 2018. The $V(N, \tau)$ and $A(N, \tau)$ are bold for $PW(N, \tau)$ ≥ 1400. The M6.3 event was on October 24, 2018 ($j$ = 3860), after the upward (northward) peak of $V(N, \tau)$ dotted-arrowed at $\tau$ = 3860. The M6.3 EQ was at a depth of 28 km, 23°58.1'N, and 122°36.1'E (North West off Ishigakijima Island) with the reverse faulting (STR 50°, DIP 77°, SLIP 41°) [B7]. A foreshock of M6.1 to M6.3 was on October 23, 2018 [B7]. The $D(N, \tau) - V(N, \tau)$ path shows the sudden northward shift of about 1 cm before the M6.3 event, with the lunar synodic oscillation. Thus, the non-co-seismic change may be a so-called slow slip event in the subduction zone [B3].

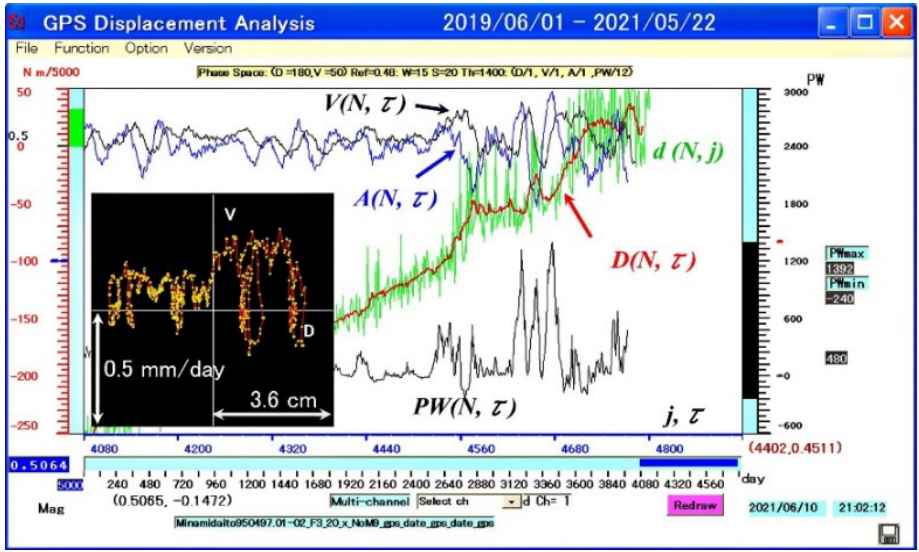

**Fig. 13-3-5c.** *PW* (*N*, *τ*) from June 1, 2019, to May 22, 2021.

Figure 13-3-5c is the update of Fig. 13-3-5b, which shows the present Philippine Sea Plate's {*N*}, as of May 22, 2021, is normal. No abnormal motion precursory to the imminent megathrust ruptures in the subduction zone.

## 13.4 Imminent megathrust events in Japan's subduction zones as of 11 March 2023

As of March 11, 2023, there are currently no signs of an imminent megathrust event based on observations of the Pacific Plate at the Chichijima-A and Hahajima GPS stations, as well as the Philippine Sea Plate at the Minamidaito-Jima GPS station [B11].

## 13.5 References on Appendix B

## 14 Appendix C (Issues in media and Japan's cabinet office)

The handling of the 2011 Tohoku M9 EQ by the media and government has raised questions about Japan's disaster preparedness and the role of reliable information in informing the public. The issues are related to the media's questionable role in the 2011 Tohoku M9 EQ and Japan's new disaster countermeasures, which can be summarized as follows:

Seismologists in Japan had models for M8 class EQs off the coast of Tohoku, but they were unable to anticipate the M9 EQ that occurred on March 11, 2011 (Fig. 3 and label 6 in Fig. 9). As a result, the Japanese government did not have any disaster countermeasures against the M9 EQ and the subsequent tsunamis that followed. Instead, they had measures in place for only M8 class events and tsunamis off the east coast of Tohoku.

On March 9, 2011, Japan had an M7.5 event in the anticipated area, one of the large foreshocks to the March 11 M9 EQ. This event caused great anxiety about the imminent M8 throughout the country. Therefore, the country and the news media needed a reliable seismological opinion on the M7.5 concerning the anticipated M8 event. Japan's national public broadcasting organization, NHK, asked only the earthquake research institute's public relations at the University of Tokyo for a seismological opinion under the obligation to broadcast emergency reporting. The view presented by the institute was based on a limited model of asperities and did not account for all the complex factors of the situation. However, NHK broadcasted the institution's view on the prime-time national evening news on March 9 as if the opinion were absolute. The public opinion was that the M7.5 event was unrelated to the anticipated M8 class event. As a result, the news relieved the great anxiety from many communities prepared for the imminent M8 event and tsunami.

Japan had been anticipating three other repeating M8 class events along the Nankai trough [C1]. Their past independent schematic fault lines are labeled 1, 2, and 3 along the green Nankai trough in Fig. 9. Similarly labeled are other significant fault lines, including those off the east coast of Hokkaido.

Seismologists and Japan's cabinet office could not anticipate the Tohoku M9 EQ. They now assume that the simultaneous chain ruptures of three independent faults and their subsequent 600 km length fault (Mw 9.1) could generate a tsunami of 34 m in height [C1] without any scientific confirmation. Japan's cabinet office has laid out disaster countermeasures against the assumed Mw 9.1 events, stirring unprecedented anxiety among many communities in Japan's western part. The concern includes an assumed 4 m height tsunami onto the city of Osaka [C2].

### 14.1 References on Appendix C